\documentclass[aps,prmat,preprint,groupedaddress]{revtex4-2}
\usepackage{color}
\usepackage{lineno}
\usepackage{graphicx}
\usepackage[english]{babel}
\begin{document}

%\preprint{}

\title{A Multiscale Investigation of the Physical Origins of Tension–Compression Asymmetry in Crystals and their Implications for Cyclic Behavior}


% \author{}
%\affiliation{}
%\thanks{}
\author{Sylvain Queyreau}
\affiliation{Université Sorbonne Paris Nord, LSPM-CNRS, UPR 3407, 93430 Villetaneuse, France}
\author{Benoit Devincre}
\affiliation{Université Paris-Saclay, LEM, CNRS - ONERA, F-92322 Chatillon, France}


\date{December 25$^{th}$ 2020}

% add 135 on DD results
% check eps and sig BE values from exp
% add abc letters in figures
% add error bars in CP model predictions
% add ref quey and PRL ref
% V check context asym in tau sat comp/tens in exp
% abstract
% V parenthesis eq
% more details fig sat comp exp
% V see how mughi present link cyclic def/fatigue
% see review Li, clarify domains of patterns
% strange no log corr on self and include rho i on logcor and storage Khkl, so 2 storing???
% check units fig 10a
% usage of <alpha> bar and a

\begin{abstract}
{Most of crystalline materials develop an hysteresis on their deformation curve when a mechanical loading is applied in alternating directions. This effect, also known as the Bauschinger effect, is intimately related to the reversibile part of the plastic deformation and controls the materials damage and ultimately their failure. In the present work, we associate mesoscale Dislocation Dynamics simulations and Finite Element simulations to identify two original dislocation mechanisms at the origin of the traction/compression asymmetry and quantify their impacts on the cyclic behaviour of FCC single-crystals. After demonstrating that no long-range internal stresses can be measured in the simulations, careful analysis of the dislocation network show that the Bauschinger effect is caused by an asymmetry in the stability of junctions formed from segments whose curvature is determined by the applied stress, and a significant portion of the stored dislocation segments is easily recovered during the backward motion of dislocations in previously explored regions of the crystal. These mechanisms are incorporated into a modified crystal plasticity framework with few parameters quantified from statistical analysis of Dislocation Dynamics simulations or from the literature. This strategy has a real predictive capability and the macroscale results are in good agreement with most of the experimental literature existing on the Bauschinger and cyclic deformation of FCC single-crystals. This work provides valuable mechanistic insight to assist in the interpretation of experiments and the design of structural components to consolidate their life under cyclic loading.}
\end{abstract}

%\keywords{}

\maketitle

\section{Introduction}

The history of past imposed solicitations dramatically affects the mechanical response of materials. This can be seen at the macroscale by a hysteresis behavior developing on the deformation curves when alternating the loading direction \cite{Bauschinger:1886, Brown:1971, Argon:2008}. The onset of plastic flow is typically well rounded on a transient, and the flow stress is often reduced compared to the end of the previous deformation stage. This tension-compression asymmetry - also known as the Bauschinger effect (BE) - exists in a multitude of crystalline materials ranging from pure and alloyed systems \cite{Brown:1971} to single \cite{Buckley:1956, Pedersen:1981,Levine:2019} and poly-crystals \cite{Buckley:1956}. Since the materials can undergo much more cumulative deformation at reduced stress levels under alternating deformation than in monotonous deformation modes, part of the plastic deformation appears to be reversible. Understanding the origins of the tension-compression asymmetry and the associated reversibility thus becomes crucial for understanding cyclic deformations and predicting the lifetime of materials for various current applications ranging from microelectronic devices to biocompatible implants and monumental structural frames of buildings \cite{Kubin:2013fk}. The tension-compression asymmetry is therefore associated with crucial technological, economic and safety issues. The modeling of the BE and cyclic behavior is a long-standing problem and a physics-based model with real predictive capability has yet to be formulated.

Dislocation dynamics and their interactions among themselves or with other crystalline defects control the plasticity of the crystalline materials and lead to the formation of organised microstructures. Among the various strengthening mechanisms, dislocation-dislocation interactions play a central role as they represent the totality of the strain hardening in pure systems, and will always remain operative even in the presence of other strengthening mechanisms. Somewhat counter-intuitively, traction-compression asymmetry is arguably the most confusing in the case of single crystals. Few experimental investigations on FCC single crystals have been attempted~\cite{Buckley:1956,Wadsworth:1963,Marukawa:1971, Yakou:1985,Pedersen:1981,Hasegawa:1982, Hsu:1984,Nasu:1984,Ebener:1991, Levine:2019}, the BE is noticeable even after very small prestrain, and  increases significantly with the amount of deformation. Most studies focus on single  \cite{Buckley:1956, Wadsworth:1963, Marukawa:1971,Hsu:1984,Yakou:1985} or double slip loading directions \cite{Buckley:1956,Wadsworth:1963,Marukawa:1971,Pedersen:1981}, but little is known about the high symmetry directions such as [001] or [111] \cite{Yakou:1985, Levine:2019}. The latter are stable multislip conditions that are essential to understanding the loading conditions found in grains of polycrystals. Besides, for seemingly identical conditions, the intensity of the observed BE may differ by half an order of magnitude from study to study, which demonstrates a very strong impact of experimental conditions and initial microstructures (which is not uncommon for single-crystal studies). The spreading of results found for BE experiments contrasts with the very reproducible  behavior observed on single-crystals deformed cyclically \cite{Cheng:1981, Mughrabi:1978,Lepisto:1986,Bretschneider:1997, Gong:1997,Li:2009, Li:2011}. The saturation stress, in particular, observed under large number of alternating deformation cycles is rather well defined for a given material and several independent teams have obtained consistent results in studies separated by several decades \cite{Li:2011}. The saturation stress appears thus to be entirely controlled by the statistical average of elementary mechanisms, rather than by more classical impacting parameters of the plastic behaviour such as initial microstructure and initial dislocation densities, or crystal purity.

Two main explanations of BE have been proposed in the literature. The first one is associated to the partial dissolution during reverse deformation of the microstructures formed during the first deformation stage \cite{Buckley:1956, Sleeswyk:1978, Depres:2008fk, Rauch:2011kx}. However, in the case of single-crystals of pure metals, there is neither  clear elementary dislocation mechanism to substantiate this proposition, nor  direct evidence of what happens within the dislocation microstructure from, say, in-situ observations in a TEM. What is more, a convincing explanation should be able to clearly justify the origins of the transient regime and the reversibility observed and provide quantitatve values. At the small scale, the motion of dislocation is not continuous but operates though intermittent strain bursts akin to avalanche phenomena \cite{Miguel:2006ly, Dimiduk:2006, Csikor:2007lr, Devincre:2008lr}. And these aspects have never been discussed in the context of the BE.

The second major explanation of the BE is found in the so-called composite model \cite{Asaro:1975eu, Mughrabi:88, Argon:2008}. According to this model, the deformation microstructures are interpreted into a hard phase (walls made of large quantities of immobile dislocations) and a soft phase (cell interiors with a lower density of mostly mobile dislocations). Dislocations accumulate on each sides of walls depending on their sign. In turn, the microstructure becomes polarized leading to resistive Long Range Internal Stresses (LRIS) or backstresses, which may assist dislocations during their reverse motion. While, there is no doubt that dislocation patterns can be found during cyclic deformation or in a Bauschinger experiments at large deformation, this is much less clear in the case of small to intermediate strains, for which patterns are not yet fully formed. For this deformation range, patterns have never been really observed in mesoscale simulations, and actual patterns may well be very different from the simplified picture of the composite model, which thus cannot constitute the BE explanation. Finally, no LRIS was found in recent Xray micro diffraction experiments at intermediate deformation \cite{Kassner:2009fk, Kassner:2013fk}.

In this work, we investigated the BE and cyclic deformation of single crystals of FCC Cu and Ni metals using multiscale simulations. Such simple systems constitute well defined and universal configurations. The associated dislocation-dislocation interactions control the plastic flow in pure systems and polycrystals with large grain sizes, and will still be present when other mechanisms are present \cite{Kubin:2013fk, Queyreau:2010}. To meet this goal, we employ Discrete Dislocation Dynamics (DDD) simulations presented in \S\ref{sec:Metho}, which is to date one of the few techniques that can relate directly realistic 3D microstructures of dislocations to the mechanical state in the crystal and the flow stress. We carried simulations of Bauschinger experiments covering a comprehensive set of conditions in terms of prestrain and loading directions (\S\ref{sec:DDDsim}). We then show that the strong BE obtained cannot be explained by backstresses that are absent from our simulations (\S\ref{sec:NoLRstress}). We then propose statistical analyses of the DDD simulations to unravel two original causes to the BE (\S\ref{sec:Identif}). From these analyses, we proposed a modified Crystal Plasticity (CP) formalism at the macroscale (\S\ref{sec:CPform}). We employ this CP model to clarify trends obtained on Bauschinger experiments through a comprehensive comparison with experimental data (\S\ref{sec:BEmacro}). Finally we show that these new elementary mechanisms are sufficient to reproduce quantitatively most of the features observed experimentally during the cyclic deformation of FCC single crystals (\S\ref{sec:cyclic}).

\section{Methodology}
\label{sec:Metho}

We simulate plastic deformation at the mesoscale using Discrete Dislocation Dynamics (DDD) simulations. We employ the microMegas DDD code \cite{Devincre:2011fk} whose parallel implementation allows for large-scale simulations with few $10^5$ segments to reproduce realistic dislocation microstructures in a large crystal volume. Dislocation lines are topologically modelled in the code with chains of interconnected segments whose orientation and length are defined on a simulation lattice that is homothetic to the real crystal lattice. The use of periodic boundary conditions allows the balancing of the net flux of dislocations across simulation boundaries and allows the use of analytical closed forms solutions for the dislocation stress field solutions of the mechanical problem in an infinite elastic isotropic media \cite{HiLo:92}. Dislocation motion is a reaction to a glide force $\mathbf{F}_g$ calculated from various contributions. i) The applied stress field drives plastic deformation; ii) the stress field of all other dislocation segments, which leads to dislocation-dislocation long-range interactions; iii) a selfstress contribution, which is a reaction force accounting for the energy variation induced by the change of curvature and length of the segment.

\begin{table}[!ht]
\caption{Material properties used in the DDD simulations.}
\begin{center}
\begin{tabular}{ c c c c c c  } 
 \hline
 \textbf{Material} & $\mathbf{\mu}$ (GPa) & $\mathbf{\nu}$ & $\mathbf{b}$ (nm) & $\mathbf{B}$ (Pa.s$^{-1}$) & $\mathbf{\tau_{III}}$ (MPa) \\
 \hline
 Ni & 94.7 & 0.276 & 0.249 & 5.5 10$^{-5}$ & 67.6 \\
 Cu & 42 & 0.431 & 0.256 & 5.5 10$^{-5}$ & 30 \\
 \hline
\end{tabular}
\end{center}
\label{tab:paramDDD}
\end{table}

Simulations of dislocation dynamics also involve mobility laws that relate the force $\mathbf{F}_g$ on the dislocation segments to their velocity $v$, which can be determined through atomistic simulations \cite{Domain:2005, Gilbert:2011, Queyreau:2011}. In FCC metals at temperatures T close to room temperature, dislocation mobility is mainly controlled by a small lattice friction force ($\mathbf{F}_f$) and phonon drag. Hence, the segment velocity law used in the present study takes the simple form, $v = ( \left| \mathbf{F}_g - \mathbf{F}_f \right|   b^2 l  ) /B  ( T  )$ where B(T) is an isotropic damping coefficient, $l$ is the segment length and $b$ the Burgers vector magnitude ($c.f.$ Table \ref{tab:paramDDD}). As the effective mass of dislocations is extremely low, inertia effects can be neglected and segment dynamics are solved with a finite difference algorithm. The updating of segment positions at each simulation time step is the occasion to detect potential collisions between dislocation lines that may be associated with reactions like junction formations or dislocation annihilations. In addition, DDD simulations also account for the dislocation cross-slip phenomenon, i.e., the ability of some dislocations with lines parallel to their Burgers vector to change their slip plane. This mechanism is a thermally activated process, which is accounted for in the simulation code through a Monte Carlo algorithm. With this approach, we assume that the probability of a cross-slip event follows an Arrhenius equation. The cross-slip activation energy $\Delta G_{CS}$ is evaluated for Ni and Cu following a procedure first defined in \cite{kubin92} and involving a critical stress amplitude $\tau_{III}$ ($c.f.$ Table \ref{tab:paramDDD}). {Since cross-slip contributions to dislocation dynamics are found to weakly affect the conclusions of this study, details regarding cross-slip calibration are not recalled and can be found elsewhere \cite{kubin92} }. Typical outputs of DDD simulations are the flow stress $(\tau)$ required to deform the simulated volume at a given strain rate, the evolution of dislocation density $(\rho)$ and the shear stain distribution on slip systems that are provided in correlation with the three-dimensional arrangement of the dislocation network.

Great care has been paid to the set-up of simulations. Since dislocation patterns and mean free path are known to be proportional to the mean spacing between dislocations \cite{Kocks:03, Zaiser:2014fk} $\lambda  \approx 1 / \sqrt{\rho}$, the dislocation discretization length and the linear dimensions of the simulated volumes have been systematically scaled with respect to $\lambda$. In all the reported simulations, the dislocation line discretization length and the simulated volume equal to $\lambda / 5$ and $(13 \times 14.4 \times 17.2 \times \lambda^3)$, respectively. Such scaling laws give for instance a simulated volume equal to $6.5 \times 7.2 \times 8.6 \: \mu\mathrm{m}^3$  for simulations with a dislocation density about $4 \; 10^{-2} \; \mathrm{m}^{-2}$ at $1 \%$ strain. Deformation is performed under a constant strain rate control that corresponds to an average velocity lower than $1 \; \mathrm{ms}^{-1}$ of mobile dislocations. This corresponds for instance to an imposed strain rate of $20 \; \mathrm{s}^{-1}$ in the simulated volume defined above.

\section{DDD simulations of the Bauschinger test}
\label{sec:DDDsim}

First, we present large scale DDD simulations of the Bauschinger test, and demonstrate that they capture most of the qualitative features and quantitative values of the BE observed experimentally. We performed a comprehensive study of the tension-compression asymmetry as a function of prestrain and single-crystal orientation, which were identified as key parameters of the BE. 

One single evaluation of the BE requires a set of three different successful large scale simulations. i) A first simulation for the prestrain, chosen in tension, is performed to build the forward dislocation microstructure. For deformation directions corresponding to stable multislip conditions, plastic activity is expected to be evenly distributed over the possible slip systems during prestrain. However, due to the micrometric dimensions of the simulation boxes and the intermittent nature of the dislocation movement, the plastic activity may not be the same on the different slip systems. This is why several initial microstructures consisting of prismatic loops have been considered for all orientations, ranging from a number of 5 to 10 (the [111] orientation proved to be the most unstable). The 'best' prestrain simulations were then selected as those associated with the most evenly distributed plastic activity over the possible slip systems. Under those conditions, the simulations reproduced the  hardening and dislocation storage rates expected from experiments for the orientation considered. The final microstructure at the end of the prestrain was then carefully relaxed to avoid missing any microplastic effects or artificially destroying/promoting metastable dislocation configurations. Two additional simulations were performed starting from that relaxed microstructure (i.e. applied stress $\sigma_{app} = 0 $). ii) A second simulation was performed in compression (or \emph{backward} direction) to observe the BE and iii) a third simulation was performed in continued tension (or \emph{forward} direction). This last curve, starting from the same accumulated strain and relaxed microstructure as ii), thus constitutes a clear reference to observe and to quantify the BE. Overall, the present manuscript covers about 20 successful large scale simulations (and required an additional 15 different unreported prestrain simulations), they were carried over a wide range of configurations with various amounts of prestrain $\gamma_{pre}$ and loading directions corresponding to single slip: [135] (with $\gamma_{pre}$ = 1 \%), and to stable multislip conditions: [112] (with $\gamma_{pre}$ = 0.25; 1 \%), [111]  ($\gamma_{pre}$ = 1 \%) and [001] ($\gamma_{pre}$ = 0.25; 0.5; 0.75; 1; 8 \%).

\begin{figure}[htbp]
\includegraphics[width = 0.24\linewidth]{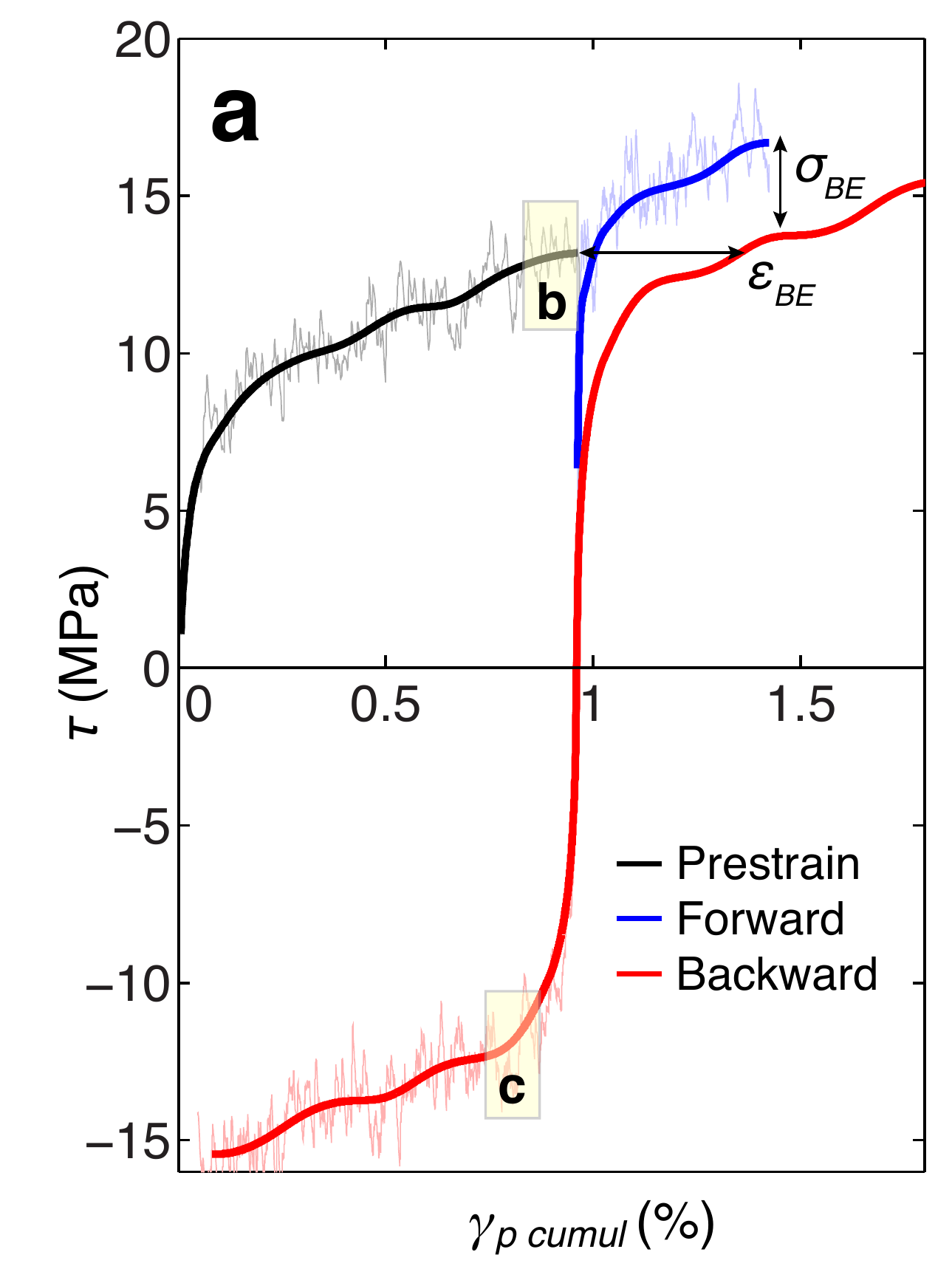}
\includegraphics[width = 0.71\linewidth]{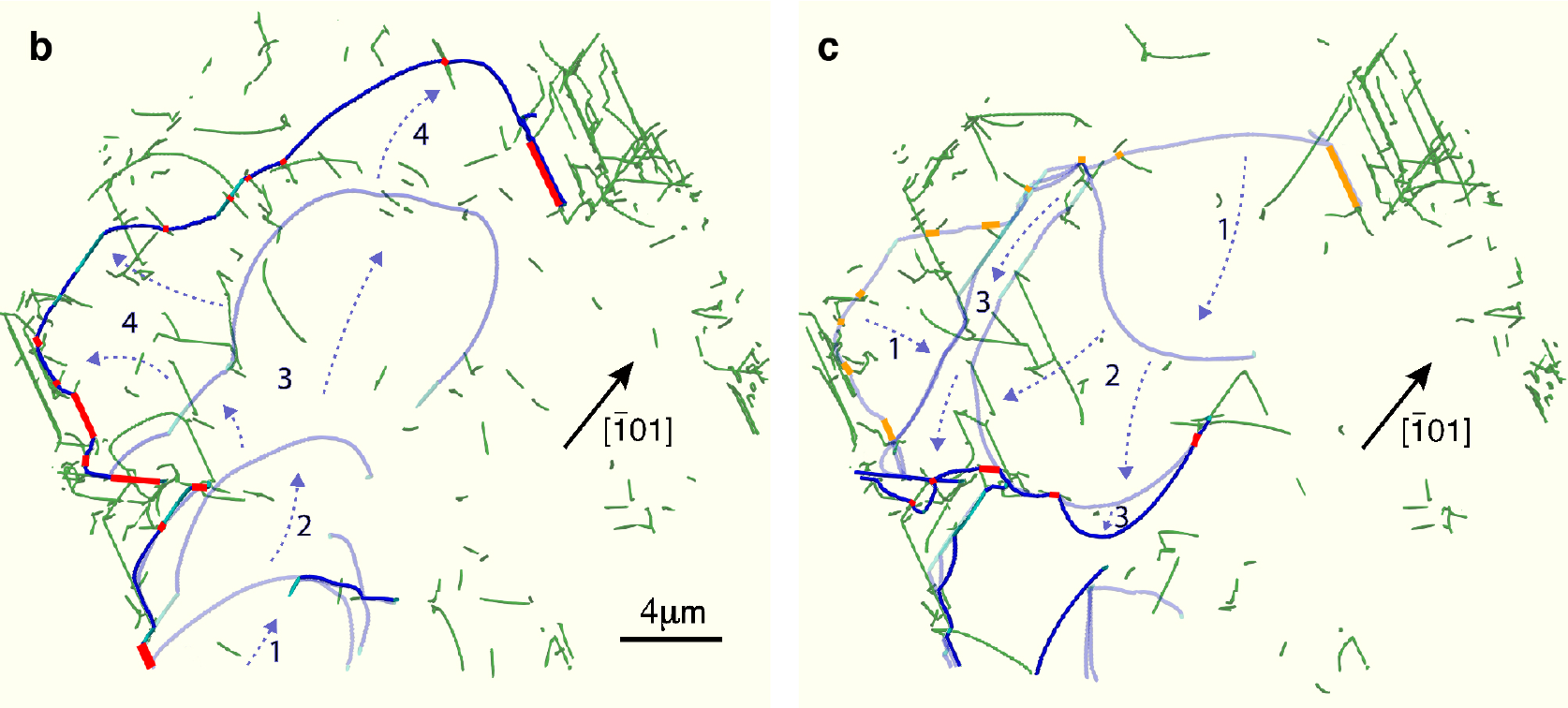}
\caption{Examples of hysteresis curve and dislocation microstructures obtained during a DDD simulation of a Bauschinger test of a Ni $[001]$ single crystal. \textbf{a}, Hysteresis behavior obtained in the flow stress as a function of cumulated plastic deformation. A first predeformation is applied in tension (in black) and then is followed by either a deformation in compression (in red) or in continued tension (in blue). Thick lines correspond to the average behavior calculated from a Gauss kernel smoothing procedure. Two typical BE parameters are also defined in the curve as the Bauschinger stress $\sigma_{BE}$ and strain $\epsilon_{BE}$. \textbf{b}, A plastic avalanche is triggered by the destruction -unzipping- of a junction at the end of the prestrain ($0.8-1\%$). This sequences corresponds to periodic snapshots of the dislocation microstructure contained in a slice of the simulated domain ($0.1\;\mu \mathrm{m}$ thick) and corresponding to the mobile dislocation systems (in blue). These segments may react and lead to binary junctions (thick red lines) with other ‘forest’ systems (in green). \textbf{c}, This second sequence corresponds to the the same region as in b) observed at the beginning of compression ($1 - 0.8\%$). The destruction of weak junctions allow remobilizing long segments of primary dislocations, and is associated to the subsequent destruction of many other junctions (in orange). The mobile segments mostly revisit regions that have been explored during prestrain. {See movies in Supplementary data.} }
\label{fig:1}
\end{figure}

DDD simulations are concerned with Ni and Cu. The motivation for focusing on these two materials is threefold. First, they constitute model materials for other fcc materials with intermediate values of stacking fault energies leading to dissociation of dislocations into Shockley partials (while not accounted for here). Second, a number of experimental investigations exist in the experimental literature regarding the cyclic deformation of these materials in the case of single crystals. Third, forest hardening associated to dislocation interactions scales with the shear modulus. Ni in particular has a relatively large shear modulus $\mu = 94.7$ GPa, forest hardening can thus be clearly seen on DDD deformation curves and this will allow for an unambiguous evaluation of the BE (in particular the BE stress parameter $\sigma_{BE}$ defined a little latter).

Fig.~\ref{fig:1}a displays a typical stress-strain curve obtained in Ni during a Bauschinger test simulation. The loading direction corresponds to [001], which exhibits the largest BE among the chosen directions. Let's focus on the forward tension behavior for now. For this high symmetry axis, at least four non-collinear slip systems are activated simultaneously and are associated to different Burgers vectors. Their interactions may lead to the formation of binary junctions that effectively minimize the overall elastic strain energy but lock mobile dislocations. The plastic flow is thus controlled by the junction destruction and is associated to an initially linear hardening with a slope close to $\mu/150$ in good agreement with experiments \cite{Kubin:2013fk}. Serrations are clearly present on all the simulated deformation curves and these can be attributed to strain bursts and dislocation avalanches. This is confirmed by the simulation sequence in Fig.~\ref{fig:1}b extracted at the end of the prestrain, where intermittent dislocation motion is observed. A strain burst typically starts from a destruction of a junction through the lateral motion -unzipping- of the mobile segment connected to the junction under the effect of the external loading \cite{Bulatov:1998}, corresponding to a stress peak on the deformation curves. This first event triggers a cascade of other junction destructions and strain production leading to a decrease in the stress applied (in a strain rate controlled simulation). Strain bursts stop when the mobile segment becomes pinned again by the collision with the dislocation microstructure and the formation of new junctions. These strain bursts are characterized by the so-called scale-free power law distribution \cite{Csikor:2007lr, Devincre:2008lr}. Fig.~\ref{fig:1}b also illustrates the large amount of immobile dislocation segments stored when dislocations become blocked and are typically left at the edge of the area swept by mobile dislocations. This mechanism is at the origin of dislocation density increase and material strengthening.

\begin{table}[!ht]
\caption{{Estimates of the BE intensity observed in DDD simulations. Error values are estimated from the fluctuations observed on the DDD deformation curves. Experimental ${\sigma_{BE}}$ values from the literature are converted for Ni using the ratios of shear modulii.}}
\begin{center}
\begin{tabular}{ c c c c } 
 \hline
\textbf{Axis} & $\mathbf{\gamma_{pre}}$ (\%) & $\mathbf{\epsilon_{BE}}$  (\%) & $\mathbf{\sigma_{BE}}$ (MPa) \\
 \hline
  \cite{Wadsworth:1963} & 0.7  & 0.2 & 0.2 \\
 DDD s.g. & 1 & 0.08 $\pm$ 0.05  & 0.4 $\pm$ 1  \\ 

 \cite{Marukawa:1971} & 2.5  & 0.6 & 0.5 \\
  \hline
 DDD $[112]$ & $0.26$ & 0.2 $\pm$ 0.1 & $1.25$ $\pm$ 1 \\ 
 \cite{Wadsworth:1963} & 0.7 & 0.5 & 0.8 \\
 DDD $[112]$ & $1.0$ & 0.3 $\pm$ 0.1 & $2.1$ $\pm$ 1  \\
 \cite{Buckley:1956} & 1.2 & 0.3 & 1.6 \\
 \hline
 DDD $[111]$ & 1.0 & 0.5 $\pm$ 0.1 & 2.8 $\pm$ 1  \\
 \cite{Yakou:1985} & 0.8 & 0.5 & 2 \\
 \hline
  DDD $[001]$ & 0.25 & 0.05 $\pm$ 0.1  & 1.2 $\pm$ 1  \\ 
 DDD $[001]$ & 0.5 & 0.1 $\pm$ 0.1 & 1.6 $\pm$ 1  \\ 
 \cite{Yakou:1985} & 0.6  & 0.5 & 3.2 \\
 DDD $[001]$ & 0.75 & 0.2 $\pm$ 0.1 & 2.3 $\pm$ 1  \\ 
 DDD $[001]$ & 1.0 & 0.4 $\pm$ 0.1  & 3.4 $\pm$ 1  \\ 
 DDD $[001]$ & 8.0 &  & 17.5   \\ 
  \hline

\end{tabular}
\end{center}
\label{tab:BEinDDD}
\end{table}

When now considering the backward respond in compression of Fig.~\ref{fig:1}a, a well defined BE can be seen with a lowered elastic limit, a well-rounded appearance of the initial plastic portion and a permanent stress softening. This compression behavior is very different from the continued tension response which is perfectly in line with the prestrain hardening. The difference between the continued tension and backward compression allows to unambiguously quantify the intensity of BE through two parameters commonly used in the experimental literature: a strain shift $\epsilon_{BE} \approx 0.4\%$ and $\sigma_{BE} \approx 3\,\mathrm{MPa}$ (see Fig.~\ref{fig:1}a). Note that serrations are also present on the compression curves and continued tension second loadings, but their investigation are left for a forthcoming investigation.

The hysteresis found in the deformation curves is not only in qualitative agreement with deformation shapes observed in experiments but also the BE intensity observed in simulations matches relatively well the experimental values. In the simulations, the BE is stronger as prestrain increases, and a hierarchy exist among the loading directions. For a given prestrain, the BE increases in intensity when considering [135], [112], [111] and [001] loading directions. A more detailed comparison covering more experimental configurations will be presented latter in the paper, but for now Table \ref{tab:BEinDDD} presents a comparison of $\epsilon_{BE}$ and $\sigma_{BE}$ obtained in our simulations with experimental values obtained for fcc single-crystals deformed in comparable conditions. Keeping in mind that experiments on single crystals are very delicate and can be impacted by many experimental conditions (see latter discussion), a relatively nice agreement can be found with our simulations. Next, we determine the origin of the BE effect obtained here.

%%\emph{table BE values in DDD}

\section{Absence of long-range internal stresses}
\label{sec:NoLRstress}

According to the composite model \cite{Asaro:1975eu, Mughrabi:88, Argon:2008}, backstresses or LRIS are thought to be the explanation of the BE and more generally to be the main cause for the hysteresis behavior observed during cyclic deformation. However, evaluating LRIS from experiments constitutes a challenging task only accessible by a handful of techniques. Existing evaluations often consist of indirect and ex-situ evaluations and there is no direct link between dislocation microstructures or dislocation mechanisms and the stress measurements. In contrast to these experimental limitations, DDD simulations offer a simple solution to map stress fields at a fine scale in correlation to both fully formed 3D dislocation microstructures and the macroscopic flow stress.  We thus employed DDD simulations to evaluate LRIS during a BE numerical experiments. The LRIS evaluation from DDD can nevertheless be performed in different but nonequivalent manners. For example, one could simply map and average stresses everywhere in the simulation domain. But this approach would however only partly capture the mechanical state of dislocations, as most of the stress map would correspond to dislocation-free regions. Instead, we rigorously evaluate the effective stress $\tau^i_{int}$ felt by the dislocations segment 'i' in the shape of the gliding component of the Peach and Koehler force $\mathbf{F}_g^{PK}$ associated to stress fields: $\tau^i_{int} =  \left| \mathbf{F}_g^{PK} \right| b L$ , with $\mathbf{F}^{PK} = \oint_L (\sum_{j} \sigma^{j \rightarrow i}) \times b \cdot d\xi$. $\sigma^{j \rightarrow i}$ is the Cauchy stress induced by a segment ‘j' onto the considered segment ‘i’, the line integral running along considered segment length $L$ and $d\xi$ its tangent vector.  Stress evaluations during simulations and for the evaluation of $\tau^i_{int}$ { are performed using the so-called non-singular theory derived in a series of papers \cite{Cai:2006, Arsenlis:2007, Queyreau:2014, Queyreau:2020}}.

\begin{figure}[htbp]
\includegraphics[width = 0.8\linewidth]{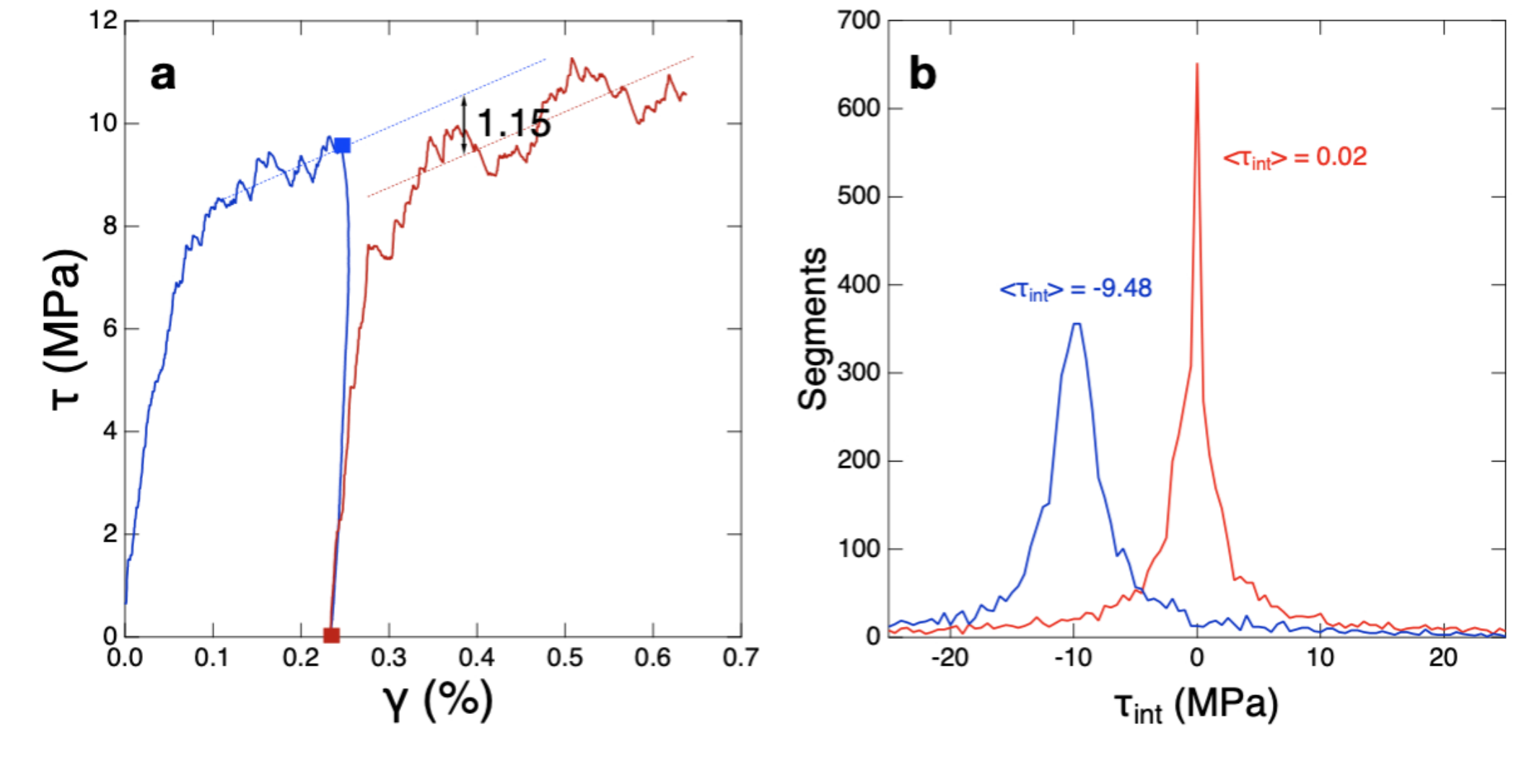}
\caption{ { Determination of internal stress in a $[112]$ BE simulation. a. Deformation curves. The blue curve corresponds to the predeformation in tension and the red curve is the subsequent deformation in compression. A clear permanent recovery (or $\sigma_{BE}$) of about 1.15 MPa in average can be seen, and often exceeds this value. b. Corresponding distributions of internal stresses felt by segments of the microstructure. The blue distribution is obtained at the end of predeformation at a stress value reported by a blue square on figure a, and distribution in red is obtained at stress relaxation as reported by a red square on the deformation curves of figure a. Note that internal stress distributions seem mostly symmetrical and the average value $<\tau_{int}>$ correspond nicely to the imposed applied stresses of 9.5 MPa.}}
\label{fig:intstress}
\end{figure}

Figure \ref{fig:intstress} displays a typical internal stress distribution obtained in a BE test simulation in Ni along $[112]$ corresponding to a double glide condition. We determined the distribution of internal stress $\tau^i_{int}$ on individual segments and the average $<\tau_{int}>$ along the entire dislocation network. The conventional explanation of BE based on polarized internal stresses would correspond to the existence of an asymmetrical $\tau_{int}$ distribution yielding a non-zero $<\tau_{int}>$ value equating half the BE stress $\sigma_{BE}$.

 \begin{figure}[htbp]
\includegraphics[width = 0.40\linewidth]{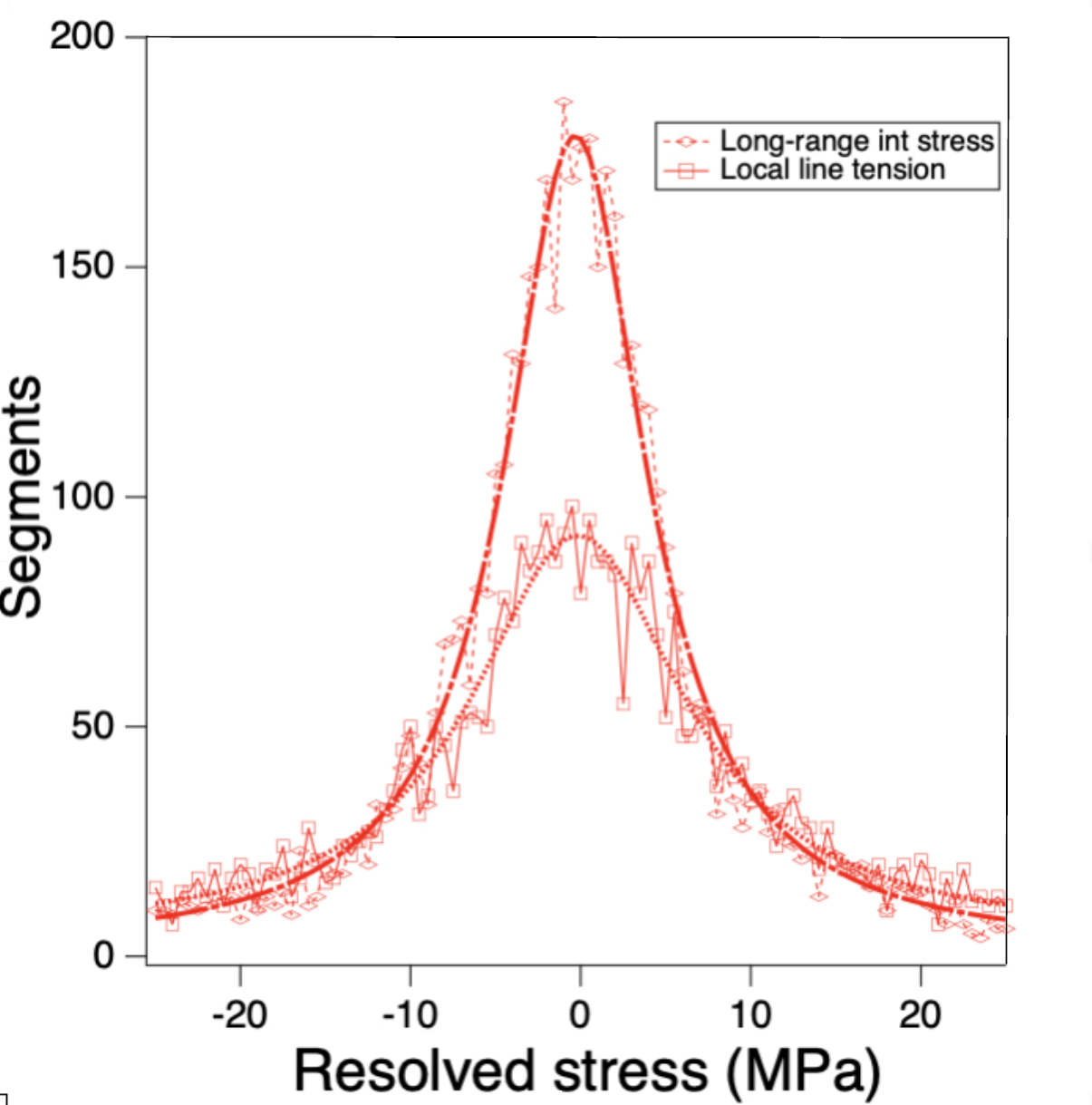}
\caption{Distributions of long-range and self stress contributions to the effective stress felt by segments in the same [112] simulation as in figure  \ref{fig:intstress}.a. This analysis was conducted on the relaxed configuration. As for the total effective stress, both distributions are symmetrical and centered onto the 0 stress value.}
\label{fig:LineTens}
\end{figure}

In our simulations, at the end of the predeformation, the distributions of $\tau_{int}$ is symmetrical and the average value $<\tau_{int}>$ corresponds exactly to the imposed stress. The latter point demonstrates the robustness of the evaluation performed. When now considering the relaxed configuration, the distribution of $\tau_{int}$ measured on all the few $10^4$ segments contained in the simulation becomes narrower but remains symmetrical, and the average value is virtually zero. Simulations allow for pushing the analysis a bit further by decomposing the stress distribution in the relaxed configuration into long range and self-stress contributions as shown in figure \ref{fig:LineTens}. Both distributions remain symmetrical and centered to zero. 

All this clearly demonstrates the absence of any polarized long range internal stresses in the simulated dislocation microstructure when unloaded that could explained the strong BE observed in our caclculations. This evaluation has been performed on other of our DDD simulations leading to identical results. In the frame of the composite model, a marked dislocation microstructure is required to justify the polarization of the internal stresses. But dislocation microstructure are weak in the early stage of deformation so it is unsurprizing that LRIS are absent here. This conclusion is also in agreement with the absence of residual curvature in any part of the dislocation microstructures at zero stress, and in agreement with the conclusions drawn by recent X-Ray micros-diffraction experiments performed on single crystals \cite{Kassner:2009fk, Kassner:2013fk}.

\section{Identification of the key mechanisms}
\label{sec:Identif}

In the absence of LRIS, the BE observed in the simulations must therefore be attributed to other physical origins. Evidences of these new mechanisms can be unveiled by the careful observation of the backward sequence of dislocation motion in Fig.~\ref{fig:1}c. The observation zone is the same as for the prestrain sequence in tension (Fig.~\ref{fig:1}b), but the dislocation behavior is now significantly different. First, junctions that were acting as strong and stable anchoring points to the dislocation microstructure are now destroyed with ease during backward motion. A first origin of the tension-compression asymmetry is thus related to the mechanical stability of junctions, which was unsuspected until now. We will see that this instability is inherited of the stress-driven curvature that mobile dislocations have before colliding with each others. The second observation has to do with the dislocation motion kinetics and the resulting dislocation density evolution. Plastic flow still occurs by strain bursts through the intermittent motion of dislocations. However, upon stress reversal, dislocations now move in the opposite -backward- direction and glide in regions of the crystal that have already been swept during the forward motion. The locked configurations of dislocations between two distinct avalanches are different as the strongest junctions -pinning points- are not located at the same positions as before and the sequence of dislocation motion is thus rather different from the one observed during the forward sequence. Backward avalanches unwind the junctions and segments stored during prestrain while forming only few new junctions as the possible forest segments are already pinned in the microstructure. The backward motion thus leads to significant decrease of dislocation density through unwinding of initially stored segments or direct annihilation with stored segments as it was observed in \cite{Queyreau:2009lr} for precipitation-hardened materials. A last observation is that since hard obstacles are placed at different locations due to the junction asymmetry, the backward motion will eventually explore new regions of the crystals. Only a part of the microstructure formed during prestrain is impacted by the backward motion and this is at the heart of the definition of the reversibility of plastic flow.

\begin{figure}[htbp]
\includegraphics[width = 0.40
\linewidth]{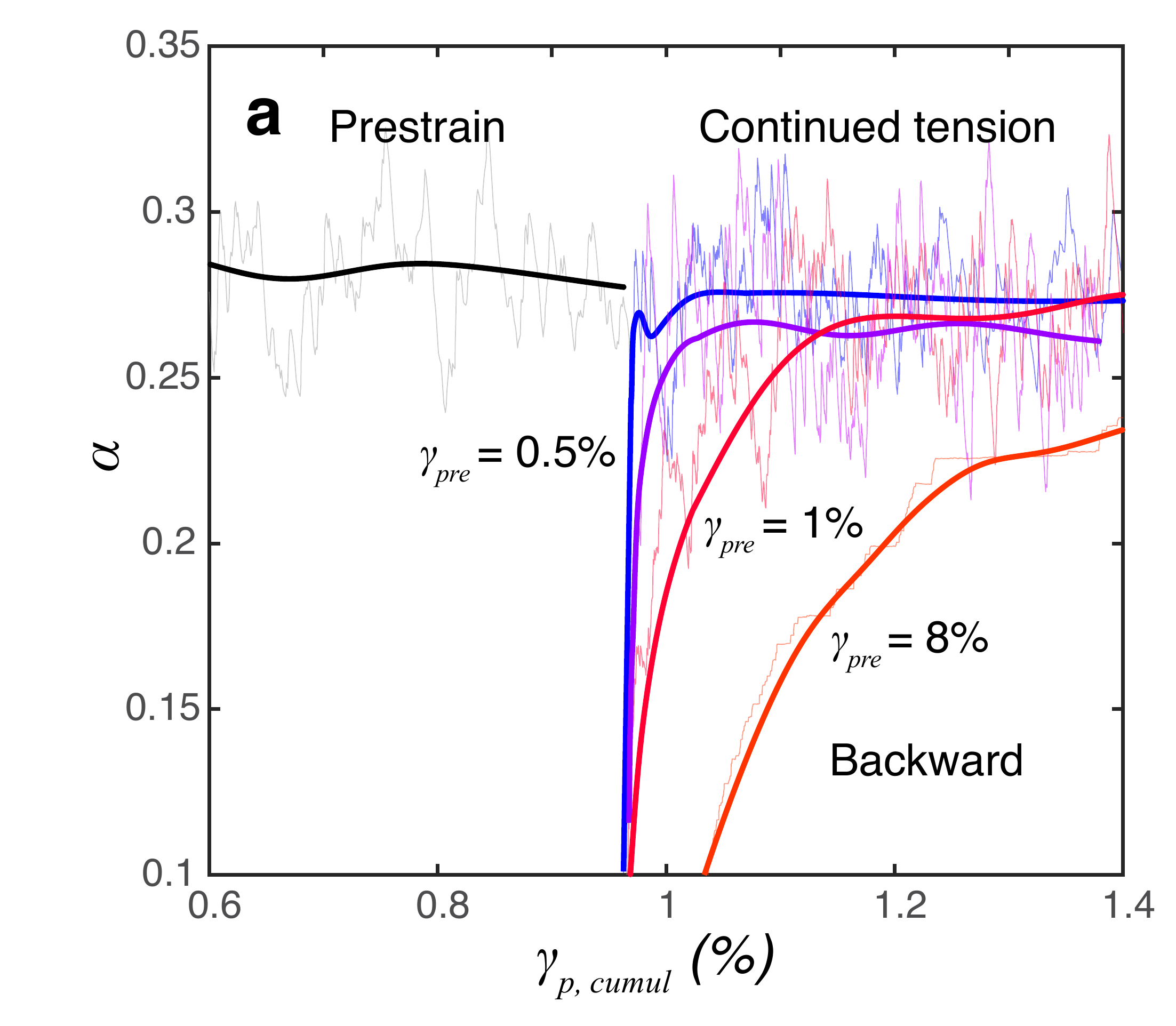}
\includegraphics[width = 0.47 \linewidth]{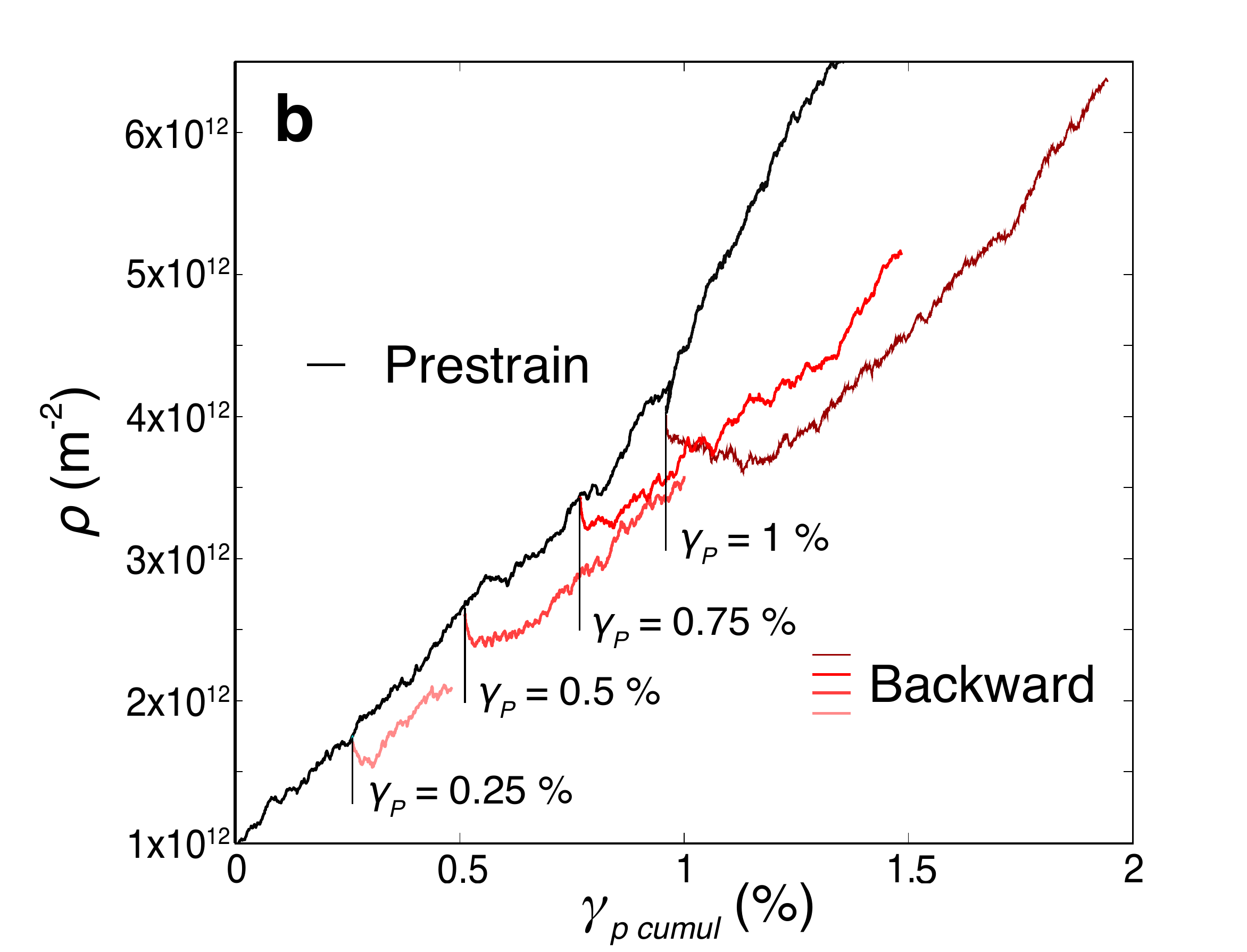}
\caption{{ Highlighting of the two original mechanisms at the origin of the traction-compression asymmetry. \textbf{a}, The reduced stability of junction in the subsequent compression sequence can be seen in the lowered average interaction coefficient simply taken as $\alpha = \tau_c / \mu b \sqrt{\rho}$. Simulations are concerned with deformations along $[001]$ axis in tension or compression (backward) following various prestrains $\gamma_{pre}$. The transient on $\alpha$ is understandably small at small prestrain (i.e. $\gamma_{pre} <$ 1\%) but becomes very significant after 8\% of prestrain. The strain values have been divided by 8 for the curve corresponding to $\gamma_{pre}$ = 8\%  so that it can be displayed alongside the other curves. \textbf{b}. Evolution of the total dislocation density over all slip systems for the same set of simulations. For curves corresponding to backward deformation, a large drop in density can be seen after the stress reversal and the rate of dislocation storage is reduced during a transient, the length of which increases with the amount of prestrain.}}
\label{fig:2causes}
\end{figure}

These mostly qualitative observations are backed up by the quantitative evolution of the key parameters shown in Fig \ref{fig:2causes}. The weaker stability under compression of junctions formed in tension can be seen in the transient lowering of the interaction coefficient $\alpha = \tau_c / \mu b \sqrt{\rho}$ with $\tau_c$ the critical shear stress (cf. Fig. \ref{fig:2causes}.a)  and in the drop in the number of junctions. The reduced storage rate is clearly seen on the total dislocation density evolution (cf. Fig. \ref{fig:2causes}.b), with a first brutal drop of density right after stress reversal, the storage rage is actually never recovered when comparing with the dislocation evolution observed for continued tension. We will see latter in the paper that experimental curves showed non-linearities that suggested long ago the existence of 2 distinct mechanisms at the origin of the BE. These two effects and their duration increase with prestrain as expected. Next, we provide a finer analysis of these two mechanisms.

\subsection{Asymmetry of the stability of junctions formed under a stress}

\begin{figure}[htbp]
\includegraphics[width = 0.95\linewidth]{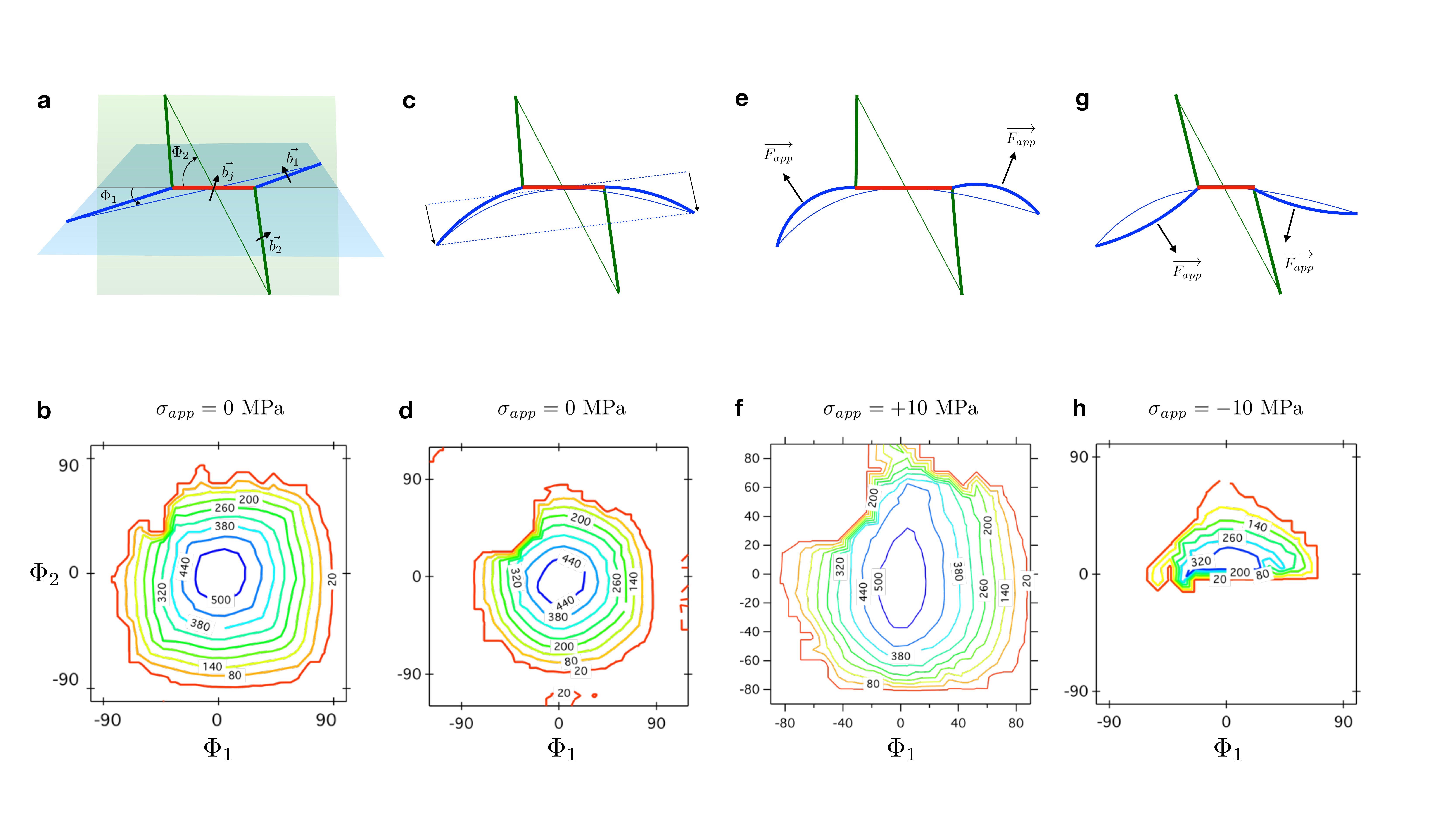}
\caption{Illustration of the stability asymmetry existing in Lomer junctions formed in extension. The various geometric configurations a, c, e and g used to construct a binary junction lead to different angular domains of junction stability b, d, f, and h, respectively. Configuration a, is the typical picture of binary junctions formed from two straight segments forming angles $\phi_1$ and  $\phi_2$  with respect to the junction line direction. This \emph{symmetric} geometry yields a junction stability domain of almost circular shape as function of $\phi_1$ and  $\phi_2$, that remains symmetric under stress. Outside the stability domain where junctions are formed, cross-state or repulsive configurations are found, but these are of little interest for the present discussion. To mimic configurations observed in large scale simulations, one of the dislocation is put in extension as in configuration c. In the absence of any applied stress, the stability domain seen in d is virtually identical to b. The asymmetry in the junction stability reveals itself when applying a non-zero applied stress $\sigma_{app}$. For a positive value for $\sigma_{app}$ that would correspond to a motion in the same direction as the initial extension e, the stability domain is considerably increased and junction lengths are increased as well in f. The angular domain of junction stability states when a junction may form, while the junction length is a good qualitative indication of its stability under stress. Finally, When applying a negative value for $\sigma_{app}$ that would correspond to the opposite backward direction to the initial extension g, the stability domain in h gets much smaller and junctions are much shorter for all angles ($\phi_1$,  $\phi_2$).}
\label{fig:carto}
\end{figure}

To illustrate the asymmetry developing in the junction stability formed during the prestrain, we investigated the stability of a single binary junction. Our choice turns toward the so-called Lomer lock as this binary reaction is one of the strongest one among junctions (just below the glissile junction), and Lomer locks may be formed among slip systems interacting for the classical loading directions considered here. For this, we revisit the methodology developed in \cite{Wickham:1999, Madec:2002, Madec:2004} where two initially straight segments are brought into collision in their middle. The initial orientation of the two segments has long been found to be one of the key parameters controlling the stability of the reaction \cite{Rodney:1999}. The various states of the reaction (unformed, formed, or a cross-state with an unzipped junction) can then be mapped as function of the two initial orientations $\phi_1$ and  $\phi_2$ (\emph{cf.} Fig. \ref{fig:carto}.a and b). In large scale simulations, the applied stress drive and thus 'polarize' the curvature of dislocations. To match the configurations observed in the large scale simulations one of the dislocation is in 'extension' as it is positioned at a small distance from the center of the other segment (\emph{cf.} Fig.~\ref{fig:carto}.c,e,g). Figure \ref{fig:carto}.b shows the angular domain of stability of the Lomer junction mostly centered at the origin $(0, 0)$ where the two dislocations are colinear. Another indication of the reaction stability is its length, which is also provided in the figure by isocurves with values up to 500 nm. When formed in extension, the stability domain increases in size and becomes elongated in the direction of elongation of the dislocation when a tensile stress of +10 MPa is applied (\emph{cf.} Fig.~\ref{fig:carto}.e and f). This anisotropy is absent and the stability domain is smaller when no stress is applied (\emph{cf.} Fig.~\ref{fig:carto}.c and d). 

The stability domain for a single Lomer lock is however dramatically diminished when the junction is still formed in extension but now subjected to a compressive applied stress of -10 MPa (\emph{cf.} Fig.~\ref{fig:carto}.g and h). The stability domain becomes really small, still slightly extended towards the initial extension direction and junctions length are dramatically smaller leading to a reduced stability. The stability of binary junctions is thus strongly anisotropic when accounting for the natural curvature of interacting dislocations. This can be seen as a self-stress effect of the parent arms of the junction that promotes or prevents the zipping of junctions. In other words, the relative extension of dislocations affects the mechanical balance of ternary nodes, which control the zipping process. 

\begin{figure}[htbp]
\includegraphics[width = 0.65\linewidth]{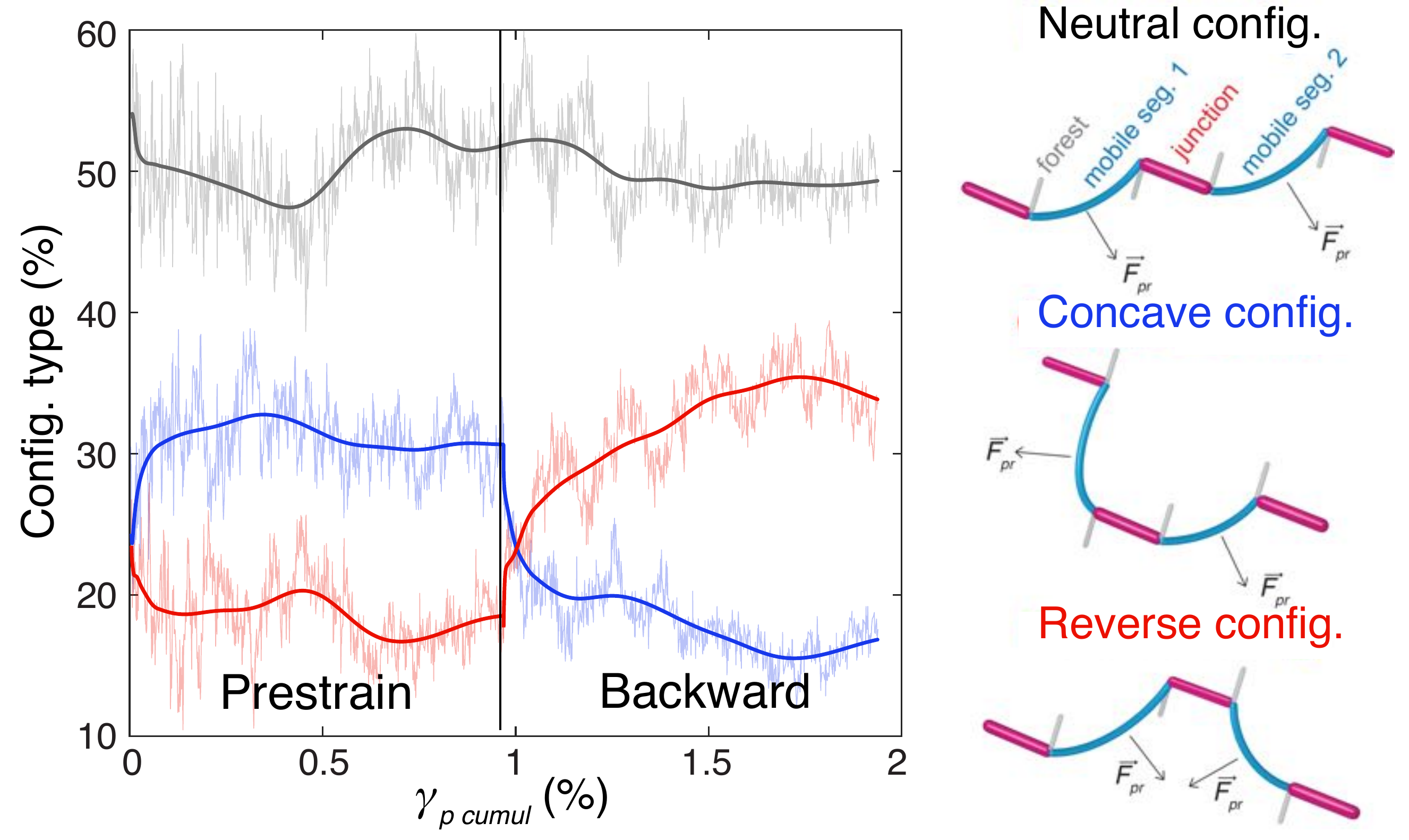}
\caption{Direct evidence of the excess of weak junctions inherited from prestrain deformation. The 3D dislocation network is analyzed in terms of ternary objects, whose schematics are shown in the right part of the figure, in the case of a $[001]$ Bauschinger test simulation. The partition of neutral, concave and convex configurations is shown as function of deformation and accounts for about two thousand junctions. Concave configurations, that are mechanically more stable, are found in greater number during the prestrain, while their contribution drops during backward deformation as they are now weaker due to the asymmetry in the junction stability. Conversely, convex configurations initially weaker during prestrain, increase in number after the stress reversal as they become more stable in compression. The transient on the junction strength will be analyzed in more varied simulation conditions a bit latter ($cf.$ Fig.~\ref{fig:fittransient} ). }
\label{fig:2}
\end{figure}

While realistic microstructures encompass more varied and complex microstructures, they can still be analyzed in terms of basic configurations resembling those of the model calculations above. We consider now ternary objects made of a first parent segment connected to a junction and a second parent segment. The curvature of parent dislocations are signed with respect to the loading direction. We then distinguish three types of configurations (see sketches in Fig.~\ref{fig:2}.a): (i) concave configurations are found when two primary segments are 'pushing' a given junction (similar to configuration Fig.~\ref{fig:carto}.e), (ii) reverse configurations correspond to two primary segments that are both 'pulling' the connected junction (akin to configuration Fig.~\ref{fig:carto}.g)  and (iii) neutral configurations where two segments are pulling the considered junction in opposite directions (similar to configuration Fig.~\ref{fig:carto}.a). The partition among these three configurations and its evolution with strain is shown in Fig.~\ref{fig:2}.a for a simulation of the BE along [001] axis. These three basic configurations are sufficient to described 100\% of the microstructure topology. Concave and convex configurations start at 25 \% each and then concave configurations take the upper hand with $\approx 32\%$ of the configurations found in the dislocation network, reverse configurations decrease to $\approx 18\%$, while neutral remain close to $\approx 50\%$. From Fig. \ref{fig:carto}, we know that concave configurations (i) are on average more stable mechanically than reverse configurations since primary segments connected to the junction promote the zipping of the junction rather than its unzipping. When loading is reversed from tension to compression, concave (i) and reverse configurations (ii) switch roles as the direction of dislocation curvature is reversed. 

An important feature of the BE is its transient nature, and this is typically poorly explained or understood by existing interpretations of the traction-compression asymmetry. But here, the explanation is rather natural. At the beginning of the compression loading, a finite pool of concave configurations for ternary objects have been inherited from the prestrain in tension. These concave configurations are unstable to different degree in compression, to a first degree this stability could be said to be proportional to the junction length. So the most unstable (short) junctions are destroyed rather early, while as the compression applied stress increases, progressively more stable (longer) junctions will be destroyed. This can be clearly seen in Fig.~\ref{fig:2}, where the number of unstable concave configuration first quickly drops from ($\approx 32\%$) then continue decreasing at a decreasing rate to stabilize to ($\approx 18\%$) of the configurations. This excess of weak junctions at stress reversal is thus the direct explanation of the lowered interaction strength observed. This transient mechanism therefore ends when most unstable configurations formed during prestrain have been destroyed or when dislocations glide in new unexplored regions of the crystals. The amount of reversibility of the plastic deformation seems then to be closely related to the competition between the easy destruction of weak junctions (inherited from the prestrain) in region already explored (reversible part) and the exploration of new regions of the crystal (unimpacted by prestrain and thus irreversible).

\subsection{Reduced storage rate}

The second origin of the BE found in DDD simulations is related to the reduced storage of dislocations. The dislocation storage rate is commonly associated to the dislocation Mean Free Path (MFP) $L_i$, which captures the average distance covered by mobile dislocations before their temporary or permanent immobilisation. In \cite{Devincre:2008lr}, the dislocation MFP was decomposed into three separate terms at the origin of the storage of a dislocation segment: (i) $p_0$ the probability to form a stable junction on a mobile dislocation when crossing other dislocations. Its value was found to be equal to $0.1 \pm 0.02$; (ii) the average length of stored dislocations $<l>$ induced by the pinning by the junction whose value is classically related to the dislocation density $\rho$ as $k_0/\sqrt{\bar{a} \rho}$ with $k_0 = 1.08 \pm 0.005$ , (iii) the average length of dislocation stored in the shape of junction $l_j$ that relates to the previous quantity as $<l_j> = \kappa <l>$ with $\kappa = 0.29 \pm 0.015$. From  geometrical arguments, it was shown that the MFP of system 'i' can be expressed in the case of symmetric slip system activity as:

\begin{equation}
L^i = \frac{\mu b^2}{\tau_c^i} \left[ \frac{\sqrt{\bar{a}} n (1 + \kappa)^{3/2}}{p_0 k_0 ( n - 1 - \kappa )} \right]
\label{eq:MFP}
\end{equation}

where $n$ is the number of activated slip systems. $\bar{a}$ is a coefficient accounting for the average slip system interaction strength $<a_{ij}>$ with other systems 'j', and will depend upon the loading axis. DDD analyses confirmed this MFP decomposition by showing that $p_0$, $\kappa$ and $k_0$ were found to be constant for various loading conditions and deformation amounts. The resulting MFP values allowed to reproduce at the macroscale with a crystal plasticity model, realistic deformation curves for fcc single-crystals and monotonic loadings. 

In the present BE simulations, the observed reduced storage of Fig.~\ref{fig:2causes}.b means that dislocations cover a longer distance before being stored, which thus is consistent with an increase of the MFP. Since the sequence in Fig.~\ref{fig:1}.c shows that the storage of dislocations is still controlled by the immobilization induced by junctions in the wake of plastic bursts, the decomposition of the MFP according to Eq .~\ref{eq:MFP} based on geometrical arguments is still relevant to the case of the reversed deformation. In \cite{Devincre:2008lr}, a statistical analysis was proposed to measure the dimensionless coefficients $p_0$, $\kappa$ and $k_0$ from DDD simulations. However, since simulations were carried under monotonic loading conditions, the analysis relied upon an integration of the dislocation density variations. Due to the transient nature of the BE, we developed a new analysis strategy based on the instantaneous density evolution to capture possible evolution of $p_0$, in particular. An obvious drawback of using instantaneous density evolution is that it yields to large fluctuations of the measured parameters. Nonetheless, the analysis made on the prestrain section of the BE simulations are very well in line with results obtained in \cite{Devincre:2008lr}.

\begin{figure}[htbp]
\includegraphics[width = 0.32\linewidth]{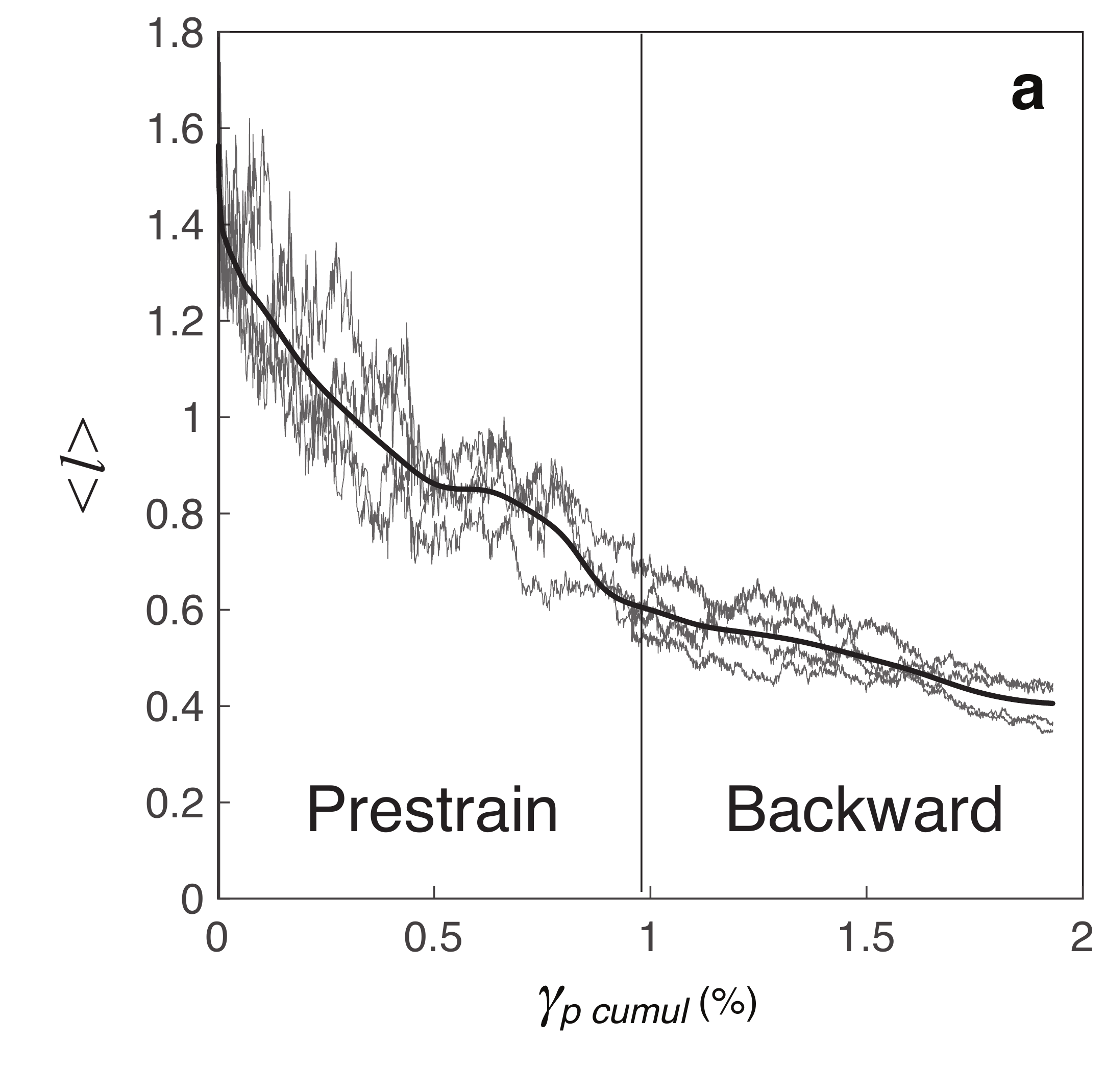}
\includegraphics[width = 0.32\linewidth]{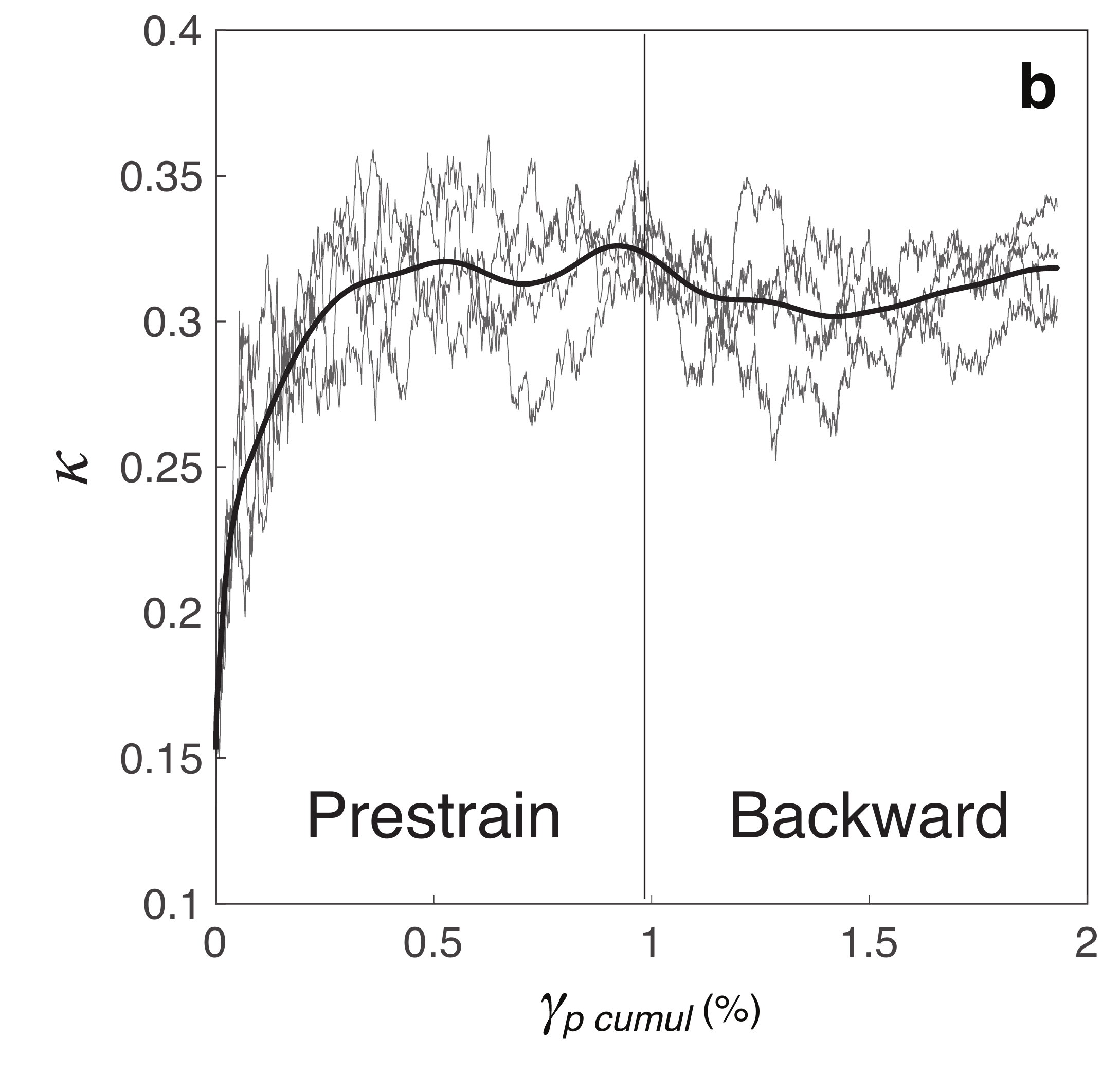}
\includegraphics[width = 0.32\linewidth]{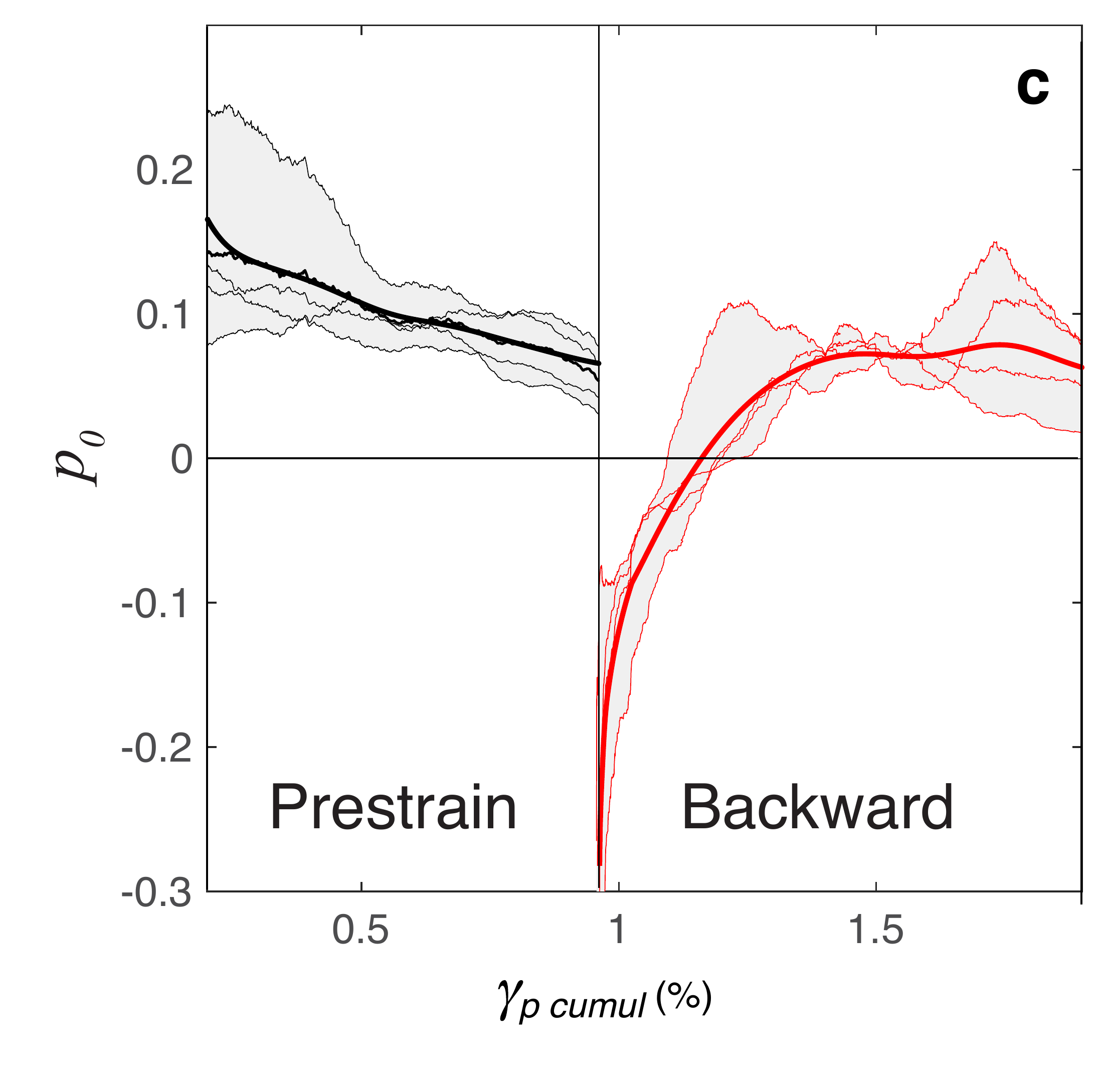}
\caption{Statistical analysis of the three physical ingredients impacting the MFP of dislocations \cite{Devincre:2008lr}. The simulations are concerned once more with the BE experiment along [001]. Gauss kernel averages (thick curves) of these parameters are shown alongside the calculations on a per system basis (thin lines), to give an appreciation of the variability of these parameters. a) Evolution of the average primary length $<l>$ in $\mu$m as function of strain. b) Evolution of the ratio of junction length defined as $\kappa = <l_j> / <l>$ with $<l_j>$ the average junction length. The expected value from monotonous loading is $\kappa = \kappa_0 / (1+\kappa_0) \approx 0.29 \pm 0.015$. Curves in a) and b) do not show any major modifications at the stress reversal and onwards, the scaling of $<l>$ and $<l_j>$ with $1/ \sqrt{\rho}$ is preserved. A small decrease of $<l>$ and $<l_j>$ is certainly present as $\rho$ decreases after stress reversal. c) Analysis of the rate of formation of junctions $p_0$ per unit of deformation as function of the deformation. After stress reversal, $p_0$ starts back to negative values and slowly goes back to its prestrain value. The expected value from monotonic calculation is $p_0 = 0.1 \pm 0.02$. The transient on $p_0$ will be analyzed in more varied simulation conditions a bit latter.}
\label{fig:MFP}
\end{figure}

The example of the results of such an analysis is displayed in Fig. \ref{fig:MFP}. Surprizingly, the average segment lengths $<l>$ and $<l_j>$ remain both mostly unchanged by the stress reversal, and their values scale simply with dislocation density as $\rho^{-1/2}$. $<l>$ and $<l_j>$ then, continue to follow the similarity principle in non monotonic loadings. On the contrary, the rate $p_0$ of formation of junctions is significantly altered after loading reversal. During, backward deformation, $p_0$ starts from negative values then slowly goes up to positive rates for a backward deformation of about ${0.5\times\gamma_{pre}}$ and only reaches its forward value after a deformation close to the amount of prestrain deformation. This effect appears to be much stronger than the previous effect on junction stability.

The transient associated to this second effect is clearly seen on fig. \ref{fig:MFP}.c. Its mechanistic explanation can be stated as follows. Segments that were stored the last, are easily unstored by the backward motion of the mobile dislocation with almost no new junction formation after stress reversal. The strongest among the junction may remain in the beginning, but the easy backward motion of mobile segment lead to arm configurations for which junctions are not stable anymore, and the primary loop collapses on itself. As backward deformation proceeds, segments on other slip systems will lead to the formation of new pinning points in different locations, helping the dislocation to propagate and store line length in new unexplored regions of the crystal.

It must be emphasized that the present mechanism associated to dislocation MFP is different from the previous mechanism regarding the stability of junctions. When considering equation ~\ref{eq:MFP}, it is however true that the MFP of dislocations is impacted by both the average strength of reactions $\bar{a}$ and the rate of junctions leading to dislocation storage $p_0$. However, these two effects are clearly separated and $p_0$ was found to be constant for the common loading directions associated to multislip and thus various interaction strengths. Indeed, the junction strength controls the initiation of a critical event (plastic burst or avalanche), while $p_0$ described the number of collisions before finding one that will ultimately stop the avalanche. As will be shown latter, the impact of the lowered $p_0$ is much a stronger effect than the lowered junctions stability, and its effect is felt long after. The reduced strength of junctions inherited from the prestrain could never explain alone the decrease in dislocation density, as a reduced $\alpha$ can at most lead to an absence of storage.

\section{Formulation of a Crystal Plasticity model} \label{sec:CP}
\label{sec:CPform}

To extend our DDD simulation results to larger accumulated plastic strain associated with cyclic loading conditions, we formulate a new Crystal Plasticity (CP) model accounting for the elementary mechanisms found in the DDD results. The formalism is derived from the dislocation density based crystal plasticity model developed to simulate monotonous loading in FCC metals \cite{Kubin:2008fk}. In the case of FCC metals at room temperature, plastic flow is controlled by dislocation-dislocation contact reactions that form junctions. This is why the critical stress required to activate a slip system 'i' takes the form:

\begin{equation}
\tau_c^i = \mu b \sqrt{ \sum_j a_{ij} \rho^j}
\label{eq:forest}
\end{equation}

{ \noindent where $\mu$ is the shear modulus, $b$ the Burgers vector magnitude and $\rho_j$ the density of dislocation on slip system 'j'. The dimensionless interaction coefficients $a_{ij}$ measure the average interaction strength between slip systems 'i' and ‘j'. In FCC crystals, for reason of symmetry, the number of independent coefficients $a_{ij}$ goes down to only six. Interaction coefficients have been successfully calculated from DDD simulations in the case of monotonous tensile tests \cite{Devincre:2006uc, Queyreau:09, Madec:2017kx}. In summary, the first four interaction coefficients which involve self and coplanar interactions, glissile junctions or Lomer locks have almost identical values with $a \approx 0.12$. The coefficient associated with interactions that form Hirth junctions have a lower value of $a \approx 0.07$. Last, the particular collinear annihilation that occurs between slip systems with identical Burgers vectors, yields to a much larger value of $a \approx 0.63$.}

\begin{figure}[htbp]
\includegraphics[width = 0.44\linewidth]{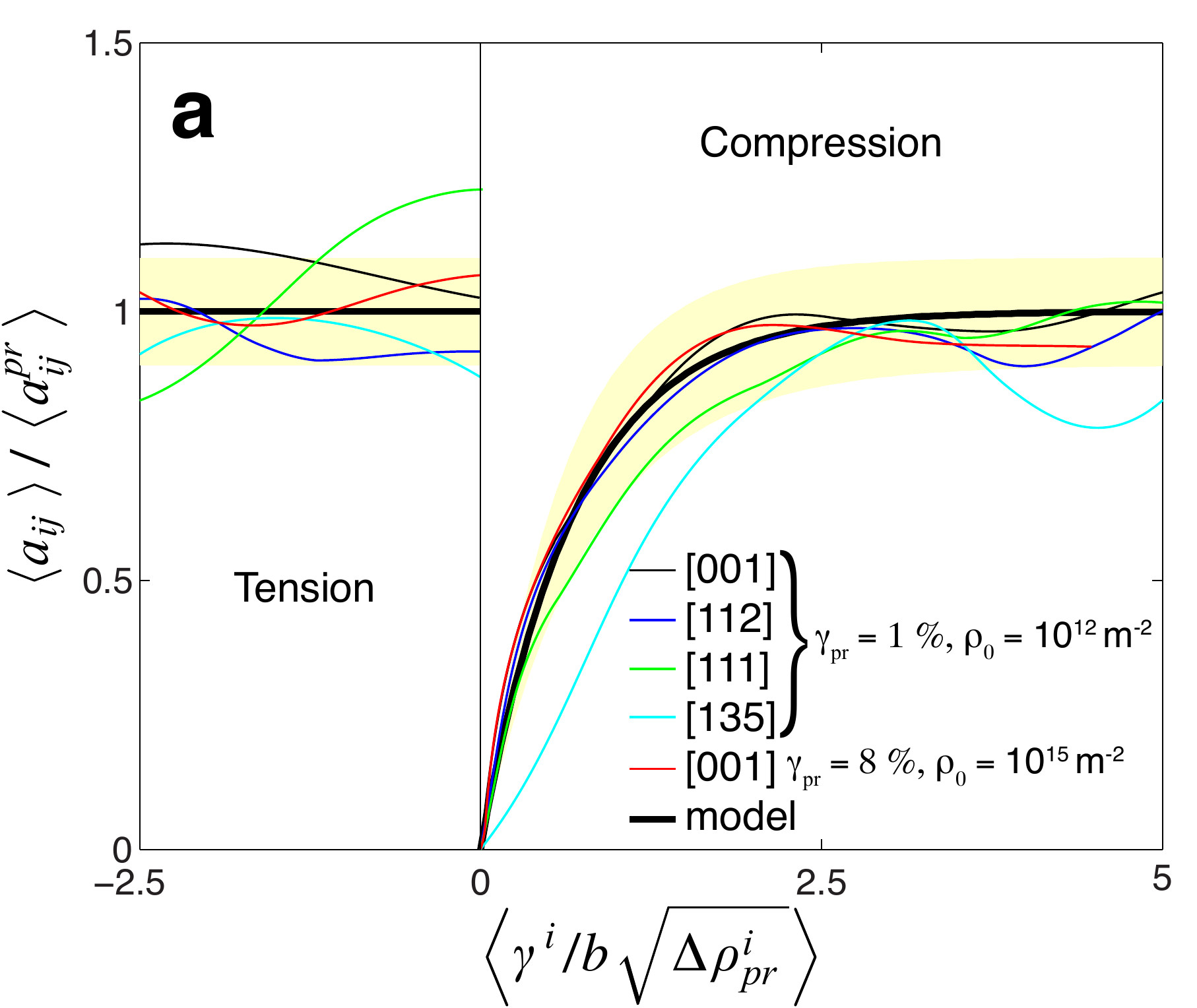}
\includegraphics[width = 0.44\linewidth]{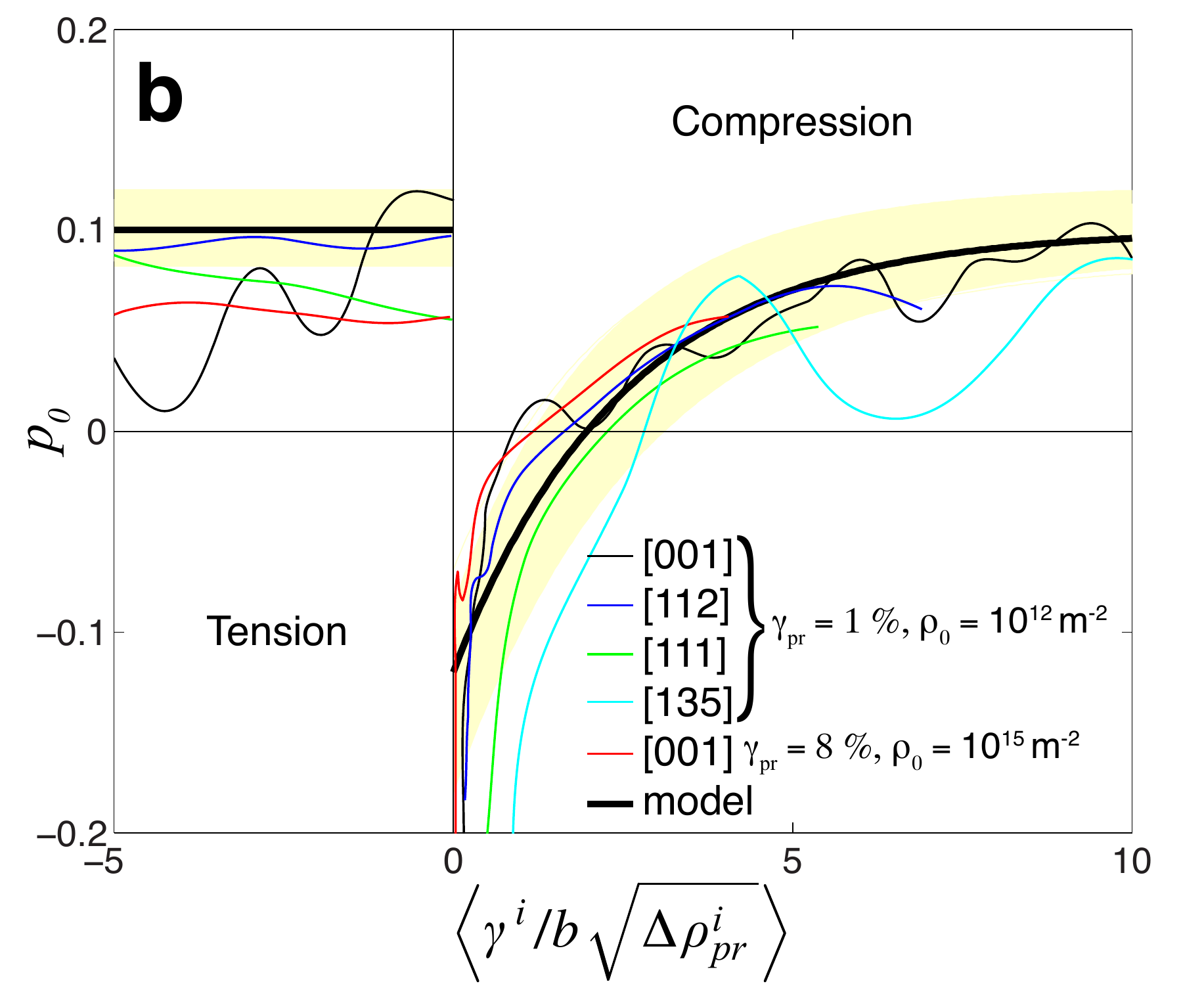}
\caption{Evolution of averaged interaction coefficient $<a_{ij}> $ (a) and rate of junction formation $p_0$ (b) during DDD simulations of selected Bauschinger tests. Curves are obtained from a Gauss kernel smoothing of the instantaneous evolutions of the physical quantities. The yellow domains represent the extend of error bars associated to the data. The rescaling of the x-axis is an average quantity per slip system and allows smearing all curves on top of each others. Comparison is also shown with the proposed model Eq. 3 and 6 in thick black lines. The details of the simulation conditions are given in the figures.}
\label{fig:fittransient}
\end{figure}

Results obtained in previous sections clearly shows that plastic flow is still controlled by dislocation-dislocation contact reactions during backward deformation. Thus, equation \ref{eq:forest} remains applicable to non-monotonic tests, but the asymmetry of junction stability we observed when reversing the deformation sign must be accounted for. This is why we propose that the interaction coefficient $a_{ij}$, defined for monotonous loadings, must be weighted by a \emph{reversibility function} $r_a$ after every strain direction changes. By a systematic investigation of the observed transient on $r_a$ after stress reversals, we obtained the following mathematical form that is consistent with all our available DDD simulations carried over a wide range of simulations:

\begin{equation}
a_{i j}^{b c k} = ( 1 - r_a ) \times a_{i j} \mathrm{, ~with ~~} r_a = \exp \left( \frac{- \gamma_{b c k}^i} {C_a b \sqrt{ \Delta \rho_{pr}^i}} \right)
\label{eq:newa}
\end{equation}

\noindent where the ‘bck’ and ‘pr’ subscripts refer to the backward strain and prestrain during a Bauschinger test, respectively. In the case of cyclic loading test, ‘bck’ subscript then refers to the current half-cycle and ‘pr’ refers to the previous half-cycle. It is important to note that the functional we define in Eq.~(\ref{eq:newa}) imposes that $a_{i j}^{b c k}$ goes back to the reference monotonous value $a_{ij}$ at the end of a transient. Hence, the terms in the exponential state the competition observed between the easy destruction of the weak junctions inherited from previous deformation stage and the formation of new junctions created during backward deformation in newly explored regions. For this reason, the length of the transitory requires the evaluation of the amount of dislocation density $\Delta \rho_{pr}^i$ (i.e., the number of junctions) stored during prestrain and $C_a$ is a dimensionless parameter accounting for the rate at which the reversibility function $r_a$ goes to zero. $\Delta \rho_{pr}^i$ is thus defined as:
\begin{equation}
{ \Delta \rho_{pr}^i} = max(\rho_{pr}^i) - min (\rho_{pr}^i)
\label{eq:deltarho}
\end{equation}
Fig. \ref{fig:fittransient} shows the evolution of  $a_{i j}^{b c k}$ we monitored for our full set of simulation conditions. Fluctuations in the curves are due to the intermittent nature of plastic flow in the relatively small simulated volume and to temporary asymmetries existing in slip system activity. In figure \ref{fig:fittransient}, it is clear that $C_a$ is not notably changed when changing the pre-strain amplitude or the loading axis direction. An average value of $C_a = 0.6 \pm 0.1$ could therefore be calculated from the entire set of our DDD simulations.

A second constitutive equation needs to be modified with respect to the crystal plasticity model initially formulated for monotonous loadings \cite{Kubin:2008fk}. This modification accounts for the drop of dislocation density observed on the active slip systems when reversing the loading axis. Following the seminal work of Kocks and Mecking \cite{Kocks:03}, the net dislocation storage rate observed during monotonic deformation test can be reproduced with the form:

\begin{equation}
%%\frac{d \rho^i}{d \gamma^i} = \frac{1}{b} \left( \frac{1}{L^i} - y \rho^i \right) \\ 
\frac{d \rho^i}{d \gamma^i} = \frac{1}{b} \left(  \frac{\sqrt{a'_0 \rho^i}}{K_{hkl}}  + \frac{1}{L^i} - y \rho^i \right)
\label{eq:KM}
\end{equation}

Equation (\ref{eq:KM}) contains terms that reflect different physical processes involved in the variation of the dislocation density with strain. The third term in the parentheses (right-hand side of the Eq.~\ref{eq:KM}) models the dynamic recovery processes occurring at large dislocation densities. It displays $y$ related to the critical distance at which two dislocations with opposite Burgers vectors can easily annihilate. Here, $y$ depends upon the loading direction to match the anisotropy observed in the onset of stage III of deformation curves. {We defined the values $y$  from experiments obtained the case of monotonous tests and in agreement with previously calculated in \cite{Devincre:2010fk}.} The first two terms in the parenthesis in Eq.~(\ref{eq:KM}) accounts for the dislocation storage processes and involves the calculation of the dislocation MFP, $L_i$ introduced earlier. 

Our results from the previous section showed that after reversing the loading direction, the dislocation MFP -through $p_0$- becomes negative and remains diminished for an extended amount of backward deformation when compared to monotonic loading conditions. This effect constitutes the physical origin of the dislocation density drop discovered in DDD simulations after strain reversal. The dislocation Mean Free Path $L_i$ can still be expressed using Eq.~(\ref{eq:MFP}) as storage is still intimately related to dislocation immobilization induced by junction pinning. Only the rate $p_0$ of formation of junctions is significantly altered after loading reversal, while the other dimensionless coefficient $\kappa$ and $k_0$ appearing in this equation remain unchanged and equal to their monotonous values. In the same manner to what we did for the evolution $a_{ij}^{bck}$, the transient on $p_0^{bck}$ can be modeled with a reversible function $r_p$ such as:

\begin{equation}
p_{0}^{b c k} = ( 1 - r_p ) \times p_{0} \mathrm{, ~with ~~} r_p = A_p \times \exp \left( \frac{- \gamma_{b c k}^i} {C_p b \sqrt{ \Delta \rho_{pr}^i}} \right)
\label{eq:P0bck}
\end{equation}

\noindent where $A_p$ and $C_p$ are two dimensionless constants: $A_p$ determines the initial negative value at loading reversal and $C_p$ controls the length of the transient regime of $p_0$. Again, in the case of a cyclic loading, ‘bck’ and ‘pr’ subscripts refer to the current and previous half-cycles, respectively. The term ${C_p b \sqrt{ \Delta \rho_{pr}^i}}$ is dimensionless and relates to the reversible part of the plastic deformation done during previous cycle. 

Figure \ref{fig:fittransient}.b shows the evolution of $p_0$ we calculated during prestrain and backward deformation for a large set of simulations including single slip (with a $[135]$ loading direction) and multiple slip system activation ($[112]$, $[111]$, $[001]$ loading directions), various initial dislocation densities $[10^{11}\;-\;10^{15}\;\mathrm{m}^{-2}]$ and different prestrain amplitudes $[0.25 - 8\%]$. It must be noted here that the fluctuations observed on Fig.~\ref{fig:fittransient}.b curves are intrinsic to the procedure of calculation of $p_0$ that requires the evaluation of the instantaneous derivative of dislocation density with respect to strain. It follows $A_p = 2 \pm 0.6$ and $C_p = 2.3 \pm 0.3$. This identification was obtained using the least square method applied to all simulation data treated as a single set and focusing on the transient part.

\begin{table}[!ht]
\caption{Physical parameters implemented in the CP simulations of the cyclic deformation of fcc single crystals. Most of theme are calculated from DDD simulations or known from the literature.}
\begin{center}
\begin{tabular}{ c c c c c c c } 
 \hline
 $\mathbf{a'_0}$ (self) &   $\mathbf{a_{ortho}}$ (Hirth)  & $\mathbf{a_2}$ (glissile) & $\mathbf{a_3}$ (Lomer) & $\mathbf{a_{colli}}$ (Collinear) & $\mathbf{\rho_{ref}}$ & $\mathbf{\rho_{0}}$\\ 
 0.122 & 0.07 & 0.137 & 0.122 & 0.625  & $10^{12}\,$ m$^{-2}$ & $10^{12}/n_{sys}$  \\
 \hline
  $\mathbf{K_{I}}$ & $\mathbf{K_{112}}$ & $\mathbf{K_{111}}$ & $\mathbf{K_{001}}$ &  $\mathbf{y_c}$ (SG, [112]) & $\mathbf{y_c}$ ([001],nm) & $\mathbf{y_c}$ ([111],nm) \\
  180 & 10.42 & 7.29 & 5 & 0.5 nm &  {3.6 (Ni), 3.4 (Cu)} & {2 (Ni), 1.5 (Cu) }\\
 \hline
 $\mathbf{\dot{\gamma}_{app}}$ & $\mathbf{\dot{\gamma}^i_{0}}$ & m & $\mathbf{C_a}$ & $\mathbf{A_p}$ & $\mathbf{C_p}$ \\
 $n_{sys} \times {\dot{\gamma}^i_{0}}$ & 10$^{-4}$ s$^{-1}$ & 35 & 0.6 $\pm$ 0.1 & 2 $\pm$ 0.6 & {2.3 $\pm$ 0.3} \\
 \hline
\end{tabular}
\end{center}
\label{tab:paramCP}
\end{table}

Lastly, a closed form for the CP model is obtained with the definition of a flow rule that relates the critical shear stress $\tau_c^{i}$ (Eq.~\ref{eq:forest}) and the strain rate $\dot{\gamma^{i}}$ to the loading resolved shear stress $\tau^{i}$ on each slip systems i:

\begin{equation}
\dot{\gamma^{i}} = \dot{\gamma_0^{i}} \left(\frac{\tau^{i}}{\tau_c^{i}} \right)^{m}
\label{eq:strainrate}
\end{equation}

\noindent where $\dot{\gamma_0^{i}}$ is a reference applied strain rate and $m$ a coefficient accounting for the material strain rate sensitivity. In the case of FCC metals at room temperature, strain rate sensitivity is rather low and is associated with the formation of jogs along dislocation lines. Here, we use typical values of $\dot{\gamma_0^{i}} = 10^{-4}\;\mathrm{s}^{-1}$ and $m > 35$. The applied strain rate was set to be $n_{sys}\times\gamma_0^{i}$, with $n_{sys}$ the number of active slip systems. Under these conditions, there is no artificial over or under Hardening induced by the strain rate sensitivity. In other words, $\tau^i$ equals $\tau^i_c$ when the system is operating. 

\begin{figure}[htbp]
\includegraphics[width = 0.31\linewidth]{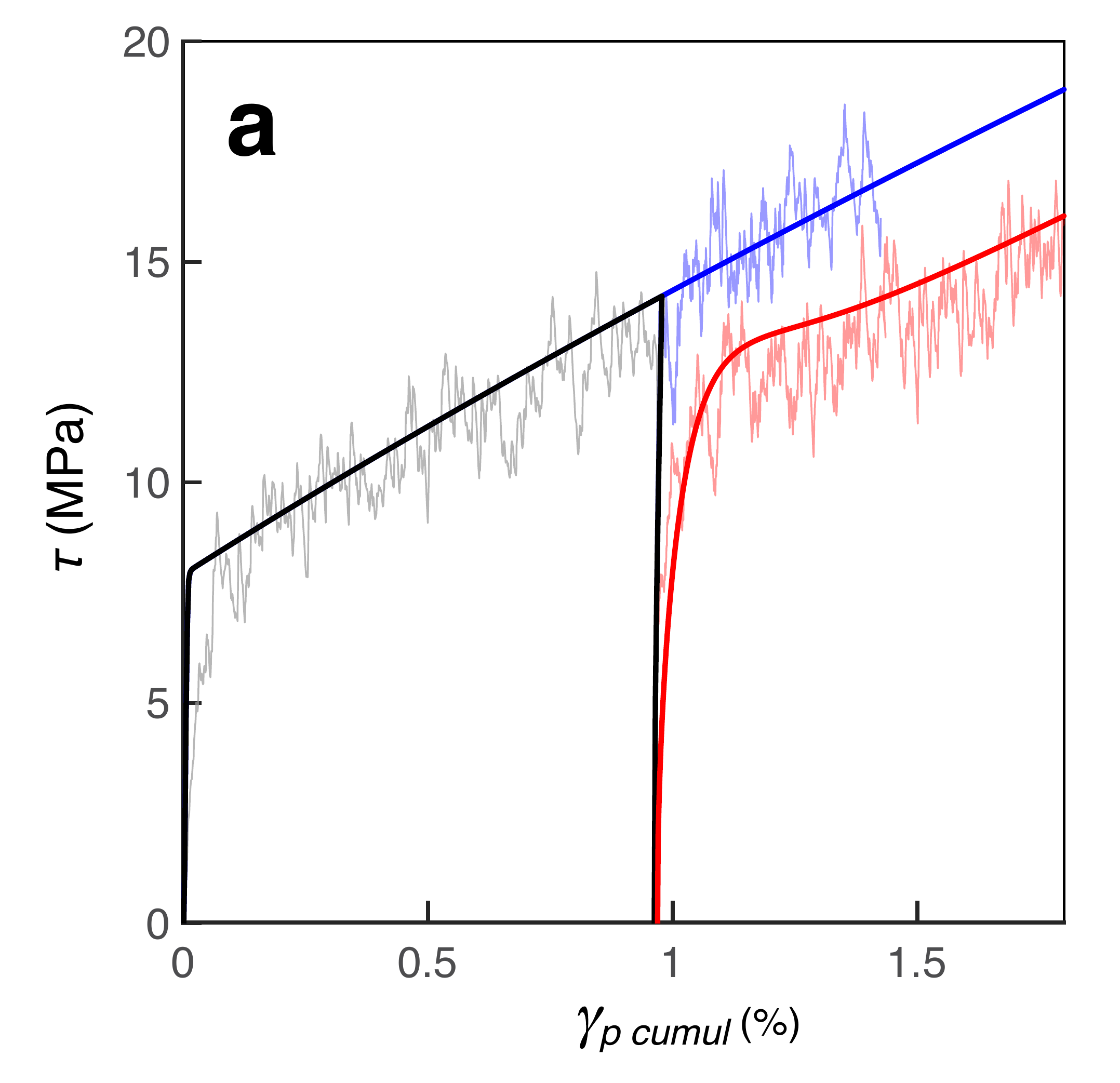}
\includegraphics[width = 0.32\linewidth]{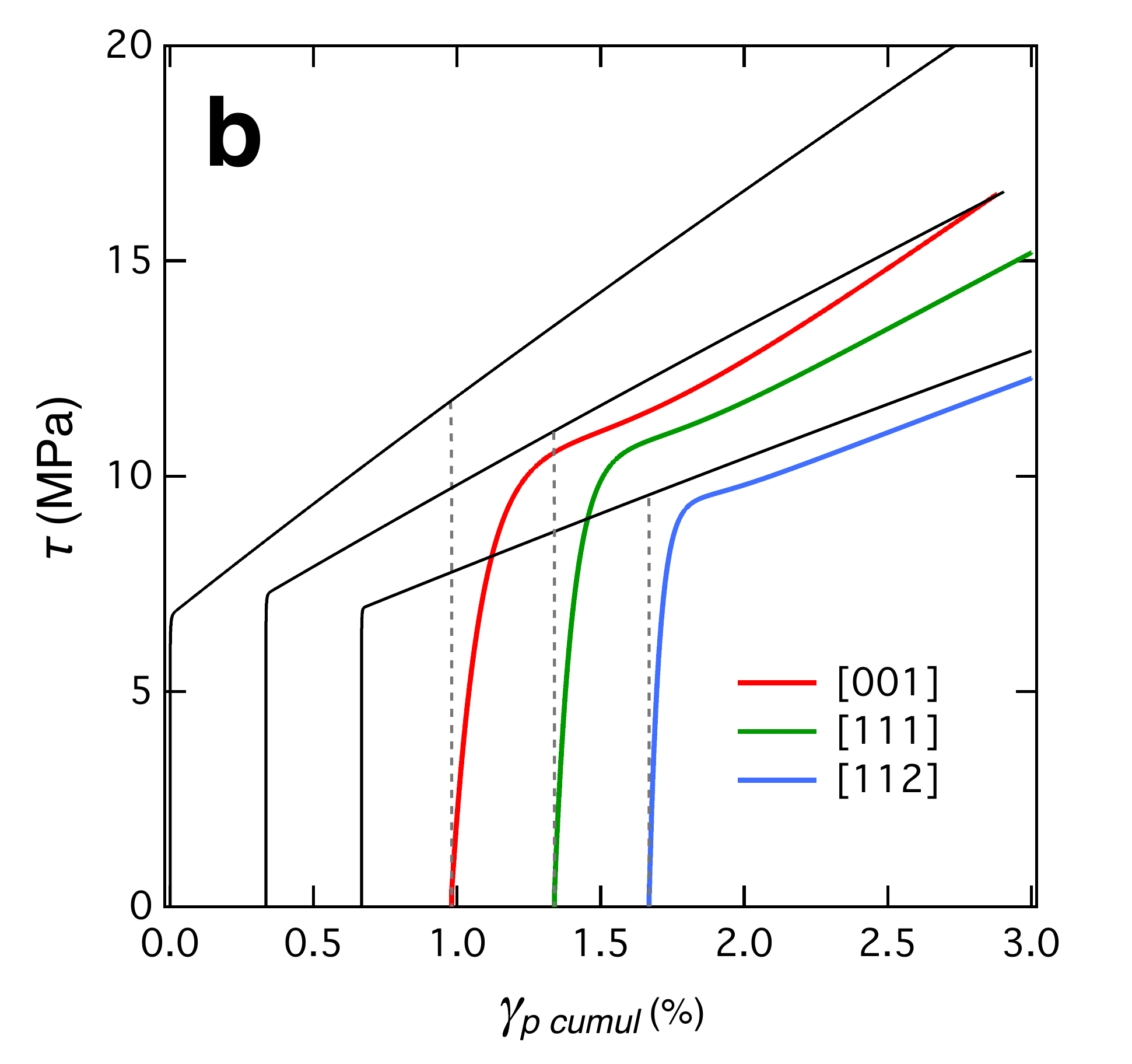}
\includegraphics[width = 0.32\linewidth]{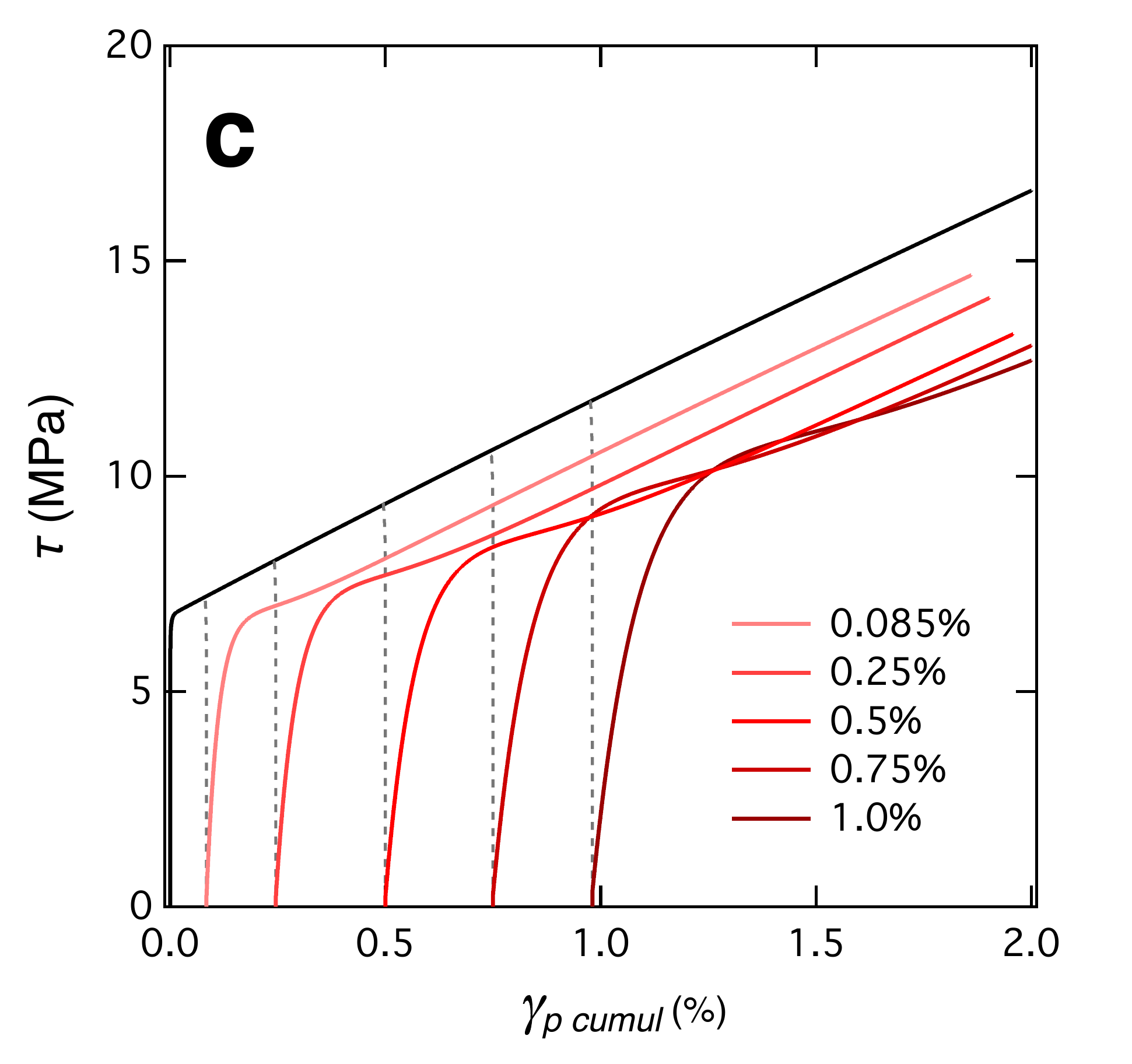}
\caption{Simulations of BE tests with our CP model physically justified from the DDD simulation results. a, direct comparison of DDD simulations (thin lines) of the BE test for the [001] axis with the CP model (thick lines), whose parameters $C_a$, $A_p$ and $C_p$ have been identified independently from the full set of simulations. b, Influence of the loading orientation on the hysteresis curves for a Ni single crystal deformed along high symmetry axis after identical prestrain. Continued tension curves in black are compared with subsequent compression curves in color. c, Effect of an increasing prestrain on the following compression curves for a $[001]$ deformation.}
\label{fig:3}
\end{figure}

The constitutive equations of the identified CP model were implemented and completed in the Z-SeT Finite Element software and Matlab using isotropic elasticity for comparisons with the DDD simulations. As the non-linear equations of the model are rather stiff and may lead to numerical instabilities for large number of deformation cycles, we employ a double nested Newton-Raphson implicit scheme for finding $\sigma_i$ and $\rho_i$. Gradients required for the Newton-Raphson algorithm are implemented analytically, with a numerical convergence criterion of $10^{-7}$. This resolution strategy yields convergence with high numerical efficiency at a reasonable numerical cost. All calculations have been made using a representative volume element with a single integration point for reason of simplicity and to be able to simulate few $10^4$ deformation cycles. The study of finite geometry effects associated to realist sample dimensions is left for a latter investigation.

To check for the self-consistency of the model and its implementation, we performed CP calculations dealing with configurations identical to those considered in the DDD simulations for a one-to-one confrontation. All DDD curves of BE simulation 
can be reproduced with a nice agreement, provided that the increase of dislocation density during prestrain $\Delta \rho_i$ is captured. Figure \ref{fig:3}.a
illustrates such a comparison in the case of the [001] BE deformation curves. The CP model captures well both the shape of the backward deformation curve and the amplitude of the BE. The same figure also shows the evolution of the shape of deformation curves with the choice of loading direction (fig.~\ref{fig:3}.b) and with the prestrain amplitude (fig.~\ref{fig:3}.c). Next, we extend the comparison of our simulations results to more varied configurations corresponding to experimental conditions.

\section{Modelling the Bauschinger effect at the macroscale}
\label{sec:BEmacro}

In this section, we demonstrate that the two mechanisms identified at the elementary dislocation scale are sufficient to reproduce the BE observed in the experimental literature on fcc single crystals. However, to achieve this goal, we need to define a quantitative metric to assess the BE in our simulations and in experiments. As mentioned earlier, several parameters have been proposed in the literature. But due to the transient nature of the BE, and as both prestrain and backward deformation curves may exhibit non-linearities, using one or a few parameters to characterize the hysteresis loop is necessarily limiting. This being said, for the purpose of the present comparison, we will simply use the two parameters presented earlier in the shape of the Bauschinger strain shift $\epsilon_{BE}$ and permanent recovery $\Delta \sigma_{BE}$. These parameters have been used in the experimental literature, and our simulations did not highlight any issues in using them. $\Delta \sigma_{BE}$ and $\epsilon_{BE}$ can be easily calculated from our simulations. In the case of experimental studies where these values were not directly reported, great care has been paid in analysing the deformation curves and in providing estimates of the $\Delta \sigma_{BE}$ and $\epsilon_{BE}$ parameters. 

\begin{figure}[htbp]
\includegraphics[width = 0.47\linewidth]{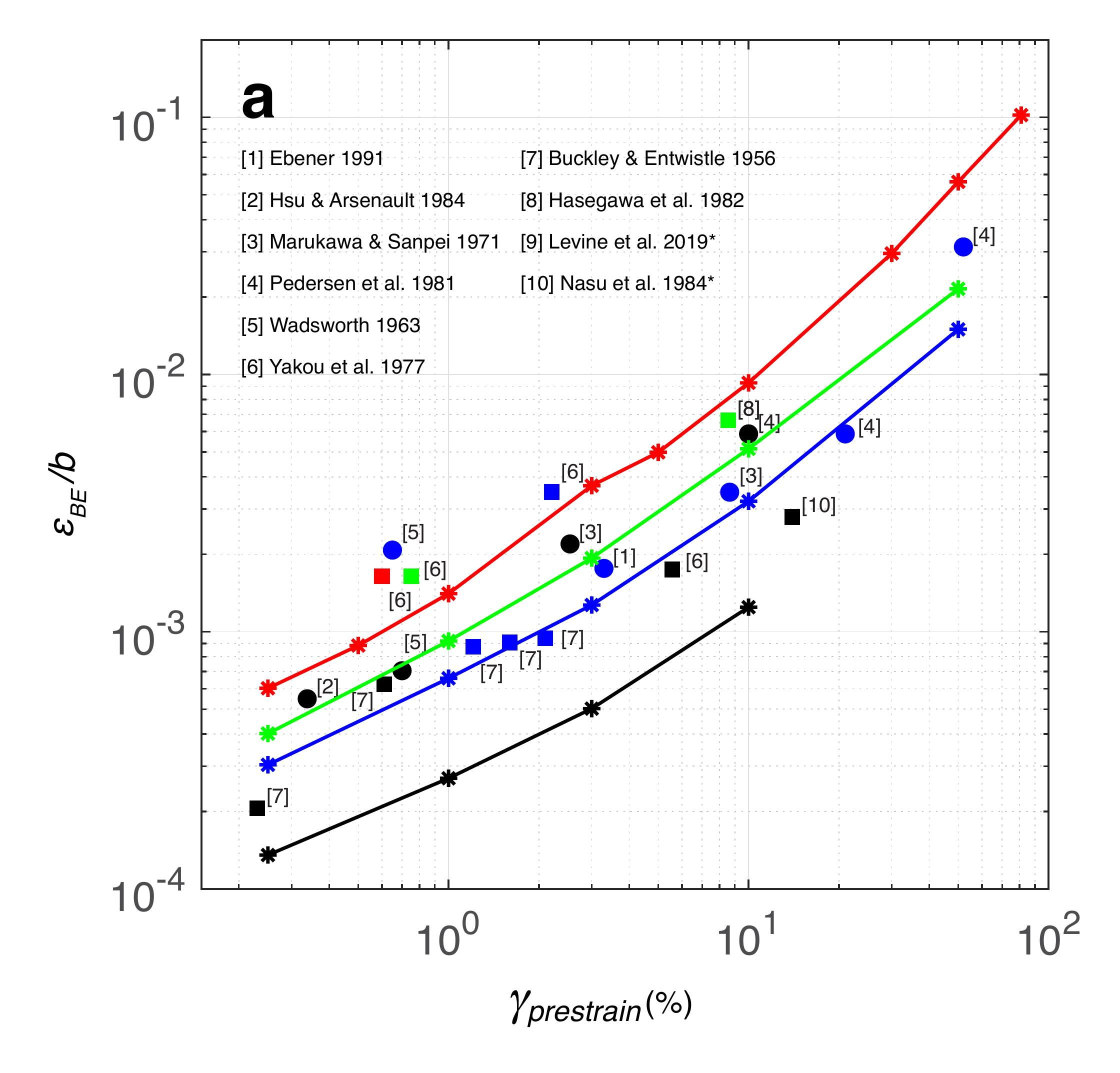}
\includegraphics[width = 0.51 \linewidth]{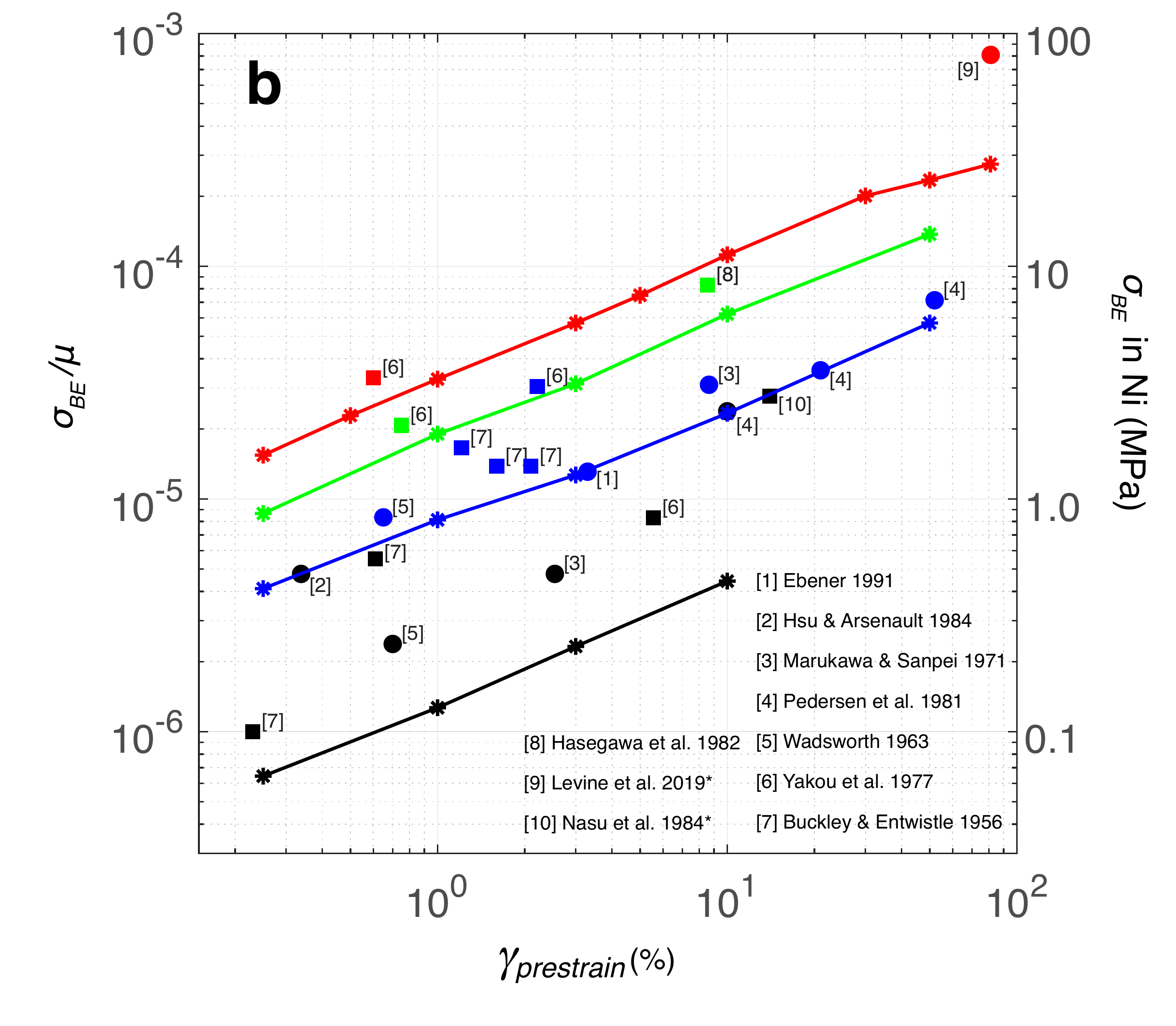}
\caption{Comparison of the experimental evaluations of the BE for Al (squares) and Cu (circles) single crystals for various slip conditions and prestrains. Data color corresponds to the slip conditions, with black points corresponding to single glide (SG), blue points to [112] or stage II double slip condition (DG), green and red points for [111] and [001] multislip conditions, respectively. For all the reported data, we checked that the hardening rate during prestrain was consistent with the reported slip conditions ($\mu /k $ with $k \approx$ 3000, 250, 200 and 150 for SG, DG, [111] and [001] loading conditions, respectively). Data marked with an asterisk do not correspond to uniaxial deformation conditions but seemed relevant here for a qualitative comparison. Provided deformation curves have been scanned and digitized into high resolution pictures to get transformed into data sets for the analysis. CP results are shown as thick lines with the same color coding. a) deformation shift $\epsilon_{BE}$ and b) permanent recovery $\sigma_{BE}$ (divided by $\mu$) as function of prestrain. The log-log scale was required to cover the large range of data. It must be emphasized that for small prestrain $< 1\%$, values for $\epsilon_{BE}$ and $\sigma_{BE}$ are rather small. $\sigma_{BE}$ values are also given in MPa for Ni to give an idea of the actual range of values. See text for more details.}
\label{fig:BEexp}
\end{figure}

The result of this analysis is presented in Fig~\ref{fig:BEexp} and regroups results from \cite{Buckley:1956,Wadsworth:1963,Marukawa:1971, Yakou:1985,Pedersen:1981,Hasegawa:1982, Hsu:1984,Nasu:1984,Ebener:1991, Levine:2019}. A first observation that can be made is that experimental values of the BE are rather scattered. As usual, experiments on single crystals are very delicate and deformation curves can be affected by several experimental conditions that may or may not be controllable, such as materials impurity content, initial dislocation density and microstructure, precise single crystal orientation or the ability of the experimental set-up to maintain a uniaxial deformation by accompanying the rotation of the crystal. We will see latter that the response of single crystals over one Bauschinger cycle is much more impacted by initial dislocation density than the cyclic deformation, among other things. As a result, the relative scatter in experimental values for $\sigma_{BE}$ and $\epsilon_{BE}$ is close to a factor of 4 for about the same prestrain and orientation. More data exist on SG \cite{Buckley:1956, Wadsworth:1963, Marukawa:1971,Hsu:1984,Yakou:1985} and DG orientations \cite{Buckley:1956,Wadsworth:1963,Marukawa:1971,Pedersen:1981}, while studies on high symmetry orientations are really seldom \cite{Yakou:1985, Levine:2019}. We nevertheless recover from this review of existing investigations some conclusions drawn in individual experimental studies. First, both strain shift $\epsilon_{BE}$ and the permanent recovery $\Delta \sigma_{BE}$ increase with the amount of deformation set in the prestrain. Besides, a clear hierarchy can be found among orientations with $\sigma_{BE}$ increasing with the number of active slip systems, for a given prestrain. The existence of hierarchy among orientation for the $\epsilon_{BE}$ is less certain from the experiemental data, and this parameters seems thus mostly controlled by the prestrain.

Despite the reservations expressed above concerning the data dispersion and the difficulties related to the experiments, the comparison with our simulation results is very favorable. The simulations have been performed using the parameters provided in Table~\ref{tab:paramCP}. The CP model predicts a monotonous -but non linear- increase of $\epsilon_{BE}$ and $\Delta \sigma_{BE}$ with prestrain. The hierarchy between orientations is also clearly recovered. But most importantly, the quantitative values predicted by the CP model are in good agreement with experimental data. It must be emphasized that this comparison is carried over 2 orders of magnitude of prestrain, and includes key single crystal orientations. To our knowledge, this has never been accomplished before.

Our CP model can provide invaluable insights about the Bauschinger behavior, which are typically inaccessible otherwise. The BE increases with strain as the stored density $\Delta \rho_i$ during prestrain increases and constitutes a pool of unstable junctions and recoverable dislocation segments. The same reasoning is the explanation for the hierarchy among orientation, as orientations associated to larger dislocation storage lead to a stronger BE. We can also now quantify the relative importance of the two elementary mechanisms at the origin of the BE. The decrease of $p_0$ contributes to a constant $\approx 90\%$ of $\sigma_{BE}$, while the asymmetry in junction stability controls the entirety of the microsplastic behavior until reaching the backward flow stress. The $\epsilon_{BE}$ which corresponds typically to a point of the deformation curves past the backward flow stress is a mixture of the two effects: with contributions 40\%-60\% coming from the -junction asymmetry or reduced MFP- effects for a prestrain of 5\%. These contributions become 25\%-75\%  for a prestrain of 50\%. We have tested the dependence of the BE on the initial dislocation density ranging from $10^{11}$ to $10^{13}$ m$^{-2}$, for small prestrain about 1\%, the observed BE as described by $\sigma_{BE}$ and $\epsilon_{BE}$ may increase by a factor of two, providing an explanation for the large spread. As prestrain increases, the impact of the initial density vanishes. 

Finally, the model is also capable to reproduce subtle effects such as the slight bump appearing at the end of the microsplastic regime on most backward curves of Fig.~\ref{fig:3}. This is also a feature that can be observed in many experimental deformation curves (see for example \cite{Ebener:1991, Daniel:1971, Nasu:1984}). This has been mostly overlooked until now but the existence of such a non linearity is the result of the existence of two distinct mechanisms. This bump is also present on the DDD curves but it is slightly masked by the serrations existing on the simulated deformation curves. In the present CP model, the bump coincides with the end of the transient on weakened junction stability as $r_a$ goes to zero. This is confirmed by the fact that this bump disappears when either of the two elementary mechanisms -junction asymmetry or reduced MFP- is turned off in the model. This bump is thus a synergy effect of the two mechanisms that are associated to different transient durations (related to $C_a$ and $C_p$ in the model). The fact that this is an effect observed experimentally is another demonstration of the reality of the effects discussed here.

\section{Consequences for the cyclic deformation of single crystals}
\label{sec:cyclic}

\begin{figure}[htbp]
\includegraphics[width = 0.45\linewidth]{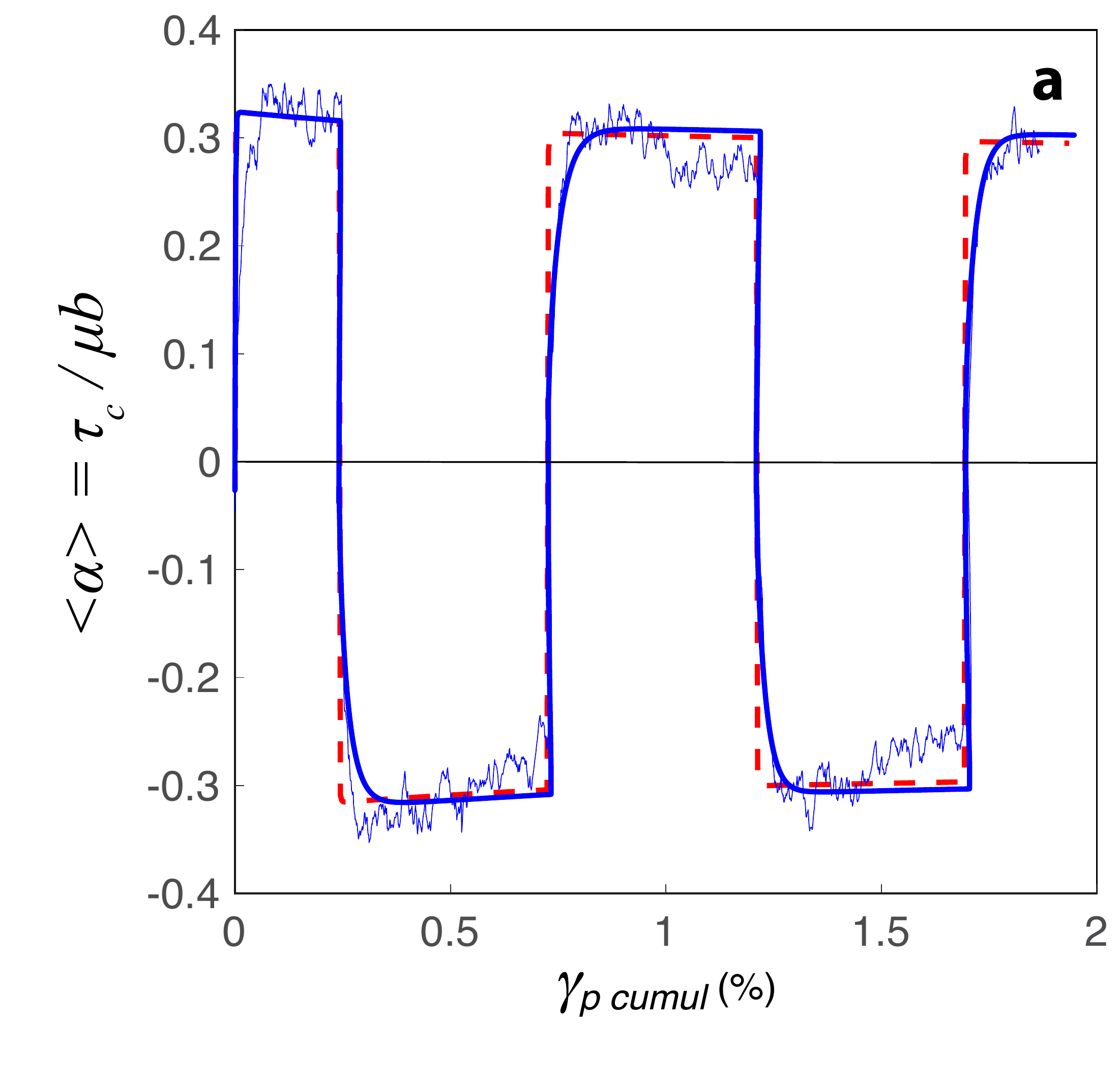}
\includegraphics[width = 0.45\linewidth]{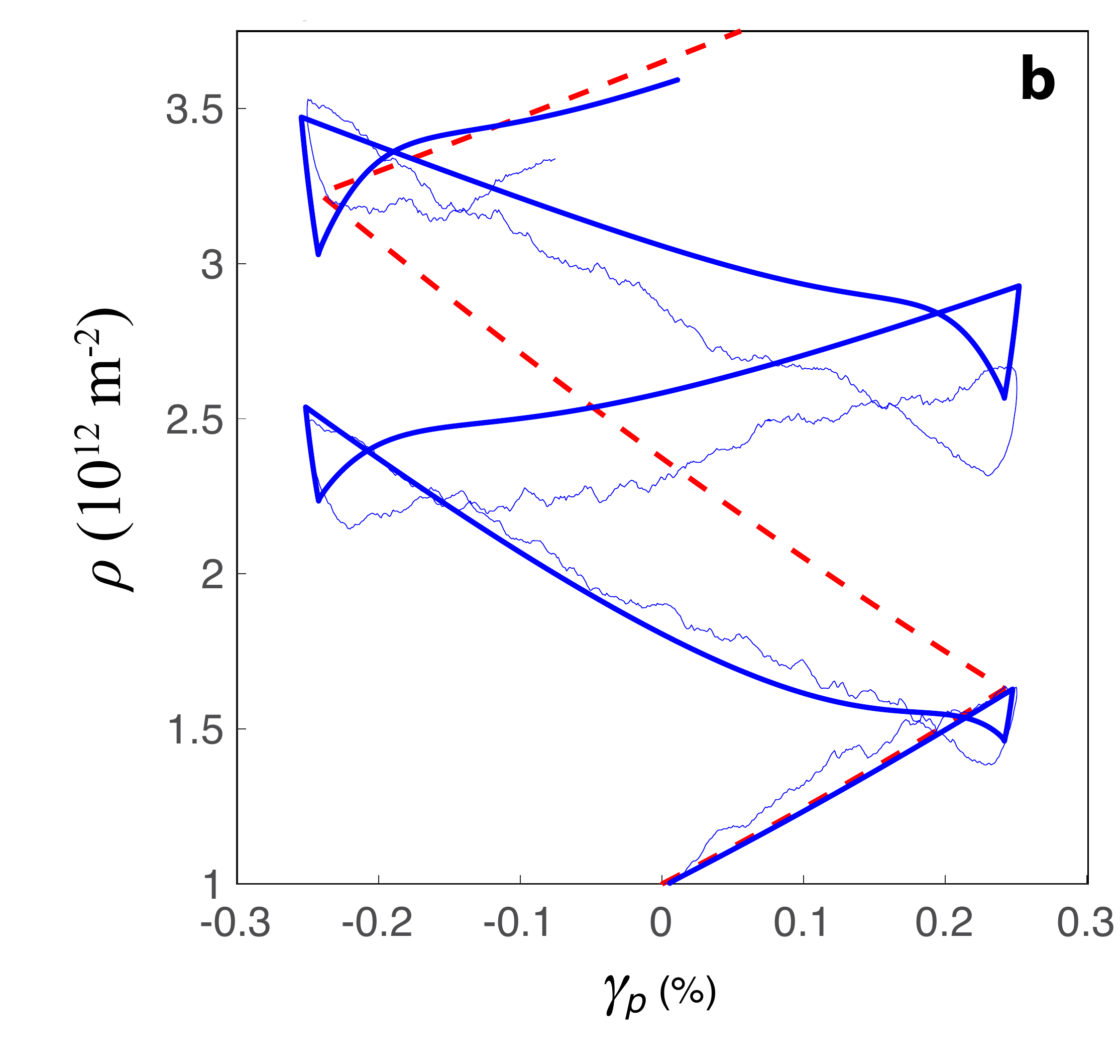}
\caption{Cyclic deformation of a [001] single crystals. The DDD simulation (thin blue lines) is compared with our CP predictions including BE (thick blue lines) or when BE are turned off (dashed red lines). a, evolution of the average interaction coefficient $\sqrt{\bar{a}}=\alpha$ as function of the accumulated strain $\gamma_{p,cumul}$. In DDD, the plastic activity on all possible slip systems is not perfectly balanced at the end of several cycles, which explain the lowering of $<\alpha>$ or the reduced storage rate on some cycles. In agreement with Fig.~\ref{fig:2causes}.a the lowering of $<\alpha>$ is rather small for such a strain amplitude of 0.25\% per half cycle. b, evolution of the total dislocation density with plastic strain $\gamma_p$. A nice agreement can be found between DDD and CP results. %\textcolor{red}{ better agreement would be found if using slip activity from DDD in the CP instead of the constitutive equation. }
}
\label{fig:cyclicdefDDD}
\end{figure}

In this final section, we assess the impact of the two elementary mechanisms identified when investigating the BE on the cyclic deformation of single crystals. For this we reuse the model presented in section \ref{sec:CP}, and the parameters from table \ref{tab:paramCP}. The prestrain corresponds now to the previous cycle of deformation, and the backward deformation is the current cycle. For comparison we also simulated using DDD the cyclic behavior of [001] and [135] single crystals for few cycles at a strain of 0.25\%, see Fig.~\ref{fig:cyclicdefDDD}. In the first few cycles of deformation, features similar to the ones observed during a single Bauschinger cycle were found with a lowered interaction strength and a drop in dislocation density after each stress reversal. These effects increase progressively in intensity following the increase of stored $\Delta{\rho_i}$ over previous cycle. Our CP model reproduces well the new evolution of $a_{ij}$ and dislocation density observed in the new DDD simulations.

\begin{figure}[htbp]
\includegraphics[width = 0.8\linewidth]{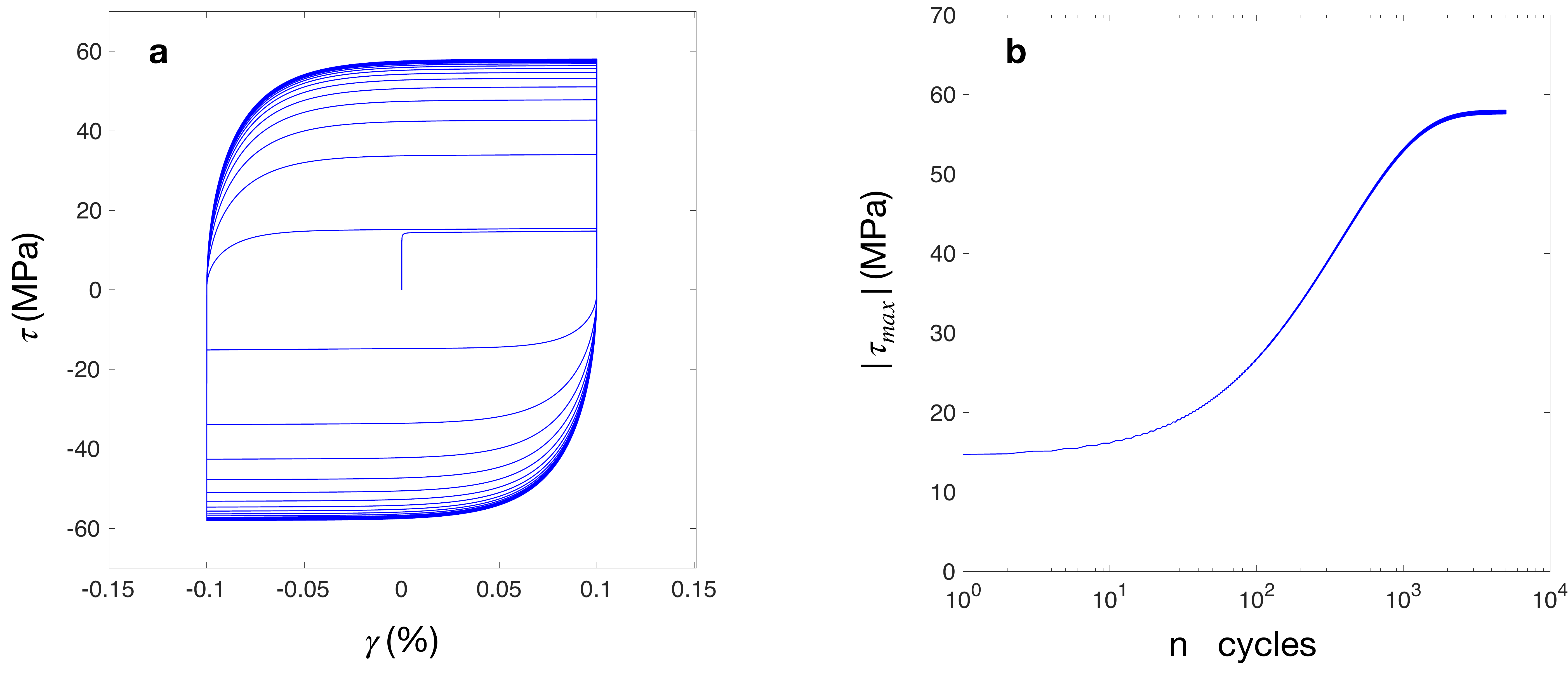}
\caption{{a. Evolution of hysteresis behavior of the deformation curves every 100 cycles. Cyclic deformation with $\gamma_{p,cy} = 0.1\%$ along the [001] axis is simulated using the CP model in the case of Ni.  b. Corresponding evolution of $|\tau_{max}|$ value found in each half-cycle as function of the number of cycles.}}
\label{fig:cyclicdefCP}
\end{figure}

The saturation stress $\tau_{sat}$ constitutes a key physical quantity for the cyclic deformation of materials for several reasons. First, by definition, $\tau_{sat}$ corresponds to the moment when the mechanical response of the deformed material does not evolve with alternating loading directions. A perfect balance is thus found between the hardening mechanism occurring during a cycle and the dynamic recovery associated to each stress reversal. Second, as the mechanical response does not evolve, the dislocation microstructure remains mostly unchanged during a very large range of cumulative deformation. Hence, in the absence of a damage processes, a deformation can theoretically be applied indefinitely and this may be exploited for engineering purposes. Understanding the saturation stress is then a necessary step in understanding the fatigue of materials and improving materials design \cite{mughrabi:2010}. Third, the value of $\tau_{sat}$ is much less impacted by the initial state of single crystals (contrary to say $\sigma_{BE}$ and $\epsilon_{BE}$) as similar $\tau_{sat}$ values have been obtained by different teams over several decades, at least for the very stable single glide condition. $\tau_{sat}$ is then closer to a materials intrinsic property than to an empirical engineer measure. The saturation  $\tau_{sat}$ has however a complex evolution, which is function of prestrain and loading direction (see review in \cite{Li:2011}). 

\begin{figure}[htbp]
\includegraphics[width = 0.98\linewidth]{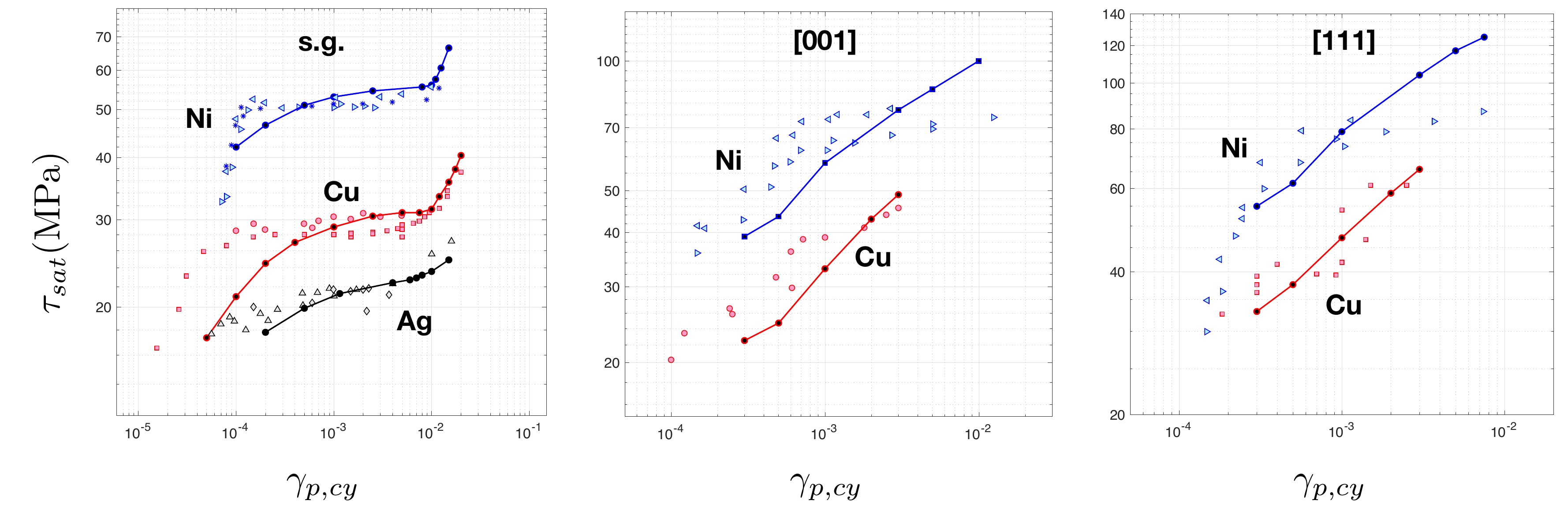}
\caption{Comparison of the CP prediction (continous lines) of the saturation stress $\tau_{sat}$ with experimental values on most of fcc single-crystal investigations (\cite{Cheng:1981, Mughrabi:1978,Lepisto:1986,Gong:1997,Bretschneider:1997, Li:2009, Li:2011}), for various loading conditions, including single glide, $[001]$ and $[111]$ axis.}
\label{fig:tausat}
\end{figure}

Figure \ref{fig:tausat} demonstrates that our CP model is capable of predicting quantitatively the evolution of $\tau_{sat}$ observed experimentally for fcc single crystals. For single glide loading directions, the curves are known to exhibit three stages with rather well defined boundaries depending upon the material \cite{Cheng:1981, Mughrabi:1978,Bretschneider:1997, Li:2009, Li:2011}. $\tau_{sat}$ first increases up to $\gamma_{cy}$ about $10^{-4}$, then $\tau_{sat}$ reaches a plateau and only slowly increases over two decades from $10^{-4}$ up to about $10^{-2}$. Interestingly the height of the quasi-plateau scales well with the shear modulus $\mu$ of materials \cite{Li:2011}. For imposed $\gamma_{cy}$ corresponding to the plateau, persistent slip bands (PSB) progressively form and their volume fraction increase until representing  100\% of the microstucture \cite{Mughrabi:1978,Li:2011}. Then, for $\gamma_{cy}$ larger than about $10^{-2}$, a third stage sees again an increase of $\tau_{sat}$ and coincides with deformation large enough to activate the conjugated slip system. When now considering multislip conditions, the cyclic behavior is a bit less documented for loading directions corresponding to [001] \cite{Bretschneider:1997,Gong:1997,Li:2011} and [111] Axis \cite{Bretschneider:1997,Lepisto:1986,Li:2011}. For these directions, $\tau_{sat}$ increases monotonically with $\gamma_{cy}$, without any plateau, the rate of increase seems however to slow down for Ni. Once again, the proposed CP predictions agree well with experimental data over a large panel of configurations covering more than 2 orders of magnitude for $\gamma_{cy}$, various loading directions and for three different materials. To the authors' knowledge this has never been accomplish before.

The CP model provides also some new mechanistic explanations of the observed cyclic behaviour. In the reported calculations, the saturation stress is controlled by the two new mechanisms discovered in the present work and also by the more classical dynamic recovery process accounted for in the simulations with the critical dislocation annihilation distance $y_c$. With the increase of cumulative plastic strain, the dislocation density increases and thus the density increment $\Delta \rho_i$ over a half cycle increases with it. With $\Delta \rho_i$, the intensity of the two effects at the origin of the BE increase in intensity and duration (through $r_a$ anf $r_p$ functions). Quite counter intuitively, the fact that lower $\tau_{sat}$ values are obtained for small $\gamma_{cy}$ do not mean that these new effects are rather stronger at small strains. Rather, it is the consequence of the competition between the transient on these effects and the length of the imposed cycle. Indeed, for small $\gamma_{cy}$, and as the cycles number is increased, the imposed strain becomes not long enough to recover the now large effects associated to $\Delta \rho_i$. The parameter $r_p$, in particular, does not vanish before the end of the cycle. This is not the case for larger $\gamma_{cy}$, that are 'long' enough to recover the normal 'monotonous' hardening rate at the end of each half-cycles. The evolution of $\tau_{sat}$ as function of $\gamma_{cy}$ is thus entirely controlled by the new mechanisms. When these effects are turned off, the saturation stress is constant and correspond logically to the monotonic theoretical saturation stress since the dependence upon deformation history -through $\Delta \rho_i$, $r_a$ and $r_p$- is removed. For single glide conditions, this asymptotic behavior is located just above the plateau observed during stage II. The competition between the transient on these new effects and the cycle duration is also the explanation of the changes in saturation hysteresis loops as function of $\gamma_{cy}$, with rounder shaped curves found for small $\gamma_{cy}$, and more trapezoidal shape found for large $\gamma_{cy}$ that remain longer that the transients of the two effects (see last figure of the joint letter \cite{queyreau:2021}). 

The model captures again very subtle effects such as a slight difference in $|\tau_{max}|$ for tension and compression as deformation starts with an extra half cycle in tension, this difference in deformation history is never recovered. This has also been observed experimentally in \cite{Kemsley:1960, Wadsworth:1963}. This slight difference in the history of deformation for cycles in tension and compression, can also be observed in the slight asymmetric shape existing on the hysteresis loops at saturation \cite{Mughrabi:1978}. At the difference of the BE observed on one tension-compression cycle, the saturation stress predicted by the CP model is virtually independent of the change in initial dislocation density. The usage of larger initial dislocation density will speed up convergence toward saturation, but will not change the saturation value. This confirms that cyclic deformation curves are more easily reproducible that single BE tests. While for monotonous loading, the shear stress of crystals oriented along [001] axis is above the flow stress of [111] oriented crystals, at least initially; here $\tau_{sat}$ for [001] deformation is always lower than that of [111] deformations. This is due to i) the Bauschinger permanent recovery that increases with the stored density and hardening rate stronger for [001], and ii) the dynamic recovery $y_c$ is much smaller for [111].

\section{Conclusions}

In this work, we have proposed comprehensive set of realistic DDD simulations of the Bauschinger experiments in single crystals of fcc metals, whose deformation is entirely controlled by dislocation-dislocation interactions. The large BE observed here is in agreement with what is observed experimentally. We proposed a detailed and precise analysis of the internal stresses felt by dislocations, through the proper integration of the Peach and Koehler force along the dislocation network. This analysis rules out the presence of backstresses in single crystals as a possible explanation of the BE observed. Instead, we identified two original causes for the BE in the shape of an excess of weak junctions due to an asymmetry in the junction stability formed under a signed applied stress, and a reduced dislocation storage due to the unwinding of stored segments, and loop collapses. These two mechanisms provide a clear and direct explanation of the BE, and allow for an unambiguous explanation of the transient and reversibility aspects of the BE. Most unstable junctions are destroyed first and most unstable loops are unstored first. These configurations are inherited from prestrain, and progressively get destroyed or mobile dislocations start exploring new regions of the crystal. 

From a statistical analysis of these mechanisms in the full set of DDD simulations, we derived a physically based CP framework including two reversibility functions and with three adimensional constants measuring the duration of the transient. This CP model allows to reproduce most of the experimental data existing on Bauschinger tests and allows to draw clear trends for the evolution of the BE in fcc single crystals. As these new mechanisms are universal, we assessed their impact on the cyclic behaviour of single crystals, and we demonstrated that these effects are sufficient to provide a quantitative prediction of the saturation stress observed for fcc single-crystals deformed in various conditions of strain increment and loading directions. The CP model provides also mechanistic explanations for key features of the cyclic deformation such as the evolution of the saturation stress with strain increment, the hierarchy existing in the cyclic behaviour observed in various loading directions and the different shapes observed in the hysteresis loops at saturation.

First, the CP model developed here has real predictive capability for macroscale simulations. It is not associated to any adhoc parameters in the sense that the functional that complement the classical constitutive equations of CP exhibit a clear physical meaning and have been suggested by mesoscale simulations. The three additional dimensionless parameters $C_A, C_p, A_p$ considered in the CP model are fitted on relevant curves from DDD simulations only. No parameters was fitted on any BE or cyclic experimental curves as it is common practice in many mechanical investigations. The only parameter coming in part from the macroscale is the critical distance of annihilation for dislocations $y_c$, but this latter parameter has been defined from monotonous experimental values following an approach from a completely different study. The present work is deliberately a bottom-up approach. Interestingly, the CP simulation results are indeed rather dependent upon the quantitative values for these parameters and slightly modified values quickly lead to unrealistic values for $\tau_{sat}$ for example. The fact that the CP model predicts so well experimental data over so many different configurations and orientations is thus not only due to the right choice of functionals but also to the proper order of magnitude for the various parameters, the largest part of these parameters are coming from the literature.

The two new mechanisms discovered here in the shape of junction stability asymmetry and reduced MFP can be classified as short range mechanisms of dislocations. They come from very general properties of dislocations (mechanical balance and constrained motion of dislocations connected in an already existing network), and thus are likely to be very general effects too. The agreement of our DDD results with experiments on cyclic deformation can thus be surprising at first. However, our set of DDD simulations of the Bauschinger experiments correspond to strain over a single cycle in the range [0.25-8\%], which turns out to cover most of the domain of plastic increment per cycle $\gamma_{cy}$ considered in the experimental literature. Therefore, if no additional effects appears, the limited strains typically simulated thanks to DDD are sufficient to rationalize macroscale results on thousands of cycles corresponding to an accumulated deformation of several hundreds of percents. A second contradiction, in appearance, is that deformed microstructures in cyclically deformed crystal exhibit very marked patterns in the shape of veines, PSB or mazes. Such patterns are commonly thought to be associated to Long-range internal stress (LRIS). However, there is no reason why the general short-range effects discovered here would not stay operative, even in such patterns. These mechanisms constitute then the first order explanation to the $\tau_{sat}$ evolution. For the larger increment of strain, there is no doubt that patterning and localisation effects becomes important, and their lack in the present study could explain the over-evaluation of our CP model of the experimental saturation stress for large prestrain and $\gamma_{cy}$.

\bibliography{Bauschinger_Suplement.bib}

\begin{thebibliography}{60}%
\makeatletter
\providecommand \@ifxundefined [1]{%
 \@ifx{#1\undefined}
}%
\providecommand \@ifnum [1]{%
 \ifnum #1\expandafter \@firstoftwo
 \else \expandafter \@secondoftwo
 \fi
}%
\providecommand \@ifx [1]{%
 \ifx #1\expandafter \@firstoftwo
 \else \expandafter \@secondoftwo
 \fi
}%
\providecommand \natexlab [1]{#1}%
\providecommand \enquote  [1]{``#1''}%
\providecommand \bibnamefont  [1]{#1}%
\providecommand \bibfnamefont [1]{#1}%
\providecommand \citenamefont [1]{#1}%
\providecommand \href@noop [0]{\@secondoftwo}%
\providecommand \href [0]{\begingroup \@sanitize@url \@href}%
\providecommand \@href[1]{\@@startlink{#1}\@@href}%
\providecommand \@@href[1]{\endgroup#1\@@endlink}%
\providecommand \@sanitize@url [0]{\catcode `\\12\catcode `\$12\catcode
  `\&12\catcode `\#12\catcode `\^12\catcode `\_12\catcode `\%12\relax}%
\providecommand \@@startlink[1]{}%
\providecommand \@@endlink[0]{}%
\providecommand \url  [0]{\begingroup\@sanitize@url \@url }%
\providecommand \@url [1]{\endgroup\@href {#1}{\urlprefix }}%
\providecommand \urlprefix  [0]{URL }%
\providecommand \Eprint [0]{\href }%
\providecommand \doibase [0]{https://doi.org/}%
\providecommand \selectlanguage [0]{\@gobble}%
\providecommand \bibinfo  [0]{\@secondoftwo}%
\providecommand \bibfield  [0]{\@secondoftwo}%
\providecommand \translation [1]{[#1]}%
\providecommand \BibitemOpen [0]{}%
\providecommand \bibitemStop [0]{}%
\providecommand \bibitemNoStop [0]{.\EOS\space}%
\providecommand \EOS [0]{\spacefactor3000\relax}%
\providecommand \BibitemShut  [1]{\csname bibitem#1\endcsname}%
\let\auto@bib@innerbib\@empty
%</preamble>
\bibitem [{\citenamefont {Bauschinger}(1886)}]{Bauschinger:1886}%
  \BibitemOpen
  \bibfield  {author} {\bibinfo {author} {\bibfnamefont {J.}~\bibnamefont
  {Bauschinger}},\ }\bibfield  {title} {\bibinfo {title} {{\"U}ber die
  ver{\"a}nderung der elastizit{\"a}tsgrenze und der festigkeit des eisens und
  stahls durch strecken und quetschen, durch erw{\"a}rmen und abk{\"u}hlen und
  durch oftmals wiederholte beanspruchung},\ }\href@noop {} {\bibfield
  {journal} {\bibinfo  {journal} {Mitteilungen des mechanisch-technischen
  Laboratoriums der K{\"o}niglich Technischen Hochschule M{\"u}nchen}\ }\textbf
  {\bibinfo {volume} {13}} (\bibinfo {year} {1886})}\BibitemShut {NoStop}%
\bibitem [{\citenamefont {Brown}\ and\ \citenamefont
  {Stobbs}(1971)}]{Brown:1971}%
  \BibitemOpen
  \bibfield  {author} {\bibinfo {author} {\bibfnamefont {L.}~\bibnamefont
  {Brown}}\ and\ \bibinfo {author} {\bibfnamefont {W.}~\bibnamefont {Stobbs}},\
  }\bibfield  {title} {\bibinfo {title} {The work-hardening of copper-silica},\
  }\href@noop {} {\bibfield  {journal} {\bibinfo  {journal} {Philosophical
  Magazine}\ }\textbf {\bibinfo {volume} {23}},\ \bibinfo {pages} {1201}
  (\bibinfo {year} {1971})}\BibitemShut {NoStop}%
\bibitem [{\citenamefont {Argon}(2008)}]{Argon:2008}%
  \BibitemOpen
  \bibfield  {author} {\bibinfo {author} {\bibfnamefont {A.~S.}\ \bibnamefont
  {Argon}},\ }\href@noop {} {{\selectlanguage {english}\emph {\bibinfo {title}
  {Strengthening mechanisms in crystal plasticity}}}},\ \bibinfo {series}
  {Oxford series on materials modelling}\ No.~\bibinfo {number} {4}\ (\bibinfo
  {publisher} {Oxford Univ. Press},\ \bibinfo {address} {Oxford},\ \bibinfo
  {year} {2008})\ \bibinfo {note} {oCLC: 255673019}\BibitemShut {NoStop}%
\bibitem [{\citenamefont {Buckley}\ and\ \citenamefont {{Entwistle,
  K.M.}}(1956)}]{Buckley:1956}%
  \BibitemOpen
  \bibfield  {author} {\bibinfo {author} {\bibfnamefont {S.}~\bibnamefont
  {Buckley}}\ and\ \bibinfo {author} {\bibnamefont {{Entwistle, K.M.}}},\
  }\bibfield  {title} {{\selectlanguage {english}\bibinfo {title} {The
  bauschinger effect in super-pure aluminum single crystals and
  polycrystals}},\ }\href {https://doi.org/10.1016/0001-6160(56)90023-2}
  {\bibfield  {journal} {\bibinfo  {journal} {Acta Metallurgica}\ }\textbf
  {\bibinfo {volume} {4}},\ \bibinfo {pages} {352} (\bibinfo {year} {1956})},\
  \bibinfo {note} {publisher: Pergamon}\BibitemShut {NoStop}%
\bibitem [{\citenamefont {Pedersen}\ \emph {et~al.}(1981)\citenamefont
  {Pedersen}, \citenamefont {Brown},\ and\ \citenamefont
  {Stobbs}}]{Pedersen:1981}%
  \BibitemOpen
  \bibfield  {author} {\bibinfo {author} {\bibfnamefont {O.}~\bibnamefont
  {Pedersen}}, \bibinfo {author} {\bibfnamefont {L.}~\bibnamefont {Brown}},\
  and\ \bibinfo {author} {\bibfnamefont {W.}~\bibnamefont {Stobbs}},\
  }\bibfield  {title} {{\selectlanguage {english}\bibinfo {title} {The
  bauschinger effect in copper}},\ }\href
  {https://doi.org/10.1016/0001-6160(81)90110-3} {\bibfield  {journal}
  {\bibinfo  {journal} {Acta Metallurgica}\ }\textbf {\bibinfo {volume} {29}},\
  \bibinfo {pages} {1843} (\bibinfo {year} {1981})}\BibitemShut {NoStop}%
\bibitem [{\citenamefont {Levine}\ \emph {et~al.}(2019)\citenamefont {Levine},
  \citenamefont {Stoudt}, \citenamefont {Creuziger}, \citenamefont {Phan},
  \citenamefont {Xu},\ and\ \citenamefont {Kassner}}]{Levine:2019}%
  \BibitemOpen
  \bibfield  {author} {\bibinfo {author} {\bibfnamefont {L.~E.}\ \bibnamefont
  {Levine}}, \bibinfo {author} {\bibfnamefont {M.~R.}\ \bibnamefont {Stoudt}},
  \bibinfo {author} {\bibfnamefont {A.}~\bibnamefont {Creuziger}}, \bibinfo
  {author} {\bibfnamefont {T.~Q.}\ \bibnamefont {Phan}}, \bibinfo {author}
  {\bibfnamefont {R.}~\bibnamefont {Xu}},\ and\ \bibinfo {author}
  {\bibfnamefont {M.~E.}\ \bibnamefont {Kassner}},\ }\bibfield  {title}
  {{\selectlanguage {english}\bibinfo {title} {Basis for the {Bauschinger}
  effect in copper single crystals: changes in the long-range internal stress
  with reverse deformation}},\ }\href
  {https://doi.org/10.1007/s10853-018-03295-6} {\bibfield  {journal} {\bibinfo
  {journal} {Journal of Materials Science}\ }\textbf {\bibinfo {volume} {54}},\
  \bibinfo {pages} {6579} (\bibinfo {year} {2019})}\BibitemShut {NoStop}%
\bibitem [{\citenamefont {Kubin}(2013)}]{Kubin:2013fk}%
  \BibitemOpen
  \bibfield  {author} {\bibinfo {author} {\bibfnamefont {L.}~\bibnamefont
  {Kubin}},\ }\bibfield  {title} {\bibinfo {title} {Dislocations, mesoscale
  simulations and plastic flow},\ }in\ \href@noop {} {\emph {\bibinfo
  {booktitle} {Oxford Series On Materials Modelling}}},\ Vol.~\bibinfo {volume}
  {5},\ \bibinfo {editor} {edited by\ \bibinfo {editor} {\bibfnamefont
  {A.}~\bibnamefont {Sutton}}\ and\ \bibinfo {editor} {\bibfnamefont
  {R.}~\bibnamefont {Rudd}}}\ (\bibinfo  {publisher} {Oxford University
  Press},\ \bibinfo {year} {2013})\BibitemShut {NoStop}%
\bibitem [{\citenamefont {Wadsworth}(1963)}]{Wadsworth:1963}%
  \BibitemOpen
  \bibfield  {author} {\bibinfo {author} {\bibfnamefont {N.}~\bibnamefont
  {Wadsworth}},\ }\bibfield  {title} {{\selectlanguage {english}\bibinfo
  {title} {Work hardening of copper crystals under cyclic straining}},\ }\href
  {https://doi.org/10.1016/0001-6160(63)90004-X} {\bibfield  {journal}
  {\bibinfo  {journal} {Acta Metallurgica}\ }\textbf {\bibinfo {volume} {11}},\
  \bibinfo {pages} {663} (\bibinfo {year} {1963})}\BibitemShut {NoStop}%
\bibitem [{\citenamefont {Marukawa}\ and\ \citenamefont
  {Sanpei}(1971)}]{Marukawa:1971}%
  \BibitemOpen
  \bibfield  {author} {\bibinfo {author} {\bibfnamefont {K.}~\bibnamefont
  {Marukawa}}\ and\ \bibinfo {author} {\bibfnamefont {T.}~\bibnamefont
  {Sanpei}},\ }\bibfield  {title} {{\selectlanguage {english}\bibinfo {title}
  {Stability of the work hardened state against stress reversal in copper
  single crystals}},\ }\href {https://doi.org/10.1016/0001-6160(71)90049-6}
  {\bibfield  {journal} {\bibinfo  {journal} {Acta Metallurgica}\ }\textbf
  {\bibinfo {volume} {19}},\ \bibinfo {pages} {1169} (\bibinfo {year}
  {1971})}\BibitemShut {NoStop}%
\bibitem [{\citenamefont {Yakou}\ \emph {et~al.}(1985)\citenamefont {Yakou},
  \citenamefont {Hasegawa},\ and\ \citenamefont {Karashima}}]{Yakou:1985}%
  \BibitemOpen
  \bibfield  {author} {\bibinfo {author} {\bibfnamefont {T.}~\bibnamefont
  {Yakou}}, \bibinfo {author} {\bibfnamefont {T.}~\bibnamefont {Hasegawa}},\
  and\ \bibinfo {author} {\bibfnamefont {S.}~\bibnamefont {Karashima}},\
  }\bibfield  {title} {{\selectlanguage {english}\bibinfo {title} {Stagnation
  of {Strain} {Hardening} {During} {Reversed} {Straining} of {Prestrained}
  {Aluminium}, {Copper} and {Iron}}},\ }\href
  {https://doi.org/10.2320/matertrans1960.26.88} {\bibfield  {journal}
  {\bibinfo  {journal} {Transactions of the Japan Institute of Metals}\
  }\textbf {\bibinfo {volume} {26}},\ \bibinfo {pages} {88} (\bibinfo {year}
  {1985})}\BibitemShut {NoStop}%
\bibitem [{\citenamefont {Hasegawa}\ \emph {et~al.}(1982)\citenamefont
  {Hasegawa}, \citenamefont {Yakou},\ and\ \citenamefont
  {Kocks}}]{Hasegawa:1982}%
  \BibitemOpen
  \bibfield  {author} {\bibinfo {author} {\bibfnamefont {T.}~\bibnamefont
  {Hasegawa}}, \bibinfo {author} {\bibfnamefont {T.}~\bibnamefont {Yakou}},\
  and\ \bibinfo {author} {\bibfnamefont {U.}~\bibnamefont {Kocks}},\ }\bibfield
   {title} {{\selectlanguage {english}\bibinfo {title} {Length changes and
  stress effects during recovery of deformed aluminum}},\ }\href
  {https://doi.org/10.1016/0001-6160(82)90061-X} {\bibfield  {journal}
  {\bibinfo  {journal} {Acta Metallurgica}\ }\textbf {\bibinfo {volume} {30}},\
  \bibinfo {pages} {235} (\bibinfo {year} {1982})}\BibitemShut {NoStop}%
\bibitem [{\citenamefont {Hsu}\ and\ \citenamefont
  {Arsenault}(1984)}]{Hsu:1984}%
  \BibitemOpen
  \bibfield  {author} {\bibinfo {author} {\bibfnamefont {R.}~\bibnamefont
  {Hsu}}\ and\ \bibinfo {author} {\bibfnamefont {R.}~\bibnamefont
  {Arsenault}},\ }\bibfield  {title} {{\selectlanguage {english}\bibinfo
  {title} {The influence of surface removal on the {Bauschinger} effect}},\
  }\href {https://doi.org/10.1016/0025-5416(84)90139-3} {\bibfield  {journal}
  {\bibinfo  {journal} {Materials Science and Engineering}\ }\textbf {\bibinfo
  {volume} {66}},\ \bibinfo {pages} {35} (\bibinfo {year} {1984})}\BibitemShut
  {NoStop}%
\bibitem [{\citenamefont {Nasu}\ \emph {et~al.}(1984)\citenamefont {Nasu},
  \citenamefont {Takeda}, \citenamefont {Tominaga},\ and\ \citenamefont
  {Kato}}]{Nasu:1984}%
  \BibitemOpen
  \bibfield  {author} {\bibinfo {author} {\bibfnamefont {Y.}~\bibnamefont
  {Nasu}}, \bibinfo {author} {\bibfnamefont {T.}~\bibnamefont {Takeda}},
  \bibinfo {author} {\bibfnamefont {T.}~\bibnamefont {Tominaga}},\ and\
  \bibinfo {author} {\bibfnamefont {O.}~\bibnamefont {Kato}},\ }\bibfield
  {title} {\bibinfo {title} {The {Bauschinger} {Effect} of {Aluminum} {Single}
  {Crystals} {Tested} under {Plane}-{Strain} {Compression}},\ }\href
  {https://doi.org/10.1299/jsme1958.27.145} {\bibfield  {journal} {\bibinfo
  {journal} {Bulletin of JSME}\ }\textbf {\bibinfo {volume} {27}},\ \bibinfo
  {pages} {145} (\bibinfo {year} {1984})}\BibitemShut {NoStop}%
\bibitem [{\citenamefont {Ebener}(1991)}]{Ebener:1991}%
  \BibitemOpen
  \bibfield  {author} {\bibinfo {author} {\bibfnamefont {H.}~\bibnamefont
  {Ebener}},\ }\bibfield  {title} {{\selectlanguage {english}\bibinfo {title}
  {Acoustic emission and the {Bauschinger} effect in {Cu} single crystals}},\
  }\href {https://doi.org/10.1016/0956-716X(91)90270-B} {\bibfield  {journal}
  {\bibinfo  {journal} {Scripta Metallurgica et Materialia}\ }\textbf {\bibinfo
  {volume} {25}},\ \bibinfo {pages} {2035} (\bibinfo {year}
  {1991})}\BibitemShut {NoStop}%
\bibitem [{\citenamefont {Cheng}\ and\ \citenamefont
  {Laird}(1981)}]{Cheng:1981}%
  \BibitemOpen
  \bibfield  {author} {\bibinfo {author} {\bibfnamefont {A.~S.}\ \bibnamefont
  {Cheng}}\ and\ \bibinfo {author} {\bibfnamefont {C.}~\bibnamefont {Laird}},\
  }\bibfield  {title} {{\selectlanguage {english}\bibinfo {title} {Mechanisms
  of fatigue hardening in copper single crystals: {The} effects of strain
  amplitude and orientation}},\ }\href
  {https://doi.org/10.1016/0025-5416(81)90112-9} {\bibfield  {journal}
  {\bibinfo  {journal} {Materials Science and Engineering}\ }\textbf {\bibinfo
  {volume} {51}},\ \bibinfo {pages} {111} (\bibinfo {year} {1981})}\BibitemShut
  {NoStop}%
\bibitem [{\citenamefont {Mughrabi}(1978)}]{Mughrabi:1978}%
  \BibitemOpen
  \bibfield  {author} {\bibinfo {author} {\bibfnamefont {H.}~\bibnamefont
  {Mughrabi}},\ }\bibfield  {title} {\bibinfo {title} {The cyclic hardening and
  saturation behaviour of copper single crystals},\ }\href
  {https://doi.org/http://dx.doi.org/10.1016/0025-5416(78)90174-X} {\bibfield
  {journal} {\bibinfo  {journal} {Materials Science and Engineering}\ }\textbf
  {\bibinfo {volume} {33}},\ \bibinfo {pages} {207} (\bibinfo {year}
  {1978})}\BibitemShut {NoStop}%
\bibitem [{\citenamefont {Lepist{\"o}}\ \emph {et~al.}(1986)\citenamefont
  {Lepist{\"o}}, \citenamefont {Kuokkala},\ and\ \citenamefont
  {Kettunen}}]{Lepisto:1986}%
  \BibitemOpen
  \bibfield  {author} {\bibinfo {author} {\bibfnamefont {T.}~\bibnamefont
  {Lepist{\"o}}}, \bibinfo {author} {\bibfnamefont {V.-T.}\ \bibnamefont
  {Kuokkala}},\ and\ \bibinfo {author} {\bibfnamefont {P.}~\bibnamefont
  {Kettunen}},\ }\bibfield  {title} {{\selectlanguage {english}\bibinfo {title}
  {Dislocation arrangements in cyclically deformed copper single crystals}},\
  }\href {https://doi.org/10.1016/0025-5416(86)90283-1} {\bibfield  {journal}
  {\bibinfo  {journal} {Materials Science and Engineering}\ }\textbf {\bibinfo
  {volume} {81}},\ \bibinfo {pages} {457} (\bibinfo {year} {1986})}\BibitemShut
  {NoStop}%
\bibitem [{\citenamefont {Bretschneider}\ \emph {et~al.}(1997)\citenamefont
  {Bretschneider}, \citenamefont {Holste},\ and\ \citenamefont
  {Tippelt}}]{Bretschneider:1997}%
  \BibitemOpen
  \bibfield  {author} {\bibinfo {author} {\bibfnamefont {J.}~\bibnamefont
  {Bretschneider}}, \bibinfo {author} {\bibfnamefont {C.}~\bibnamefont
  {Holste}},\ and\ \bibinfo {author} {\bibfnamefont {B.}~\bibnamefont
  {Tippelt}},\ }\bibfield  {title} {{\selectlanguage {english}\bibinfo {title}
  {Cyclic plasticity of nickel single crystals at elevated temperatures}},\
  }\href {https://doi.org/10.1016/S1359-6454(97)00030-X} {\bibfield  {journal}
  {\bibinfo  {journal} {Acta Materialia}\ }\textbf {\bibinfo {volume} {45}},\
  \bibinfo {pages} {3775} (\bibinfo {year} {1997})}\BibitemShut {NoStop}%
\bibitem [{\citenamefont {Gong}\ \emph {et~al.}(1997)\citenamefont {Gong},
  \citenamefont {Wang},\ and\ \citenamefont {Wang}}]{Gong:1997}%
  \BibitemOpen
  \bibfield  {author} {\bibinfo {author} {\bibfnamefont {B.}~\bibnamefont
  {Gong}}, \bibinfo {author} {\bibfnamefont {Z.}~\bibnamefont {Wang}},\ and\
  \bibinfo {author} {\bibfnamefont {Z.}~\bibnamefont {Wang}},\ }\bibfield
  {title} {\bibinfo {title} {Cyclic deformation behavior and dislocation
  structures of [001] copper single crystals---{I} {Cyclic} stress-strain
  response and surface feature},\ }\href
  {https://doi.org/10.1016/S1359-6454(96)00288-1} {\bibfield  {journal}
  {\bibinfo  {journal} {Acta Materialia}\ }\textbf {\bibinfo {volume} {45}},\
  \bibinfo {pages} {1365} (\bibinfo {year} {1997})}\BibitemShut {NoStop}%
\bibitem [{\citenamefont {Li}\ \emph {et~al.}(2009)\citenamefont {Li},
  \citenamefont {Zhang}, \citenamefont {Li}, \citenamefont {Li},\ and\
  \citenamefont {Wang}}]{Li:2009}%
  \BibitemOpen
  \bibfield  {author} {\bibinfo {author} {\bibfnamefont {P.}~\bibnamefont
  {Li}}, \bibinfo {author} {\bibfnamefont {Z.}~\bibnamefont {Zhang}}, \bibinfo
  {author} {\bibfnamefont {X.}~\bibnamefont {Li}}, \bibinfo {author}
  {\bibfnamefont {S.}~\bibnamefont {Li}},\ and\ \bibinfo {author}
  {\bibfnamefont {Z.}~\bibnamefont {Wang}},\ }\bibfield  {title}
  {{\selectlanguage {english}\bibinfo {title} {Effect of orientation on the
  cyclic deformation behavior of silver single crystals: {Comparison} with the
  behavior of copper and nickel single crystals}},\ }\href
  {https://doi.org/10.1016/j.actamat.2009.06.048} {\bibfield  {journal}
  {\bibinfo  {journal} {Acta Materialia}\ }\textbf {\bibinfo {volume} {57}},\
  \bibinfo {pages} {4845} (\bibinfo {year} {2009})}\BibitemShut {NoStop}%
\bibitem [{\citenamefont {Li}\ \emph {et~al.}(2011)\citenamefont {Li},
  \citenamefont {Li}, \citenamefont {Wang},\ and\ \citenamefont
  {Zhang}}]{Li:2011}%
  \BibitemOpen
  \bibfield  {author} {\bibinfo {author} {\bibfnamefont {P.}~\bibnamefont
  {Li}}, \bibinfo {author} {\bibfnamefont {S.}~\bibnamefont {Li}}, \bibinfo
  {author} {\bibfnamefont {Z.}~\bibnamefont {Wang}},\ and\ \bibinfo {author}
  {\bibfnamefont {Z.}~\bibnamefont {Zhang}},\ }\bibfield  {title}
  {{\selectlanguage {english}\bibinfo {title} {Fundamental factors on formation
  mechanism of dislocation arrangements in cyclically deformed fcc single
  crystals}},\ }\href {https://doi.org/10.1016/j.pmatsci.2010.12.001}
  {\bibfield  {journal} {\bibinfo  {journal} {Progress in Materials Science}\
  }\textbf {\bibinfo {volume} {56}},\ \bibinfo {pages} {328} (\bibinfo {year}
  {2011})}\BibitemShut {NoStop}%
\bibitem [{\citenamefont {Sleeswyk}\ \emph {et~al.}(1978)\citenamefont
  {Sleeswyk}, \citenamefont {James}, \citenamefont {Plantinga},\ and\
  \citenamefont {Maathuis}}]{Sleeswyk:1978}%
  \BibitemOpen
  \bibfield  {author} {\bibinfo {author} {\bibfnamefont {A.}~\bibnamefont
  {Sleeswyk}}, \bibinfo {author} {\bibfnamefont {M.}~\bibnamefont {James}},
  \bibinfo {author} {\bibfnamefont {D.}~\bibnamefont {Plantinga}},\ and\
  \bibinfo {author} {\bibfnamefont {W.}~\bibnamefont {Maathuis}},\ }\bibfield
  {title} {{\selectlanguage {english}\bibinfo {title} {Reversible strain in
  cyclic plastic deformation}},\ }\href
  {https://doi.org/10.1016/0001-6160(78)90011-1} {\bibfield  {journal}
  {\bibinfo  {journal} {Acta Metallurgica}\ }\textbf {\bibinfo {volume} {26}},\
  \bibinfo {pages} {1265} (\bibinfo {year} {1978})}\BibitemShut {NoStop}%
\bibitem [{\citenamefont {D{\'e}pr{\'e}s}\ \emph {et~al.}(2008)\citenamefont
  {D{\'e}pr{\'e}s}, \citenamefont {Fivel},\ and\ \citenamefont
  {Tabourot}}]{Depres:2008fk}%
  \BibitemOpen
  \bibfield  {author} {\bibinfo {author} {\bibfnamefont {C.}~\bibnamefont
  {D{\'e}pr{\'e}s}}, \bibinfo {author} {\bibfnamefont {M.}~\bibnamefont
  {Fivel}},\ and\ \bibinfo {author} {\bibfnamefont {L.}~\bibnamefont
  {Tabourot}},\ }\bibfield  {title} {\bibinfo {title} {A dislocation-based
  model for low-amplitude fatigue behaviour of face-centred cubic single
  crystals},\ }\href
  {http://www.sciencedirect.com/science/article/B6TY2-4RY8SWD-3/2/15c31e2952dfe2f1101156c5db20975a}
  {\bibfield  {journal} {\bibinfo  {journal} {Scripta Materialia}\ }\textbf
  {\bibinfo {volume} {58}},\ \bibinfo {pages} {1086} (\bibinfo {year}
  {2008})}\BibitemShut {NoStop}%
\bibitem [{\citenamefont {Rauch}\ \emph {et~al.}(2011)\citenamefont {Rauch},
  \citenamefont {Gracio}, \citenamefont {Barlat},\ and\ \citenamefont
  {Vincze}}]{Rauch:2011kx}%
  \BibitemOpen
  \bibfield  {author} {\bibinfo {author} {\bibfnamefont {E.}~\bibnamefont
  {Rauch}}, \bibinfo {author} {\bibfnamefont {J.~J.}\ \bibnamefont {Gracio}},
  \bibinfo {author} {\bibfnamefont {F.}~\bibnamefont {Barlat}},\ and\ \bibinfo
  {author} {\bibfnamefont {G.}~\bibnamefont {Vincze}},\ }\bibfield  {title}
  {\bibinfo {title} {Modelling the plastic behaviour of metals under complex
  loading conditions},\ }\href
  {http://stacks.iop.org/0965-0393/19/i=3/a=035009} {\bibfield  {journal}
  {\bibinfo  {journal} {Modelling and Simulation in Materials Science and
  Engineering}\ }\textbf {\bibinfo {volume} {19}} (\bibinfo {year}
  {2011})}\BibitemShut {NoStop}%
\bibitem [{\citenamefont {Miguel}\ and\ \citenamefont
  {Zapperi}(2006)}]{Miguel:2006ly}%
  \BibitemOpen
  \bibfield  {author} {\bibinfo {author} {\bibfnamefont {M.~C.}\ \bibnamefont
  {Miguel}}\ and\ \bibinfo {author} {\bibfnamefont {S.}~\bibnamefont
  {Zapperi}},\ }\bibfield  {title} {\bibinfo {title} {Fluctuations in
  plasticity at the microscale},\ }\href
  {http://science.sciencemag.org/content/312/5777/1151.abstract} {\bibfield
  {journal} {\bibinfo  {journal} {Science}\ }\textbf {\bibinfo {volume}
  {312}},\ \bibinfo {pages} {1151} (\bibinfo {year} {2006})}\BibitemShut
  {NoStop}%
\bibitem [{\citenamefont {Dimiduk}\ \emph {et~al.}(2006)\citenamefont
  {Dimiduk}, \citenamefont {Woodward}, \citenamefont {Le{S}ar},\ and\
  \citenamefont {Uchic}}]{Dimiduk:2006}%
  \BibitemOpen
  \bibfield  {author} {\bibinfo {author} {\bibfnamefont {D.}~\bibnamefont
  {Dimiduk}}, \bibinfo {author} {\bibfnamefont {C.}~\bibnamefont {Woodward}},
  \bibinfo {author} {\bibfnamefont {M.}~\bibnamefont {Le{S}ar}},\ and\ \bibinfo
  {author} {\bibfnamefont {M.}~\bibnamefont {Uchic}},\ }\bibfield  {title}
  {\bibinfo {title} {Scale-free intermittent flow in crystal plasticity},\
  }\href@noop {} {\bibfield  {journal} {\bibinfo  {journal} {Science}\ }\textbf
  {\bibinfo {volume} {312}},\ \bibinfo {pages} {1188} (\bibinfo {year}
  {2006})}\BibitemShut {NoStop}%
\bibitem [{\citenamefont {Csikor}\ \emph {et~al.}(2007)\citenamefont {Csikor},
  \citenamefont {Motz}, \citenamefont {Weygand}, \citenamefont {Zaiser},\ and\
  \citenamefont {Zapperi}}]{Csikor:2007lr}%
  \BibitemOpen
  \bibfield  {author} {\bibinfo {author} {\bibfnamefont {F.}~\bibnamefont
  {Csikor}}, \bibinfo {author} {\bibfnamefont {C.}~\bibnamefont {Motz}},
  \bibinfo {author} {\bibfnamefont {D.}~\bibnamefont {Weygand}}, \bibinfo
  {author} {\bibfnamefont {M.}~\bibnamefont {Zaiser}},\ and\ \bibinfo {author}
  {\bibfnamefont {S.}~\bibnamefont {Zapperi}},\ }\bibfield  {title} {\bibinfo
  {title} {Dislocation avalanches, strain bursts, and the problem of plastic
  forming at the micromeer scale},\ }\href@noop {} {\bibfield  {journal}
  {\bibinfo  {journal} {Science}\ }\textbf {\bibinfo {volume} {318}},\ \bibinfo
  {pages} {251} (\bibinfo {year} {2007})}\BibitemShut {NoStop}%
\bibitem [{\citenamefont {Devincre}\ \emph {et~al.}(2008)\citenamefont
  {Devincre}, \citenamefont {Hoc},\ and\ \citenamefont
  {Kubin}}]{Devincre:2008lr}%
  \BibitemOpen
  \bibfield  {author} {\bibinfo {author} {\bibfnamefont {B.}~\bibnamefont
  {Devincre}}, \bibinfo {author} {\bibfnamefont {T.}~\bibnamefont {Hoc}},\ and\
  \bibinfo {author} {\bibfnamefont {L.}~\bibnamefont {Kubin}},\ }\bibfield
  {title} {\bibinfo {title} {Dislocation mean free paths and strain hardening
  of crystals},\ }\href@noop {} {\bibfield  {journal} {\bibinfo  {journal}
  {Science}\ }\textbf {\bibinfo {volume} {320}},\ \bibinfo {pages} {1745}
  (\bibinfo {year} {2008})}\BibitemShut {NoStop}%
\bibitem [{\citenamefont {Asaro}(1975)}]{Asaro:1975eu}%
  \BibitemOpen
  \bibfield  {author} {\bibinfo {author} {\bibfnamefont {R.~J.}\ \bibnamefont
  {Asaro}},\ }\bibfield  {title} {\bibinfo {title} {Elastic-plastic memory and
  kinematic-type hardening},\ }\href
  {https://doi.org/http://dx.doi.org/10.1016/0001-6160(75)90044-9} {\bibfield
  {journal} {\bibinfo  {journal} {Acta Metallurgica}\ }\textbf {\bibinfo
  {volume} {23}},\ \bibinfo {pages} {1255} (\bibinfo {year}
  {1975})}\BibitemShut {NoStop}%
\bibitem [{\citenamefont {Mughrabi}(1988)}]{Mughrabi:88}%
  \BibitemOpen
  \bibfield  {author} {\bibinfo {author} {\bibfnamefont {H.}~\bibnamefont
  {Mughrabi}},\ }\bibfield  {title} {\bibinfo {title} {Dislocation clustering
  and long-range internal stresses in monotonically and cyclically deformed
  metal crystals},\ }\href@noop {} {\bibfield  {journal} {\bibinfo  {journal}
  {Revue de Physique Appliquee}\ }\textbf {\bibinfo {volume} {23}},\ \bibinfo
  {pages} {367} (\bibinfo {year} {1988})}\BibitemShut {NoStop}%
\bibitem [{\citenamefont {Kassner}\ \emph {et~al.}(2009)\citenamefont
  {Kassner}, \citenamefont {Geantil}, \citenamefont {Levine},\ and\
  \citenamefont {Larson}}]{Kassner:2009fk}%
  \BibitemOpen
  \bibfield  {author} {\bibinfo {author} {\bibfnamefont {M.~E.}\ \bibnamefont
  {Kassner}}, \bibinfo {author} {\bibfnamefont {P.}~\bibnamefont {Geantil}},
  \bibinfo {author} {\bibfnamefont {L.~E.}\ \bibnamefont {Levine}},\ and\
  \bibinfo {author} {\bibfnamefont {B.~C.}\ \bibnamefont {Larson}},\ }\bibfield
   {title} {\bibinfo {title} {Mapping mesoscale heterogeneity in the plastic
  deformation of a copper single crystal},\ }\href@noop {} {\bibfield
  {journal} {\bibinfo  {journal} {Int. J. Mech. Sci.}\ }\textbf {\bibinfo
  {volume} {100}},\ \bibinfo {pages} {333} (\bibinfo {year}
  {2009})}\BibitemShut {NoStop}%
\bibitem [{\citenamefont {Kassner}\ \emph {et~al.}(2013)\citenamefont
  {Kassner}, \citenamefont {Geantil},\ and\ \citenamefont
  {Levine}}]{Kassner:2013fk}%
  \BibitemOpen
  \bibfield  {author} {\bibinfo {author} {\bibfnamefont {M.~E.}\ \bibnamefont
  {Kassner}}, \bibinfo {author} {\bibfnamefont {P.}~\bibnamefont {Geantil}},\
  and\ \bibinfo {author} {\bibfnamefont {L.~E.}\ \bibnamefont {Levine}},\
  }\bibfield  {title} {\bibinfo {title} {Long range internal stresses in
  single-phase crystalline materials},\ }\bibfield  {booktitle} {\emph
  {\bibinfo {booktitle} {In Honor of Rob Wagoner}},\ }\href
  {https://doi.org/http://dx.doi.org/10.1016/j.ijplas.2012.10.003} {\bibfield
  {journal} {\bibinfo  {journal} {International Journal of Plasticity}\
  }\textbf {\bibinfo {volume} {45}},\ \bibinfo {pages} {44} (\bibinfo {year}
  {2013})}\BibitemShut {NoStop}%
\bibitem [{\citenamefont {Queyreau}\ \emph {et~al.}(2010)\citenamefont
  {Queyreau}, \citenamefont {Monnet},\ and\ \citenamefont
  {Devincre}}]{Queyreau:2010}%
  \BibitemOpen
  \bibfield  {author} {\bibinfo {author} {\bibfnamefont {S.}~\bibnamefont
  {Queyreau}}, \bibinfo {author} {\bibfnamefont {G.}~\bibnamefont {Monnet}},\
  and\ \bibinfo {author} {\bibfnamefont {B.}~\bibnamefont {Devincre}},\
  }\bibfield  {title} {{\selectlanguage {english}\bibinfo {title} {Orowan
  strengthening and forest hardening superposition examined by dislocation
  dynamics simulations}},\ }\href
  {https://doi.org/10.1016/j.actamat.2010.06.028} {\bibfield  {journal}
  {\bibinfo  {journal} {Acta Materialia}\ }\textbf {\bibinfo {volume} {58}},\
  \bibinfo {pages} {5586} (\bibinfo {year} {2010})}\BibitemShut {NoStop}%
\bibitem [{\citenamefont {Devincre}\ \emph {et~al.}(2011)\citenamefont
  {Devincre}, \citenamefont {Madec}, \citenamefont {Monnet}, \citenamefont
  {Queyreau}, \citenamefont {Gatti},\ and\ \citenamefont
  {Kubin}}]{Devincre:2011fk}%
  \BibitemOpen
  \bibfield  {author} {\bibinfo {author} {\bibfnamefont {B.}~\bibnamefont
  {Devincre}}, \bibinfo {author} {\bibfnamefont {R.}~\bibnamefont {Madec}},
  \bibinfo {author} {\bibfnamefont {G.}~\bibnamefont {Monnet}}, \bibinfo
  {author} {\bibfnamefont {S.}~\bibnamefont {Queyreau}}, \bibinfo {author}
  {\bibfnamefont {R.}~\bibnamefont {Gatti}},\ and\ \bibinfo {author}
  {\bibfnamefont {L.}~\bibnamefont {Kubin}},\ }\bibinfo {title} {Mechanics of
  nano-objects}\ (\bibinfo  {publisher} {Presses de l'Ecole des Mines de
  Paris},\ \bibinfo {year} {2011})\ Chap.\ \bibinfo {chapter} {Modeling crystal
  plasticity with dislocation dynamics simulations: The 'microMegas'
  code}\BibitemShut {NoStop}%
\bibitem [{\citenamefont {Hirth}\ and\ \citenamefont {Lothe}(1992)}]{HiLo:92}%
  \BibitemOpen
  \bibfield  {author} {\bibinfo {author} {\bibfnamefont {J.~P.}\ \bibnamefont
  {Hirth}}\ and\ \bibinfo {author} {\bibfnamefont {J.}~\bibnamefont {Lothe}},\
  }\href@noop {} {\emph {\bibinfo {title} {Theory of Dislocations}}}\ (\bibinfo
   {publisher} {Krieger},\ \bibinfo {address} {Malabar (Florida)},\ \bibinfo
  {year} {1992})\BibitemShut {NoStop}%
\bibitem [{\citenamefont {Domain}\ and\ \citenamefont
  {Monnet}(2005)}]{Domain:2005}%
  \BibitemOpen
  \bibfield  {author} {\bibinfo {author} {\bibfnamefont {C.}~\bibnamefont
  {Domain}}\ and\ \bibinfo {author} {\bibfnamefont {G.}~\bibnamefont
  {Monnet}},\ }\bibfield  {title} {{\selectlanguage {english}\bibinfo {title}
  {Simulation of {Screw} {Dislocation} {Motion} in {Iron} by {Molecular}
  {Dynamics} {Simulations}}},\ }\href
  {https://doi.org/10.1103/PhysRevLett.95.215506} {\bibfield  {journal}
  {\bibinfo  {journal} {Physical Review Letters}\ }\textbf {\bibinfo {volume}
  {95}},\ \bibinfo {pages} {215506} (\bibinfo {year} {2005})}\BibitemShut
  {NoStop}%
\bibitem [{\citenamefont {Gilbert}\ \emph {et~al.}(2011)\citenamefont
  {Gilbert}, \citenamefont {Queyreau},\ and\ \citenamefont
  {Marian}}]{Gilbert:2011}%
  \BibitemOpen
  \bibfield  {author} {\bibinfo {author} {\bibfnamefont {M.~R.}\ \bibnamefont
  {Gilbert}}, \bibinfo {author} {\bibfnamefont {S.}~\bibnamefont {Queyreau}},\
  and\ \bibinfo {author} {\bibfnamefont {J.}~\bibnamefont {Marian}},\
  }\bibfield  {title} {{\selectlanguage {english}\bibinfo {title} {Stress and
  temperature dependence of screw dislocation mobility in $\alpha$-{Fe} by
  molecular dynamics}},\ }\href {https://doi.org/10.1103/PhysRevB.84.174103}
  {\bibfield  {journal} {\bibinfo  {journal} {Physical Review B}\ }\textbf
  {\bibinfo {volume} {84}},\ \bibinfo {pages} {174103} (\bibinfo {year}
  {2011})}\BibitemShut {NoStop}%
\bibitem [{\citenamefont {Queyreau}\ \emph {et~al.}(2011)\citenamefont
  {Queyreau}, \citenamefont {Marian}, \citenamefont {Gilbert},\ and\
  \citenamefont {Wirth}}]{Queyreau:2011}%
  \BibitemOpen
  \bibfield  {author} {\bibinfo {author} {\bibfnamefont {S.}~\bibnamefont
  {Queyreau}}, \bibinfo {author} {\bibfnamefont {J.}~\bibnamefont {Marian}},
  \bibinfo {author} {\bibfnamefont {M.~R.}\ \bibnamefont {Gilbert}},\ and\
  \bibinfo {author} {\bibfnamefont {B.~D.}\ \bibnamefont {Wirth}},\ }\bibfield
  {title} {{\selectlanguage {english}\bibinfo {title} {Edge dislocation
  mobilities in bcc {Fe} obtained by molecular dynamics}},\ }\bibfield
  {journal} {\bibinfo  {journal} {Physical Review B}\ }\textbf {\bibinfo
  {volume} {84}},\ \href {https://doi.org/10.1103/PhysRevB.84.064106}
  {10.1103/PhysRevB.84.064106} (\bibinfo {year} {2011})\BibitemShut {NoStop}%
\bibitem [{\citenamefont {Kubin}\ \emph {et~al.}(1992)\citenamefont {Kubin},
  \citenamefont {Canova}, \citenamefont {Condat}, \citenamefont {Devincre},
  \citenamefont {Pontikis},\ and\ \citenamefont {Br{\'e}chet}}]{kubin92}%
  \BibitemOpen
  \bibfield  {author} {\bibinfo {author} {\bibfnamefont {L.}~\bibnamefont
  {Kubin}}, \bibinfo {author} {\bibfnamefont {G.}~\bibnamefont {Canova}},
  \bibinfo {author} {\bibfnamefont {M.}~\bibnamefont {Condat}}, \bibinfo
  {author} {\bibfnamefont {B.}~\bibnamefont {Devincre}}, \bibinfo {author}
  {\bibfnamefont {V.}~\bibnamefont {Pontikis}},\ and\ \bibinfo {author}
  {\bibfnamefont {Y.}~\bibnamefont {Br{\'e}chet}},\ }\bibfield  {title}
  {\bibinfo {title} {Dislocation microstructures and plastic flow: A
  3$\mathrm{D}$ simulation},\ }\href@noop {} {\bibfield  {journal} {\bibinfo
  {journal} {Solid State Phenom.}\ }\textbf {\bibinfo {volume} {23-24}},\
  \bibinfo {pages} {455} (\bibinfo {year} {1992})}\BibitemShut {NoStop}%
\bibitem [{\citenamefont {Kocks}\ and\ \citenamefont
  {Mecking}(2003)}]{Kocks:03}%
  \BibitemOpen
  \bibfield  {author} {\bibinfo {author} {\bibfnamefont {U.}~\bibnamefont
  {Kocks}}\ and\ \bibinfo {author} {\bibfnamefont {H.}~\bibnamefont
  {Mecking}},\ }\bibfield  {title} {\bibinfo {title} {Physics and phenomenology
  of strain hardening: the fcc case},\ }\href@noop {} {\bibfield  {journal}
  {\bibinfo  {journal} {Progress in Materials Science}\ }\textbf {\bibinfo
  {volume} {48}},\ \bibinfo {pages} {171} (\bibinfo {year} {2003})}\BibitemShut
  {NoStop}%
\bibitem [{\citenamefont {Zaiser}\ and\ \citenamefont
  {Sandfeld}(2014)}]{Zaiser:2014fk}%
  \BibitemOpen
  \bibfield  {author} {\bibinfo {author} {\bibfnamefont {M.}~\bibnamefont
  {Zaiser}}\ and\ \bibinfo {author} {\bibfnamefont {S.}~\bibnamefont
  {Sandfeld}},\ }\bibfield  {title} {\bibinfo {title} {Scaling properties of
  dislocation simulations in the similitude regime},\ }\href
  {http://stacks.iop.org/0965-0393/22/i=6/a=065012} {\bibfield  {journal}
  {\bibinfo  {journal} {Modelling and Simulation in Materials Science and
  Engineering}\ }\textbf {\bibinfo {volume} {22}},\ \bibinfo {pages} {065012}
  (\bibinfo {year} {2014})}\BibitemShut {NoStop}%
\bibitem [{\citenamefont {Bulatov}\ \emph {et~al.}(1998)\citenamefont
  {Bulatov}, \citenamefont {Abraham}, \citenamefont {Kubin}, \citenamefont
  {Devincre},\ and\ \citenamefont {Yip}}]{Bulatov:1998}%
  \BibitemOpen
  \bibfield  {author} {\bibinfo {author} {\bibfnamefont {V.}~\bibnamefont
  {Bulatov}}, \bibinfo {author} {\bibfnamefont {F.~F.}\ \bibnamefont
  {Abraham}}, \bibinfo {author} {\bibfnamefont {L.}~\bibnamefont {Kubin}},
  \bibinfo {author} {\bibfnamefont {B.}~\bibnamefont {Devincre}},\ and\
  \bibinfo {author} {\bibfnamefont {S.}~\bibnamefont {Yip}},\ }\bibfield
  {title} {\bibinfo {title} {Connecting atomistic and mesoscale simulations of
  crystal plasticity},\ }\href {https://doi.org/10.1038/35577} {\bibfield
  {journal} {\bibinfo  {journal} {Nature}\ }\textbf {\bibinfo {volume} {391}},\
  \bibinfo {pages} {669} (\bibinfo {year} {1998})}\BibitemShut {NoStop}%
\bibitem [{\citenamefont {Cai}\ \emph {et~al.}(2006)\citenamefont {Cai},
  \citenamefont {Arsenlis}, \citenamefont {Weinberger},\ and\ \citenamefont
  {Bulatov}}]{Cai:2006}%
  \BibitemOpen
  \bibfield  {author} {\bibinfo {author} {\bibfnamefont {W.}~\bibnamefont
  {Cai}}, \bibinfo {author} {\bibfnamefont {A.}~\bibnamefont {Arsenlis}},
  \bibinfo {author} {\bibfnamefont {C.}~\bibnamefont {Weinberger}},\ and\
  \bibinfo {author} {\bibfnamefont {V.}~\bibnamefont {Bulatov}},\ }\bibfield
  {title} {{\selectlanguage {english}\bibinfo {title} {A non-singular continuum
  theory of dislocations}},\ }\href
  {https://doi.org/10.1016/j.jmps.2005.09.005} {\bibfield  {journal} {\bibinfo
  {journal} {Journal of the Mechanics and Physics of Solids}\ }\textbf
  {\bibinfo {volume} {54}},\ \bibinfo {pages} {561} (\bibinfo {year}
  {2006})}\BibitemShut {NoStop}%
\bibitem [{\citenamefont {Arsenlis}\ \emph {et~al.}(2007)\citenamefont
  {Arsenlis}, \citenamefont {Cai}, \citenamefont {Tang}, \citenamefont {Rhee},
  \citenamefont {Oppelstrup}, \citenamefont {Hommes}, \citenamefont {Pierce},\
  and\ \citenamefont {Bulatov}}]{Arsenlis:2007}%
  \BibitemOpen
  \bibfield  {author} {\bibinfo {author} {\bibfnamefont {A.}~\bibnamefont
  {Arsenlis}}, \bibinfo {author} {\bibfnamefont {W.}~\bibnamefont {Cai}},
  \bibinfo {author} {\bibfnamefont {M.}~\bibnamefont {Tang}}, \bibinfo {author}
  {\bibfnamefont {M.}~\bibnamefont {Rhee}}, \bibinfo {author} {\bibfnamefont
  {T.}~\bibnamefont {Oppelstrup}}, \bibinfo {author} {\bibfnamefont
  {G.}~\bibnamefont {Hommes}}, \bibinfo {author} {\bibfnamefont {T.~G.}\
  \bibnamefont {Pierce}},\ and\ \bibinfo {author} {\bibfnamefont {V.~V.}\
  \bibnamefont {Bulatov}},\ }\bibfield  {title} {\bibinfo {title} {Enabling
  strain hardening simulations with dislocation dynamics},\ }\href
  {https://doi.org/10.1088/0965-0393/15/6/001} {\bibfield  {journal} {\bibinfo
  {journal} {Modelling and Simulation in Materials Science and Engineering}\
  }\textbf {\bibinfo {volume} {15}},\ \bibinfo {pages} {553} (\bibinfo {year}
  {2007})}\BibitemShut {NoStop}%
\bibitem [{\citenamefont {Queyreau}\ \emph {et~al.}(2014)\citenamefont
  {Queyreau}, \citenamefont {Marian}, \citenamefont {Wirth},\ and\
  \citenamefont {Arsenlis}}]{Queyreau:2014}%
  \BibitemOpen
  \bibfield  {author} {\bibinfo {author} {\bibfnamefont {S.}~\bibnamefont
  {Queyreau}}, \bibinfo {author} {\bibfnamefont {J.}~\bibnamefont {Marian}},
  \bibinfo {author} {\bibfnamefont {B.~D.}\ \bibnamefont {Wirth}},\ and\
  \bibinfo {author} {\bibfnamefont {A.}~\bibnamefont {Arsenlis}},\ }\bibfield
  {title} {\bibinfo {title} {Analytical integration of the forces induced by
  dislocations on a surface element},\ }\href
  {https://doi.org/10.1088/0965-0393/22/3/035004} {\bibfield  {journal}
  {\bibinfo  {journal} {Modelling and Simulation in Materials Science and
  Engineering}\ }\textbf {\bibinfo {volume} {22}},\ \bibinfo {pages} {035004}
  (\bibinfo {year} {2014})}\BibitemShut {NoStop}%
\bibitem [{\citenamefont {Queyreau}\ \emph {et~al.}(2020)\citenamefont
  {Queyreau}, \citenamefont {Hoang}, \citenamefont {Shi}, \citenamefont
  {Aubry},\ and\ \citenamefont {Arsenlis}}]{Queyreau:2020}%
  \BibitemOpen
  \bibfield  {author} {\bibinfo {author} {\bibfnamefont {S.}~\bibnamefont
  {Queyreau}}, \bibinfo {author} {\bibfnamefont {K.}~\bibnamefont {Hoang}},
  \bibinfo {author} {\bibfnamefont {X.}~\bibnamefont {Shi}}, \bibinfo {author}
  {\bibfnamefont {S.}~\bibnamefont {Aubry}},\ and\ \bibinfo {author}
  {\bibfnamefont {A.}~\bibnamefont {Arsenlis}},\ }\bibfield  {title} {\bibinfo
  {title} {Analytical integration of the tractions induced by non-singular
  dislocations on an arbitrary shaped triangular quadratic element},\ }\href
  {https://doi.org/10.1088/1361-651x/aba736} {\bibfield  {journal} {\bibinfo
  {journal} {Modelling and Simulation in Materials Science and Engineering}\
  }\textbf {\bibinfo {volume} {28}},\ \bibinfo {pages} {075001} (\bibinfo
  {year} {2020})}\BibitemShut {NoStop}%
\bibitem [{\citenamefont {Queyreau}\ and\ \citenamefont
  {Devincre}(2009)}]{Queyreau:2009lr}%
  \BibitemOpen
  \bibfield  {author} {\bibinfo {author} {\bibfnamefont {S.}~\bibnamefont
  {Queyreau}}\ and\ \bibinfo {author} {\bibfnamefont {B.}~\bibnamefont
  {Devincre}},\ }\bibfield  {title} {\bibinfo {title} {Bauschinger effect in
  precipitation-strengthened materials: A dislocation dynamics investigation},\
  }\href {http://www.informaworld.com/10.1080/09500830903005433} {\bibfield
  {journal} {\bibinfo  {journal} {Philosophical Magazine Letters}\ }\textbf
  {\bibinfo {volume} {89}},\ \bibinfo {pages} {419} (\bibinfo {year}
  {2009})}\BibitemShut {NoStop}%
\bibitem [{\citenamefont {Wickham}\ \emph {et~al.}(1999)\citenamefont
  {Wickham}, \citenamefont {Schwarz},\ and\ \citenamefont
  {St\"olken}}]{Wickham:1999}%
  \BibitemOpen
  \bibfield  {author} {\bibinfo {author} {\bibfnamefont {L.~K.}\ \bibnamefont
  {Wickham}}, \bibinfo {author} {\bibfnamefont {K.~W.}\ \bibnamefont
  {Schwarz}},\ and\ \bibinfo {author} {\bibfnamefont {J.~S.}\ \bibnamefont
  {St\"olken}},\ }\bibfield  {title} {\bibinfo {title} {Rules for forest
  interactions between dislocations},\ }\href
  {https://doi.org/10.1103/PhysRevLett.83.4574} {\bibfield  {journal} {\bibinfo
   {journal} {Phys. Rev. Lett.}\ }\textbf {\bibinfo {volume} {83}},\ \bibinfo
  {pages} {4574} (\bibinfo {year} {1999})}\BibitemShut {NoStop}%
\bibitem [{\citenamefont {Madec}\ \emph {et~al.}(2002)\citenamefont {Madec},
  \citenamefont {Devincre},\ and\ \citenamefont {Kubin}}]{Madec:2002}%
  \BibitemOpen
  \bibfield  {author} {\bibinfo {author} {\bibfnamefont {R.}~\bibnamefont
  {Madec}}, \bibinfo {author} {\bibfnamefont {B.}~\bibnamefont {Devincre}},\
  and\ \bibinfo {author} {\bibfnamefont {L.~P.}\ \bibnamefont {Kubin}},\
  }\bibfield  {title} {\bibinfo {title} {From dislocation junctions to forest
  hardening},\ }\href {https://doi.org/10.1103/PhysRevLett.89.255508}
  {\bibfield  {journal} {\bibinfo  {journal} {Phys. Rev. Lett.}\ }\textbf
  {\bibinfo {volume} {89}},\ \bibinfo {pages} {255508} (\bibinfo {year}
  {2002})}\BibitemShut {NoStop}%
\bibitem [{\citenamefont {Madec}\ and\ \citenamefont
  {Kubin}(2004)}]{Madec:2004}%
  \BibitemOpen
  \bibfield  {author} {\bibinfo {author} {\bibfnamefont {R.}~\bibnamefont
  {Madec}}\ and\ \bibinfo {author} {\bibfnamefont {L.~P.}\ \bibnamefont
  {Kubin}},\ }\bibfield  {title} {{\selectlanguage {english}\bibinfo {title}
  {Dislocation {Interactions} and {Symmetries} in {BCC} {Crystals}}},\ }in\
  \href {https://doi.org/10.1007/978-1-4020-2111-4_7} {{\selectlanguage
  {english}\emph {\bibinfo {booktitle} {{IUTAM} {Symposium} on {Mesoscopic}
  {Dynamics} of {Fracture} {Process} and {Materials} {Strength}}}}},\ Vol.\
  \bibinfo {volume} {115},\ \bibinfo {editor} {edited by\ \bibinfo {editor}
  {\bibfnamefont {G.~M.~L.}\ \bibnamefont {Gladwell}}, \bibinfo {editor}
  {\bibfnamefont {H.}~\bibnamefont {Kitagawa}},\ and\ \bibinfo {editor}
  {\bibfnamefont {Y.}~\bibnamefont {Shibutani}}}\ (\bibinfo  {publisher}
  {Springer Netherlands},\ \bibinfo {address} {Dordrecht},\ \bibinfo {year}
  {2004})\ pp.\ \bibinfo {pages} {69--78},\ \bibinfo {note} {series Title:
  Solid Mechanics and its Applications}\BibitemShut {NoStop}%
\bibitem [{\citenamefont {Rodney}\ and\ \citenamefont
  {Phillips}(1999)}]{Rodney:1999}%
  \BibitemOpen
  \bibfield  {author} {\bibinfo {author} {\bibfnamefont {D.}~\bibnamefont
  {Rodney}}\ and\ \bibinfo {author} {\bibfnamefont {R.}~\bibnamefont
  {Phillips}},\ }\bibfield  {title} {{\selectlanguage {english}\bibinfo {title}
  {Structure and {Strength} of {Dislocation} {Junctions}: {An} {Atomic} {Level}
  {Analysis}}},\ }\href {https://doi.org/10.1103/PhysRevLett.82.1704}
  {\bibfield  {journal} {\bibinfo  {journal} {Physical Review Letters}\
  }\textbf {\bibinfo {volume} {82}},\ \bibinfo {pages} {1704} (\bibinfo {year}
  {1999})}\BibitemShut {NoStop}%
\bibitem [{\citenamefont {Kubin}\ \emph {et~al.}(2008)\citenamefont {Kubin},
  \citenamefont {Devincre},\ and\ \citenamefont {Hoc}}]{Kubin:2008fk}%
  \BibitemOpen
  \bibfield  {author} {\bibinfo {author} {\bibfnamefont {L.}~\bibnamefont
  {Kubin}}, \bibinfo {author} {\bibfnamefont {B.}~\bibnamefont {Devincre}},\
  and\ \bibinfo {author} {\bibfnamefont {T.}~\bibnamefont {Hoc}},\ }\bibfield
  {title} {\bibinfo {title} {Modeling dislocation storage rates and mean free
  paths in face-centered cubic crystals},\ }\href
  {http://www.sciencedirect.com/science/article/B6TW8-4TK1TRH-1/2/36a61f402fbf44e8b6d3548be10fb752}
  {\bibfield  {journal} {\bibinfo  {journal} {Acta Materialia}\ }\textbf
  {\bibinfo {volume} {56}},\ \bibinfo {pages} {6040} (\bibinfo {year}
  {2008})}\BibitemShut {NoStop}%
\bibitem [{\citenamefont {Devincre}\ \emph {et~al.}(2006)\citenamefont
  {Devincre}, \citenamefont {Kubin},\ and\ \citenamefont
  {Hoc}}]{Devincre:2006uc}%
  \BibitemOpen
  \bibfield  {author} {\bibinfo {author} {\bibfnamefont {B.}~\bibnamefont
  {Devincre}}, \bibinfo {author} {\bibfnamefont {L.}~\bibnamefont {Kubin}},\
  and\ \bibinfo {author} {\bibfnamefont {T.}~\bibnamefont {Hoc}},\ }\bibfield
  {title} {\bibinfo {title} {Physical analyses of crystal plasticity by dd
  simulations},\ }\href@noop {} {\bibfield  {journal} {\bibinfo  {journal}
  {Scripta Materialia}\ }\textbf {\bibinfo {volume} {54}},\ \bibinfo {pages}
  {741} (\bibinfo {year} {2006})}\BibitemShut {NoStop}%
\bibitem [{\citenamefont {Queyreau}\ \emph {et~al.}(2009)\citenamefont
  {Queyreau}, \citenamefont {Monnet},\ and\ \citenamefont
  {Devincre}}]{Queyreau:09}%
  \BibitemOpen
  \bibfield  {author} {\bibinfo {author} {\bibfnamefont {S.}~\bibnamefont
  {Queyreau}}, \bibinfo {author} {\bibfnamefont {G.}~\bibnamefont {Monnet}},\
  and\ \bibinfo {author} {\bibfnamefont {B.}~\bibnamefont {Devincre}},\
  }\bibfield  {title} {\bibinfo {title} {Slip systems interactions in
  {$[$}alpha{$]$}-iron determined by dislocation dynamics simulations},\ }\href
  {http://www.sciencedirect.com/science/article/B6TWX-4RP0MK0-1/2/48d7b229150c25ce6d7976b1ee9733a9}
  {\bibfield  {journal} {\bibinfo  {journal} {International Journal of
  Plasticity}\ }\textbf {\bibinfo {volume} {25}},\ \bibinfo {pages} {361}
  (\bibinfo {year} {2009})}\BibitemShut {NoStop}%
\bibitem [{\citenamefont {Madec}\ and\ \citenamefont
  {Kubin}(2017)}]{Madec:2017kx}%
  \BibitemOpen
  \bibfield  {author} {\bibinfo {author} {\bibfnamefont {R.}~\bibnamefont
  {Madec}}\ and\ \bibinfo {author} {\bibfnamefont {L.~P.}\ \bibnamefont
  {Kubin}},\ }\bibfield  {title} {\bibinfo {title} {Dislocation strengthening
  in fcc metals and in bcc metals at high temperatures},\ }\href
  {https://doi.org/http://dx.doi.org/10.1016/j.actamat.2016.12.040} {\bibfield
  {journal} {\bibinfo  {journal} {Acta Materialia}\ }\textbf {\bibinfo {volume}
  {126}},\ \bibinfo {pages} {166} (\bibinfo {year} {2017})}\BibitemShut
  {NoStop}%
\bibitem [{\citenamefont {Devincre}\ and\ \citenamefont
  {Kubin}(2010)}]{Devincre:2010fk}%
  \BibitemOpen
  \bibfield  {author} {\bibinfo {author} {\bibfnamefont {B.}~\bibnamefont
  {Devincre}}\ and\ \bibinfo {author} {\bibfnamefont {L.}~\bibnamefont
  {Kubin}},\ }\bibfield  {title} {\bibinfo {title} {Scale transitions in
  crystal plasticity by dislocation dynamics simulations},\ }\bibfield
  {booktitle} {\emph {\bibinfo {booktitle} {Computational metallurgy and scale
  transitions}},\ }\href
  {http://www.sciencedirect.com/science/article/B6X19-50W11S7-2/2/1b092e299425c344236eb4ca12b9ee7f}
  {\bibfield  {journal} {\bibinfo  {journal} {Comptes Rendus Physique}\
  }\textbf {\bibinfo {volume} {11}},\ \bibinfo {pages} {274} (\bibinfo {year}
  {2010})}\BibitemShut {NoStop}%
\bibitem [{\citenamefont {Daniel}\ and\ \citenamefont
  {Horne}(1971)}]{Daniel:1971}%
  \BibitemOpen
  \bibfield  {author} {\bibinfo {author} {\bibfnamefont {R.~C.}\ \bibnamefont
  {Daniel}}\ and\ \bibinfo {author} {\bibfnamefont {G.~T.}\ \bibnamefont
  {Horne}},\ }\bibfield  {title} {{\selectlanguage {english}\bibinfo {title}
  {The {Bauschinger} effect and cyclic hardening in copper}},\ }\href
  {https://doi.org/10.1007/BF02664248} {\bibfield  {journal} {\bibinfo
  {journal} {Metallurgical Transactions}\ }\textbf {\bibinfo {volume} {2}},\
  \bibinfo {pages} {1161} (\bibinfo {year} {1971})}\BibitemShut {NoStop}%
\bibitem [{\citenamefont {Mughrabi}(2010)}]{mughrabi:2010}%
  \BibitemOpen
  \bibfield  {author} {\bibinfo {author} {\bibfnamefont {H.}~\bibnamefont
  {Mughrabi}},\ }\bibfield  {title} {{\selectlanguage {english}\bibinfo {title}
  {Fatigue, an everlasting materials problem - still en vogue}},\ }\href
  {https://doi.org/10.1016/j.proeng.2010.03.003} {\bibfield  {journal}
  {\bibinfo  {journal} {Procedia Engineering}\ }\textbf {\bibinfo {volume}
  {2}},\ \bibinfo {pages} {3} (\bibinfo {year} {2010})}\BibitemShut {NoStop}%
\bibitem [{\citenamefont {Queyreau}\ and\ \citenamefont
  {Devincre}(2020)}]{queyreau:2021}%
  \BibitemOpen
  \bibfield  {author} {\bibinfo {author} {\bibfnamefont {S.}~\bibnamefont
  {Queyreau}}\ and\ \bibinfo {author} {\bibfnamefont {B.}~\bibnamefont
  {Devincre}},\ }\bibfield  {title} {\bibinfo {title} {{On the Origins of
  Tension--Compression Asymmetry in Crystals and Implications for Cyclic
  Behavior}}} (\bibinfo {year} {2020}),\ \bibinfo {note} {working paper or
  preprint}\BibitemShut {NoStop}%
\bibitem [{\citenamefont {Kemsley}\ and\ \citenamefont
  {Paterson}(1960)}]{Kemsley:1960}%
  \BibitemOpen
  \bibfield  {author} {\bibinfo {author} {\bibfnamefont {D.}~\bibnamefont
  {Kemsley}}\ and\ \bibinfo {author} {\bibfnamefont {M.}~\bibnamefont
  {Paterson}},\ }\bibfield  {title} {{\selectlanguage {english}\bibinfo {title}
  {The influence of strain amplitude on the work hardening of copper crystals
  in alternating tension and compression}},\ }\href
  {https://doi.org/10.1016/0001-6160(60)90032-8} {\bibfield  {journal}
  {\bibinfo  {journal} {Acta Metallurgica}\ }\textbf {\bibinfo {volume} {8}},\
  \bibinfo {pages} {453} (\bibinfo {year} {1960})}\BibitemShut {NoStop}%
\end{thebibliography}%

\end{document}